\documentclass{aa}  
\usepackage{graphicx}
	
\usepackage{txfonts}
\usepackage{xcolor}
\usepackage{mathrsfs}  
\usepackage{amssymb, amsmath, amsbsy}
\usepackage{longtable}
\usepackage{multirow}
\usepackage[T1]{fontenc}
\usepackage{currvita}
\usepackage{tablefootnote}
\usepackage{hyperref}
\usepackage{soul}

\def\ha{H$\alpha$}
\def\hb{H$\beta$}

\def\l5100{L$_{\rm 5100\AA}$}
\def\mbh{$M\mathrm{_{BH}}$}
\def\kms{\,km\,s$^{-1}$}

\newenvironment{myfont}{\fontfamily{lmtt}\selectfont}{\par}
\DeclareTextFontCommand{\textmyfont}{\myfont}
\begin{document}

\title{Spectropolarimetry and spectral decomposition of high-accreting Narrow Line Seyfert 1 galaxies\thanks{Based on data collected at ESO under programme 098.B-0426(B)}}

\titlerunning{Spectropolarimetry of high-accreting NLSy1s}
\authorrunning{M. \'Sniegowska, S. Panda, B. Czerny et al.}

\author{Marzena {\'S}niegowska\thanks{msniegowska@camk.edu.pl}\inst{1,2,3} \and
Swayamtrupta Panda\inst{2,1,4,}\thanks{CNPq Fellow, spanda@lna.br} \and
Bo{\.z}ena Czerny\inst{2} \and\\
%\vspace{0.25cm}
\DJ orge Savi\'c\inst{5,6,7} \and
Mary Loli Mart\'{\i}nez-Aldama\inst{2,8,9} \and
Paola Marziani\inst{10} \and
Jian-Min Wang\inst{11,12,13} \and
Pu Du\inst{11} \and
Luka \v{C}. Popovi\'c\inst{5,14} \and
Chandra Shekhar Saraf\inst{1}
}

\institute{Nicolaus Copernicus Astronomical Center, Polish Academy of Sciences, ul. Bartycka 18, 00-716 Warsaw, Poland
    \and Center for Theoretical Physics, Polish Academy of Sciences, Al. Lotnik\'ow 32/46, 02-668 Warsaw, Poland
    \and School of Physics and Astronomy, Tel Aviv University, Tel Aviv 69978, Israel
    \and Laborat\'orio Nacional de Astrof\'isica - MCTI, R. dos Estados Unidos, 154 - Na\c{c}\~oes, Itajub\'a - MG, 37504-364, Brazil
    \and Astronomical Observatory Belgrade Volgina 7, P.O. Box 74 11060, Belgrade, 11060, Serbia
    \and Universit\'e de Strasbourg, CNRS, Observatoire Astronomique de Strasbourg, UMR 7550, 11 rue de l’Universit\'e, F-67000 Strasbourg, France 
    \and Institut d’Astrophysique et de Géophysique, Université de Liège, Allée du 6 Août 19c, 4000 Liège, Belgium
    \and Departamento de Astronomia, Universidad de Chile, Camino del Observatorio 1515, Santiago, Chile
    \and Instituto de Fisica y Astronomía, Facultad de Ciencias, Universidad de Valparaíso, Gran Bretaña 1111, Valparaiso, Chile
    \and Istituto Nazionale di Astrofisica (INAF), Osservatorio Astronomico di Padova, 35122 Padova, Italy
    \and  Key Laboratory for Particle Astrophysics, Institute of High Energy Physics, Chinese Academy of Sciences, 19B Yuquan Road, Beijing 100049, People’s Republic of China
    \and School of Astronomy and Space Sciences, University of Chinese Academy of Sciences, 19A Yuquan Road, Beijing 100049, China
    \and National Astronomical Observatories of China, Chinese Academy of Sciences, 20A Datun Road, Beijing 100020, China
    \and Department of Astronomy, Faculty of Mathematics, University of Belgrade, Studentski trg 16, 11000 Belgrade, Serbia \\}

   %\date{Received September 15, 1996; accepted March 16, 1997}

% \abstract{}{}{}{}{} 
% 5 {} token are mandatory
 
  \abstract
   {Narrow-line Seyfert 1 (NLSy1) galaxies have been shown to have high Eddington ratios and relatively small black hole mass. The measurement of these black hole masses is based on the virial relation that is dependent on the distribution of the line-emitting gas and the viewing angle to the source. Spectropolarimetry enables us to probe the geometry of this line-emitting gas and independently estimate the viewing angle of the source by comparing the spectrum viewed under natural light and polarized light.}
  {We aim to (i) estimate the virial factor using the viewing angles inferred from spectropolarimetric measurements for a sample of NLSy1s which influences the measurement of the black hole masses; (ii) model the natural and polarized spectra around the \ha{} region using spectral decomposition and spectral fitting techniques; (iii) infer the physical conditions (e.g., density and optical depth) of the broad-line region and the scattering medium responsible for the polarization of the \ha{} emission line (and continuum); and (iv) model the Stokes parameters using the polarization radiative transfer code {\sc STOKES}.}
   {Using the FORS2 instrument at the European Southern Observatory's (ESO) Very Large Telescope, We performed spectropolarimetric observations of three NLSy1: Mrk 1044, SDSS J080101.41+184840.7, and IRAS 04416+1215. We used the {\sc ESO Reflex} workflow to perform a standard data reduction and extract the natural and polarized spectra. We then modeled the \ha{} region in the reduced spectra using {\sc IRAF} spectral fitting procedures and estimated the Stokes parameters and the viewing angles of the three sources. We modeled the Stokes parameters, inferred the properties of the scattering media located in the equatorial and polar regions, and simulated the spectra observed both in natural light and in polarized light using the polarization radiative transfer code {\sc STOKES}. }
   {The viewing angles recovered for the three sources indicate that they occupy separate locations in the viewing angle plane, from an almost face-on (IRAS 04416+1215) to an intermediate (SDSS J080101.41+184840.7), to a highly inclined (Mrk 1044) orientation. Nevertheless, we confirm that all three sources are high Eddington ratio objects. We were successful in recovering the observed \ha{} line profile in both the natural and polarized light using the {\sc STOKES} modeling. We recovered the polarization fractions of the order of 0.2-0.5\% for the three sources although the recovery of the phase angle is sub-optimal, mainly due to the noise in the observed data. Our principal component analysis shows that the sample of 25 sources,  collected from the literature and including our sources,
   %, which include our sources, Fairall 9 from \citet{2021bowei}, and sources from \citet{2021capetti}
   are mainly driven by the black hole mass and Eddington ratio. We reaffirm the connection of the strength of the optical FeII emission with the Eddington ratio, but the dependence on the viewing angle is moderate and resembles more of a secondary effect.}
   {}


   \maketitle
%
%-------------------------------------------------------------------

\section{Introduction}
\label{sec:intro}

Narrow-line Seyfert 1 (NLSy1) galaxies and quasars with relatively narrow permitted lines are generally believed to have relatively high Eddington ratios \citep[e.g.,][]{mathur2000, wangNetzer2003, 2004grupe}. The NLSy1 galaxies are contextualized as a sub-population within a larger population of sources accreting at relatively high rates (Eddington ratio, $\lambda_{\rm Edd} >$ 0.2 for Population A, following \citealt{2000sulentic}). However, uncertainty as to  how high these ratios can be remains. This is an important issue from the theoretical point of view, as we have a basic understanding of the accretion when the Eddington ratio is moderate, but the models are unreliable at the extremely high accretion rates. Moreover, strong outflows might, in principle, prevent too high accretion rates from happening in active galactic nuclei (AGN). Therefore, the study of sources considered as highly super Eddington is of extreme importance.

The determination of the Eddington ratio requires both the measurement of the bolometric luminosity of a source and its black hole mass. Currently, the reverberation mapping method is considered the most reliable method for black hole mass measurement \citep{Peterson_1993,peterson2004,cackett2021}. It measures the profiles of the broad emission lines coming from the broad-line region (BLR) and the delay between the variable continuum and the line response, which constrains the velocity and the BLR size. The method is based on the assumption that the BLR is predominantly in Keplerian motion, and the results generally agree with the results based on the black hole mass - bulge velocity relation \citep[see e.g.,][]{ferrarese_merritt2000,gultekin+09}, which supports the underlying assumption.
However, some objects with an \hb{} full width at half maximum (FWHM) of about 2000 km s$^{-1}$ show \hb{} lags much shorter than the well-known radius-luminosity relation \citep{bentz2013}  when we compared objects with the same luminosity \citep{dupu2014,wang2014,dupu2015,dupu2016_alt,dupu2018}. This implies a surprisingly small black hole mass and, for a measured bolometric luminosity, a very high Eddington ratio. 

Such high Eddington objects challenge the theoretical models of the accretion process, and they are important for understanding the rapid growth of the black hole mass at high redshifts. Theoretically, these objects should be modeled with slim accretion disks \citep{abramowicz1988}. The innermost part of such a disk is geometrically puffed up, and a significant fraction of the radiation is expected to be released in a collimating funnel. In such a picture, the irradiation of the outer parts of the disk, where the BLR is located, is geometrically less efficient \citep{wang_shield2014}. Thus, the BLR may be located elsewhere, such as in the biconical flow \citep{corbett2000}. If so, the BLR we observe would not be in Keplerian
motion, and the black hole mass measurement may be highly biased. In addition, in extreme cases of a source seen along the symmetry axis, the lines can also be much narrower due to a purely geometrical factor \citep[e.g.,][]{baldi2016}. A comparison of black hole masses estimated using the reverberation mapping measurements with the masses estimated from the stellar dispersion may not give the final agreement. Due to an evolutionary effect, NLSy1 galaxies are sometimes argued to be outliers from the black hole mass-stellar dispersion relation. 
Their black holes are too small for their bulge masses, and they accrete vigorously, increasing mass and moving toward the standard relation appropriate for mature AGN \citep[e.g.,][]{mathur2000,mathur2001}. {\citet{Robinson2011ASPC..449..431R}, with the spectropolarimetric measurements for 16 NLSy1, concluded that NLSy1 galaxies represent an extreme realization of one or more physical parameters, such as black hole mass and/or accretion rate, rather than simply being preferentially oriented close to face-on viewing angles. Thus, NLSy1 galaxies are not simply a sub-population that is preferentially viewed close to the AGN symmetry axis, and hence, the orientation of the accretion disk cannot be the main parameter governing the broad-line widths in these sources.

Using spectropolarimetry, {it is possible to} obtain new and independent insight into the geometry and full velocity field of the line-emitting material as well as into the location of the fully ionized scattering medium. Spectropolarimetry revealed the nature of type 2 AGN as active galaxies with the BLR hidden by the dusty-molecular torus \citep{antonucci1985}. Moreover, \citet{smith2004} and \citet{smith2005} later showed the advantage of using this technique to study type 1 AGN as well. The
wavelength-dependent polarization angle also indicates whether the scattering takes place in the polar or the equatorial region and measures the viewing angle of the system. In low viewing angle sources, emission lines in
polarized light are much broader than in the unpolarized spectrum. Even simply comparing the kinematic line width in total and polarized light can verify the statement about the actual value of the black hole mass and the Eddington ratio of a source \citep{baldi2016}. A new independent method of black hole mass measurement has been introduced, and it is based on the change of the polarization angle across the broad emission line. 
This change depends on the rotational velocity in the BLR, and it allows for direct measurement of the black hole mass \citep{AfaPop2015}.
The method was proposed by \citet{Afanasiev2014,Afanasiev2015} and applied to mostly \ha-line based observations \citep[e.g.,][]{AfaPop2015,Afanasiev2015,2019afanasiev}
%\textcolor{green}{all of them applied this method???} 
using the spectropolarimetry data from the 
%6m SAO RAS telescope, or 
Very Large Telescope (VLT)/FORS2. {Meanwhile,} \citet{2021bowei} has shown that for Fairall 9, the mass measurements from \ha{} and \hb{} lines are consistent.

The first spectropolarimetric black hole mass measurement with the MgII line using the rotation of the polarization angle was recently performed by \citet{2021savic}. Their result is in good agreement with other black hole mass estimation techniques \citep[see][]{AfaPop2015,2019afanasiev}.

We present the spectropolarimetric measurements and spectral decomposition of the \ha{} spectral region of three NLSy1 galaxies, Mrk 1044, SDSS J080101.41+184840.7, and IRAS 04416+1215,\footnote{Also known as SDSS J044428.77+122111.7} performed with the VLT/FORS2 instrument. We also show the spectral decomposition of the archival spectra for these three sources in \hb{} and \ha{} regions.  %\textcolor{green}{Furthermore, } 
All three sources are part of the  super-Eddington accreting massive black holes (SEAMBH) Project sample \citep{wang2014}, selected for reverberation monitoring as candidates for super-Eddington accreting sources.
%The object has been monitored by \citet{wang2014}. 
In the following paragraphs, we briefly describe the archival information of the three chosen sources.

The source Mrk 1044 ($z = 0.016451 \pm 0.000037$; RA 02:30:05.5, DEC -08:59:53 from NED\footnote{\href{https://ned.ipac.caltech.edu/}{https://ned.ipac.caltech.edu/}})
is one of the nearest and brightest Seyfert galaxies \citep[V mag = 14.5, from NED;][]{veronCetty2006}.
The line width of the Balmer lines was already measured by \citet{rafanelli1983}. The authors reported an FWHM of 3600 km s$^{-1}$ for the broad components (BCs)\ of both the \ha{} and \hb{} lines. They stressed the comparable intensity of the narrow components (NCs), with an FWHM width of 1000 km s$^{-1}$, and the very low value of the [OIII]/\hb{} ratio (below 0.6). The source was later reclassified as an NLSy1 galaxy \citep{osterbrock1985}. 
This source was also observed in X-rays, and the viewing angle obtained by \citet{mallick2018} is $i$ = $47.2^{+1.0}_{-2.5}$, from the joint fitting of Swift, XMM-Newton, and NuSTAR X-ray spectra.  

The source SDSS J080101.41+184840.7 ($z = 0.13954 \pm 0.00001$; RA 08:01:01.41, DEC +18:48:40.78 from NED)  with V mag = 16.88, from \citet{2010A&A...518A..10V}, is the second object in our sample.
\citet{Liu_Henzen2021} modeled XMM-Newton spectrum of SDSS J080101.41+184840.7 and obtained a steep photon index (2.33 $\pm$0.06) using a power-law model modified by Galactic absorption. However, in the model used by \citet{Liu_Henzen2021}, the viewing angle was not one of the model parameters.

\cite{tortosa_etal_2022} recently reported a detailed analysis of the broadband observations of IRAS 04416+1215 (r mag = 16.24, z = 0.089). This source shows a narrow \hb{} line \citep[FWHM = 1670 \kms{};][]{moran_etal_1996} and a very broad [OIII] lines \citep[FWHM = 1150 \kms{};][]{veron_cetty_2001}.
In their work, \citet{tortosa_etal_2022} showed that the best-fitting model in the X-ray band is composed of a soft excess, three ionized outflows, neutral absorption, and a reflection component. Prominent soft X-ray excess supports the claim for a super-Eddington accretion rate in this source. Through a dedicated monitoring campaign using the Lijiang 2.4m telescope under the SEAMBH Project, \citet{dupu2016_alt} reported the measurements of the \hb{} FWHMs for the three sources (see Table 5 in their paper):  Mrk 1044 (1178$\pm$22 \kms{}), SDSS J080101.41+184840.7 (1930$\pm$18 \kms{}), and IRAS 04416+1215 (1522$\pm$44 \kms{}).

This paper is organized as follows:
Details on the performed observations and data reduction are reported in Section \ref{sec:reduction}. In Section \ref{sec:dataanalysis}, we present the analysis of spectropolarimetric data obtained using
the VLT and the archival spectroscopic data for each source. In Section \ref{sec:results}, we show the spectral decomposition for natural and polarized light for our objects and spectral modeling. Next, we perform polarization radiative transfer modeling using {\sc STOKES} and compare it with our observed results in Section \ref{sec:STOKES}. We discuss the implications of these results, outline the physical parameters responsible for the trends observed in our sample using principal component analysis, and summarize our findings from this study in Section \ref{sec:discussion}.
In this work, we use $\Lambda$CDM cosmology ($\Omega_{\Lambda}$ = 0.7, $\Omega_{\mathrm M}$ = 0.3, $H_{0}$ = 70 km s$^{-1}$ Mpc$^{-1}$).

\section{Observations and data reduction}\label{sec:reduction}

Our spectropolarimetric observations were performed with the FORS2 instrument mounted on the UT1 telescope of the 8.2m ESO VLT. The observations were taken using the GRISM-300I in combination with the blocking filter OG590. The spectral range of our data is 6000 - 10000\AA. The 0.7-arcsecond wide slit was oriented along the parallactic angle, and the multi-object spectropolarimetry observations were performed with a $2048\times 2048$ pixel CCD with a spatial resolution of 0.126 arcsec pixel$^{-1}$.

The spectropolarimetric observation of Mrk 1044 was accomplished on 2016 October 27. The source SDSS J080101.41+184840.7 was observed on 2016 December 26, 2016 December 30, and 2017 January 02. Finally, the IRAS 04416+1215 was observed on 2016 November 06. For all observational runs, we used the \citet{2006patat} prescription of observing with four quarter-wave plate angles (0, 22.5, 45, 67.5 deg.). For Mrk 1044, two exposures were taken at each angle ($4\times 2$ spectra), with each exposure lasting 206 s. For IRAS 04416+1215, three exposures lasting 455 s were taken at each angle. For SDSS J080101.41+184840.7, three exposures of 250 s each were taken on 2017 January 02, and three exposures of 280 s each were taken during the observations in December 2016. 

Together with the sources, two standard stars were observed: one star that is highly polarized and one star that is unpolarized (Ve6-23 and HD62499, respectively). Observations were performed in the service mode (program ID: 098.B-0426(B), PI: B. Czerny), and observing conditions were good, with an atmospheric seeing $\sim 0.8$. Since only one unpolarized star was observed and it was not located close to any of our sources in the sky, we decided not to subtract the effect of the interstellar medium from our measurements. This is certainly justified for Mrk 1044 and SDSS J080101.41+184840.7, although not quite for IRAS 04416+1215 (see Appendix~\ref{sec:contamination}).

To reduce data, we used {\sc ESO Reflex}\footnote{\href{https://www.eso.org/sci/software/esoreflex/}{https://www.eso.org/sci/software/esoreflex/}}  \citep{2008hook}. The workflow combines the bias frames into a master bias that is then subtracted from the science image. The same procedure is performed with the lamp flats, and the science image is flat-fielded. The workflow then removes the cosmic ray events and performs the wavelength calibration using the standard He-Ar arc lamps. As final products, among others, we obtained the extracted 1D spectra with the Stokes parameters ($U$ and $Q$), total linear polarization ($L$), and flux in ADU/s ($I$) as a function of wavelength with the associated uncertainties. 

A comparison between the natural light spectrum and the one from the SDSS catalog showed that the AGN continuum form was different in both spectra. The same effect was observed in the reduced standard stars obtained with {\sc ESO Reflex}, where an expected blackbody continuum was not recovered. This difference suggests that one correction was not applied by the pipeline. To correct the continuum form in the natural light spectrum, we used the IRAF routines ``standard'' and ``sensfunc'', and the standard star HD~64299. After the correction, a good agreement was obtained between the natural light spectrum and the one from the SDSS catalog.

\section{Data analysis} \label{sec:dataanalysis}

\subsection{Standard star}

In Table \ref{tab:standars-stars}, we present the measurements for the standard stars HD 64299 and Ve 6-23, used in our work. \cite{2017MNRAS.464.4146C} provide measurements for Ve 6-23\footnote{Ve 6-23 in \cite{2017MNRAS.464.4146C} is marked as Vela1 95 in Table 1 in \cite{2017MNRAS.464.4146C}.} between 7.20\% and 7.03\% for the I band. In our case, we obtained 7.12\%, which is comparable to the average value of Ve 6-23 polarization (7.13\%) from \cite{2017MNRAS.464.4146C}. \citet{10.1117/1.JATIS.5.2.028002} obtained a 6.23$\pm$ 0.03\% polarization level for this star, but observations were performed with a different instrument.\footnote{The instrument was CasPol, a dual-beam polarimeter mounted at the 2.15-m Jorge Sahade Telescope, located at the Complejo Astronómico El Leoncito, Argentina.} The polarization angle for Ve 6-23 is 170, which is consistent with the angle \citet{2017MNRAS.464.4146C} calculated (171.95$\pm$0.01). In the case of HD 64299, \citet{10.1117/1.JATIS.5.2.028002} performed the measurements for this star in the I band and obtained $P$ = 0.16 $\pm$ 0.01\% (Table 3 therein). Our measurements for this star, $P$ = 0.018\%, differ from \citet{10.1117/1.JATIS.5.2.028002}, which may be connected with different CasPol instrumental polarization in the I band. \citet{10.1117/1.JATIS.5.2.028002} report the inconsistent measurements in the I band between CasPol literature. However, results obtained by CasPol in B and V bands, seem to be consistent with the ones from the literature, including FORS2, \citep[for details see][]{10.1117/1.JATIS.5.2.028002}. {The instrumental polarization is small, and the standard measurements were reproduced within errors, in particular the angle that depends on the half-wave plate (HWP) achromaticity.}

\begin{table*}[h]
\center
\caption{Observed standard stars. The star type is indicated as ``Pol.'' if the star is polarized or as ``unPol.'' if it is unpolarized, {following} \cite{2017MNRAS.464.4146C}. }
\label{tab:standars-stars}
\begin{tabular}{@{}lcccccc@{}}
%\toprule
\hline
\hline

  NAME & RA & DEC & $P $& $\chi$ & $Q$ & $U$\\ 
     & [J2000] & [J2000] & [\%] & [$^{\circ}$] & [\%] & [\%]\\ \hline
HD64299 & 07:52:25.4 & -23:17:47.5   & 0.02$\pm$0.02 & 89$\pm$66  & -0.017$\pm$0.009 & 0.006$\pm$0.007   \\
    (unPol.) & &  &  &  &  & \\ 
Ve6-23 & 	09:06:00.01 & -47:18:58.2  & 7.12$\pm$0.03 & 170$\pm$1 & 6.84$\pm$0.06 & -1.99$\pm$0.04\\ 
     (Pol.) & &  &  &  &  & \\ \hline
\end{tabular}
\tablefoot{Columns show the name of the source, the coordinates (taken from the \href{http://SIMBAD.cds.unistra.fr/SIMBAD/}{SIMBAD Astronomical Database}), the averaged {degree of polarization ($P$)}, the averaged {polarization angle ($\chi$)}, and the averaged Stokes parameters: $Q$ and $U$.}
\end{table*}

\subsection{Spectropolarimetry of NLSy1}
\label{sec:spectropol-corrections}

In this work, we present three types of polarization measurements: the mean polarization of each source, the polarization of the continuum, and the polarization of the line corrected for continuum polarization.
By a mean polarization ($P_{mean}$), we refer to the polarization calculated from the mean $Q$ and $U$ of the whole wavelength range of the spectra. As a continuum polarization ($P_{cont}$), we consider the polarization calculated from the mean $Q$ and $U$ at two windows surrounding \ha{} from its blue and red sides. Each window has a width of 100\AA\ approximately, and they are separated from the center of \ha{} by at least 300\AA. The continuum polarization angle ($\chi_{cont}$) was also calculated from the mean $Q$ and $U$ of the same windows. For the line polarization measurement ($P_{line}$), we considered the polarization of the region with a width of approximately 50\AA\ surrounding the center of the emission line.

In addition, we corrected for the contamination of the continuum underneath the emission line (from the Stokes parameters, which we mark as $\widetilde U_{line}$ and $\widetilde Q_{line}$). We estimated the Stokes parameters, $Q$ and $U$ ({normalized to the intensity, I}), using the two windows (width = 100\AA) surrounding \ha{} as described above. Then, we calculated the wavelength-averaged vector of  $ Q_{cont} \cdot I_{cont}$ and $U_{cont} \cdot I_{cont}$ for the continuum in these two windows. Finally, we calculated the mean value of $ Q_{cont} \cdot I_{cont}$ and $ U_{cont} \cdot I_{cont}$ for the two windows and subtracted this mean contribution from each point of the \ha{} line vector, as shown in formulas below:

\begin{equation}
   Q_{line}=( \widetilde Q_{line} \cdot I_{line} - Q_{cont} \cdot I_{cont})/I_{line}
       \label{eq:substractionq}
\end{equation}
\begin{equation}
     U_{line}=( \widetilde U_{line} \cdot I_{line} - U_{cont} \cdot I_{cont})/I_{line}
     \label{eq:substractionu}
\end{equation}
 
The further procedure of calculating the fraction of linearly polarized radiation is the same as for the $P_{mean}$ and $P_{cont}$. 

\subsection{Archival spectra of NLSy1}
\label{sec:archival-spectra}

We decided to explore the physical parameters of our sample in the optical band starting from archival reduced data. For Mrk1044, we used spectrum from \cite{2009MNRAS.399..683J}, and for SDSS J080101.41+184840.7 and IRAS 04416+1215, we used data from SDSS \citep{2017AJ....154...28B}). For these spectra, we performed spectral decomposition of the \hb{} and \ha{} regions. Then, we used values obtained for the \ha{} region from archival spectra as initial values for the decomposition of non-polarized spectra of \ha{} from FORS2/PMOS.

To analyze all spectra we used the {\tt specfit} task from IRAF \citep{1994ASPC...61..437K}, including various components in two ranges: 4400-5200\AA~ for the \hb{} region and 6200-6800\AA~ for the \ha{} region. In the \hb{} range, we followed the method presented in \citet{refId0}. In both regions, we modeled the accretion disk emission and the starlight contribution to the spectrum assuming a power-law shape without performing a more complicated procedure for the starlight component, since the ranges are relatively narrow. To model the continuum, we used the continuum windows around 4430, 4760, and 5100\AA\ in case of \hb{} \citep[e.g.][]{1991ApJ...373..465F}, and in case of \ha{}, we used 5700-6000\AA\ \citep[e.g.][]{2002ApJS..143..257K}.
In both regions, we considered the Fe {\sc ii} template from \cite{marziani2009}.

The [OIII]{$\lambda\lambda$4959,5007} doublet lines in the \hb{} region were modeled with two Gaussians that represent narrow and semi-broad components. We assumed the theoretical ratio of 1:3 between the strengths of the components \citep{2007MNRAS.374.1181D}. The [N II]{$\lambda\lambda$6548,6584} and [S II]{$\lambda\lambda$6716,6731} doublet lines in the \ha{} region were modeled with Gaussian profiles, with the assumption of the theoretical ratio 1:3 between their strengths and the FWHM of lines fixed to the [OIII] narrow component (NC). For each Balmer line, we decided to use the Lorentzian shape for the broad component (BC) since it is considered to be more suitable for the NLSy1 than the Gaussian \citep[e.g.][]{1997ApJ...489..656L}. We modeled the Balmer emission lines taking into account two components: one Lorentzian component and one narrow Gaussian component (to reproduce the peak of the line). Moreover, we kept the FWHM of the narrow Gaussian component fixed to the FWHM of the NCs for all forbidden lines and relaxed it only for the last iteration of the fitting. To estimate errors, we used the maximum and minimum continuum levels, and for each case, we performed the fit of the spectrum. We assumed that the distribution of errors follows the triangular distribution  \citep{DAgostini2003BAYESIANRI}. For each line measurement, we calculated the variance using the following formula for the triangular distribution:
\begin{equation}
    \sigma^2(X) = \frac{\Delta^2 x_+ + \Delta^2 x_- + \Delta x_+ + \Delta x_-}{18}
\end{equation}
where $\Delta x_+$ and $\Delta x_-$ are the differences between measurements of the maximum and best continuum and between the best and minimum continuum, respectively. This formulation can be explained as the linear decrease in either side of the maximum of the distribution (which is the best fit) to the values obtained for maximum and minimum contributions of the continuum. We find this analytical method easy and sufficiently precise. We propagated uncertainties using standard formulas of error propagation for the reported values in this work.

\section{Results} \label{sec:results}

\subsection{NLSy1 spectra decomposition in natural light}
We fit each spectrum individually, and for each, we followed the same steps to keep the same accuracy.
As a first step, we decomposed the \hb{} and \ha{} emission line profiles in natural light from archival spectra. 
Then, we fit \ha{} in natural light obtained from our FORS2/VLT observations. To remain consistent, we kept the same values for the \ha{} components obtained from archival spectra as input and then fit the \ha{} in natural light from VLT.
%with \textcolor{green}{keeping the rest of the fitted parameters free.}
We present our measurements of FWHM of Balmer lines' components in Table \ref{tab:specfit}. Each line was modeled with three components: broad (Lorentzian), narrow (Gaussian), and blue (Gaussian). In all cases, the FWHM of the \ha{} BC is slightly narrower than the \hb{} one, whereas the blue components are broader.
 We present the spectral decompositions together with those done for
archival spectra in Figure \ref{fig:specfit-unpol}. 
%All 3 objects show characteristic features for NLSy1 galaxies:...
We did not see any significant differences between \ha{} fits from archival spectra and the VLT spectra. 
%Thus, we treat the archival spectra of \ha{} just as input parameters.
For further analysis, we used the archival spectra of \hb{} (since we did not perform observations for this line) and the VLT \ha{} spectra. The Balmer lines are slightly asymmetric, which may indicate the presence of an outflowing component. The presence of a blue, outflowing component in Balmer lines is visible for all three objects. In the case of IRAS 04416+1215, it is visible for both lines and Mrk 1044, it is visible for \hb{}. However, the \hb{} blue component in Mrk 1044 is almost negligible. For SDSS J080101.41+184840.7, the blue components are the weakest but are still needed to obtain the lowest residual values. All objects from our sample show prominent Fe {\sc ii} emission around the \hb{} region, but the [OIII] doublet and \hb{} NC differ considerably between objects.  The Fe {\sc ii} emission around the \ha{} region is not significant, and the lower contamination in this region is a general property observed in other sources \citep[e.g., in I Zw 1][Figure 7 therein]{2004A&A...417..515V} and in {\sc CLOUDY} photoionization models \citep{2021ApJ...907...12S}.

%The objects differ considerably 
In the case of Mrk 1044, the \hb{} and \ha{} NC intensities are similar to BCs of those lines. We reproduced relatively well the profile of the [OIII] doublet using just one component, NC. The noticeable asymmetry in [OIII]5007\AA\ was nicely reproduced by fitting the Fe {\sc ii} template.
In the case of SDSS J080101.41+184840.7, the \hb{} and \ha{} NC of those lines is around one-third of the BC. The intensity of the semi-broad component of [OIII] is comparable to the NC of this line.
In IRAS 04416+1215, the NC of \hb{} and \ha{} play the least significant role in the modeling of the full profile. 
The most prominent, in comparison to the other two sources, asymmetry in the red side of \ha{} is caused by the [NII] doublet, whose intensity is the highest for this source.
The semi-broad component of [OIII] is dominant, and it is the only source from our sample with that feature.

\begin{table*}[]
\centering
\caption{Measurements in the \hb{} (upper panel) and \ha{} (bottom panel) region for archival spectra.}
\label{tab:specfit}
\begin{tabular}{lcccc}
\hline
\hline
NAME                     & FWHM (\hb{}) NC  & FWHM (\hb{}) BC  & FWHM (\hb{}) BLUE  & c($\frac{1}{2}$) BLUE\\ \hline
Mrk 1044                 & 500$\pm$3        & 1660$\pm$61      & 2550$\pm$57      & -2070$\pm$41 \\
SDSS J080101.41+184840.7 & 570$\pm$43       & 1680$\pm$65      & 2540$\pm$16     & -2070$\pm$15  \\
IRAS 04416+1215          & 490$\pm$5        & 1400$\pm$43      & 2580$\pm$2      &  -1480$\pm$7 \\
\hline
\hline
NAME                     & FWHM (\ha{}) NC & FWHM (\ha{}) BC & FWHM (\ha{}) BLUE & c($\frac{1}{2}$) BLUE\\ \hline
Mrk 1044                 & 500$\pm$2        & 1290$\pm$5       & 2590$\pm$2      & -1290$\pm$11  \\
SDSS J080101.41+184840.7 & 630$\pm$18       & 1530$\pm$14      & 2990$\pm$5       &  -1280$\pm$21  \\
IRAS 04416+1215          & 400$\pm$7        & 1300$\pm$31      & 2790$\pm$9       &-1420$\pm$62\\ \hline 
\end{tabular}
\tablefoot{Columns are as follows: (1) Object's name, (2), (3) and (4) report the FWHM of narrow, broad, and blue components in km s$^{-1}$ for Balmer lines, {and (5) reports} the centroid at the half intensity in km s$^{-1}$, respectively.}
\end{table*}

\begin{sidewaysfigure*}[htp!]
\includegraphics[scale=0.65]{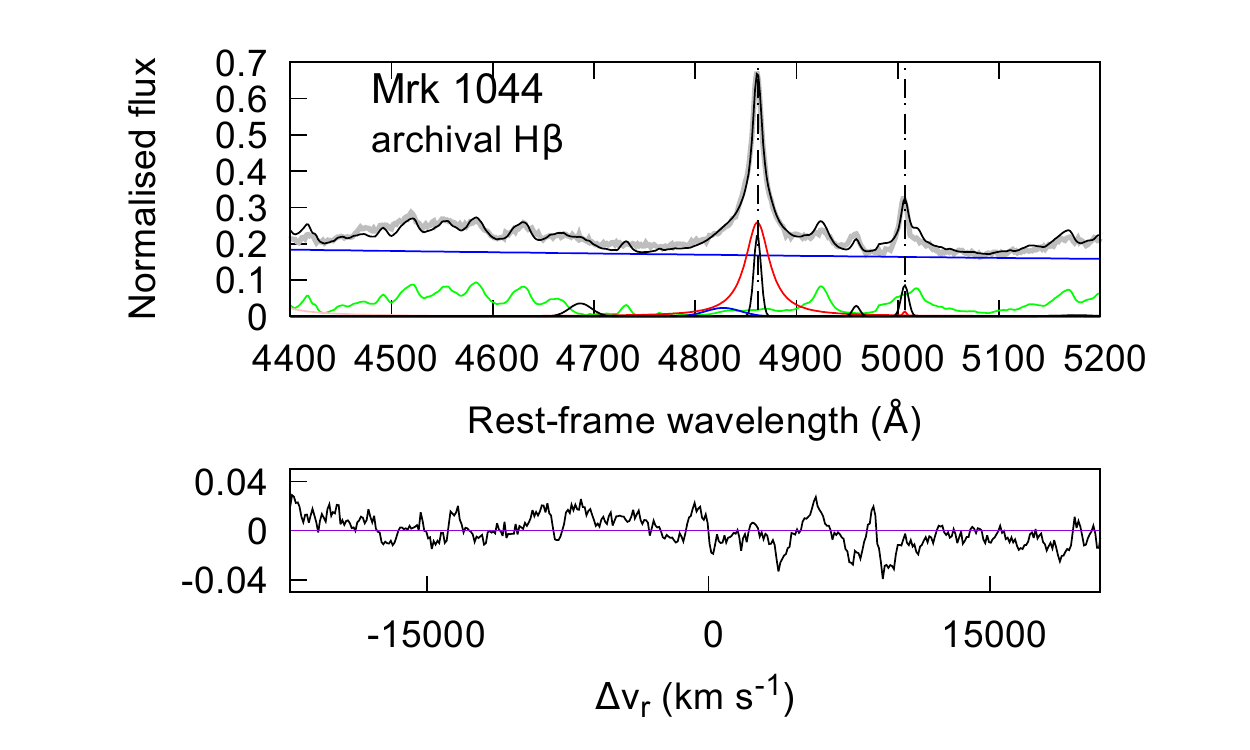}\hspace{-0.1cm}
\includegraphics[scale=0.65]{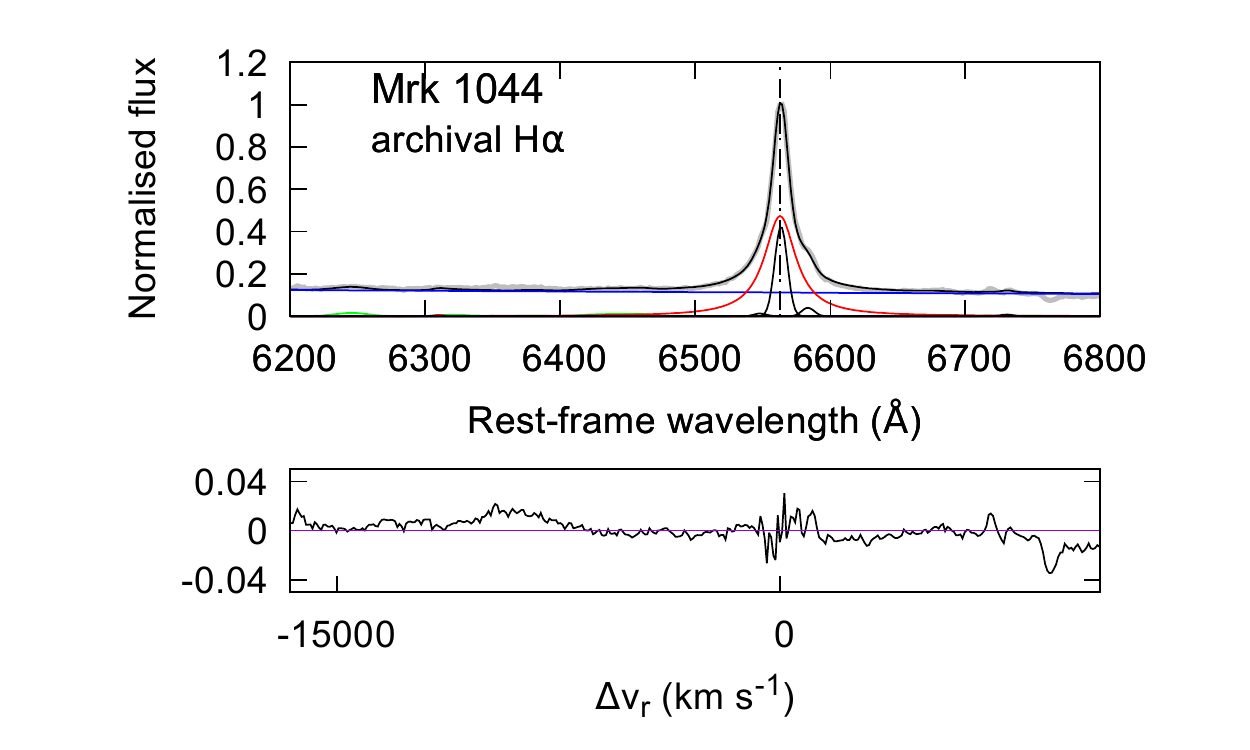}\hspace{-0.1cm}
\includegraphics[scale=0.65]{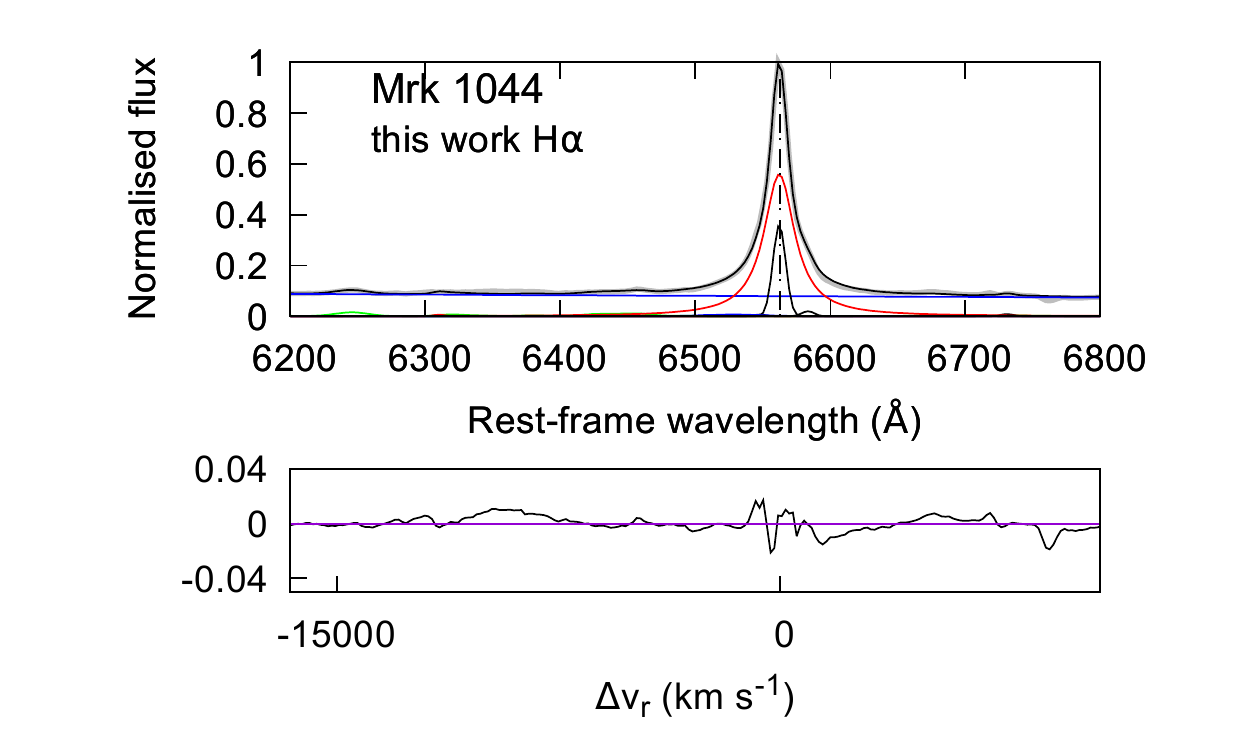} \\
\includegraphics[scale=0.65]{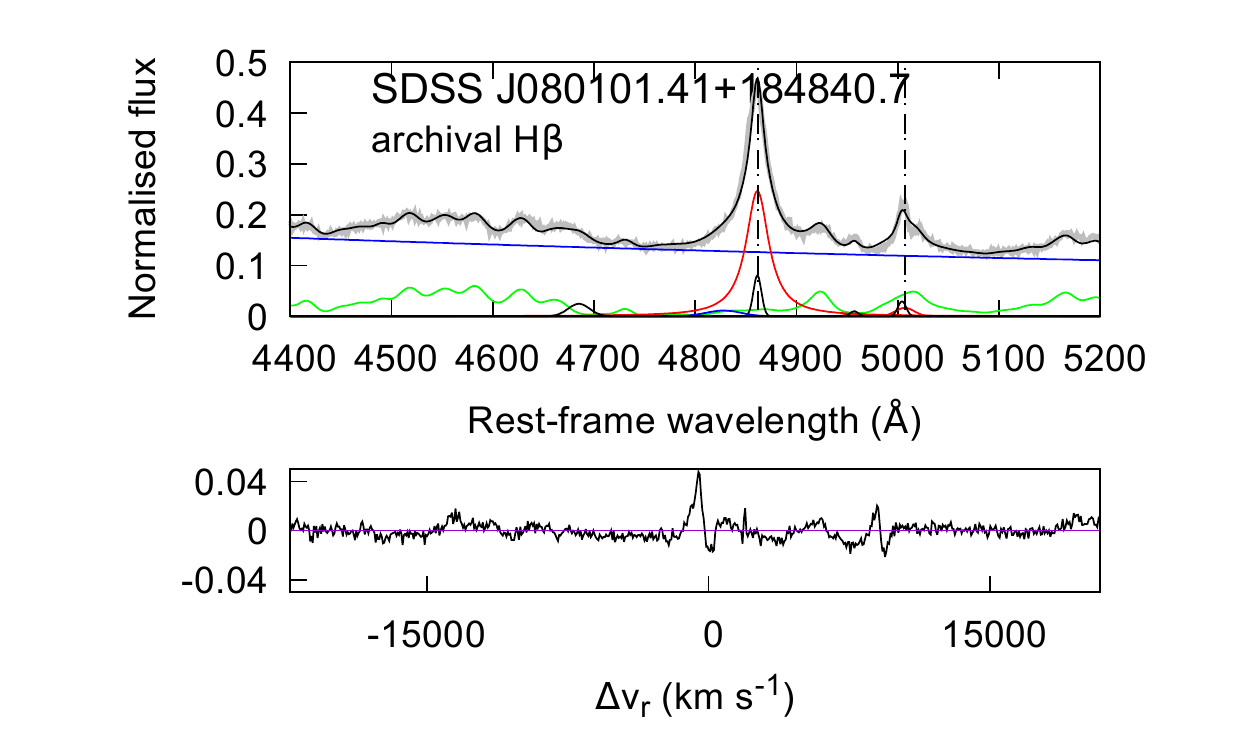}\hspace{-0.1cm}
\includegraphics[scale=0.65]{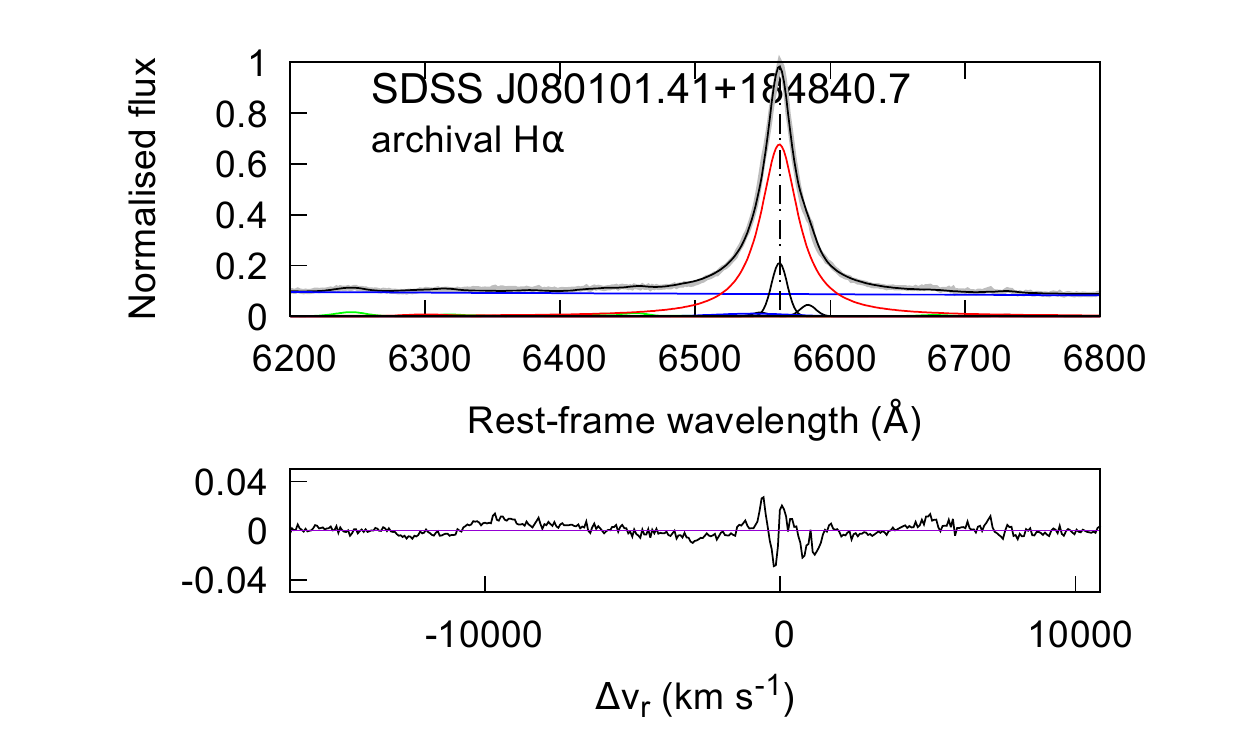}\hspace{-0.1cm}
\includegraphics[scale=0.65]{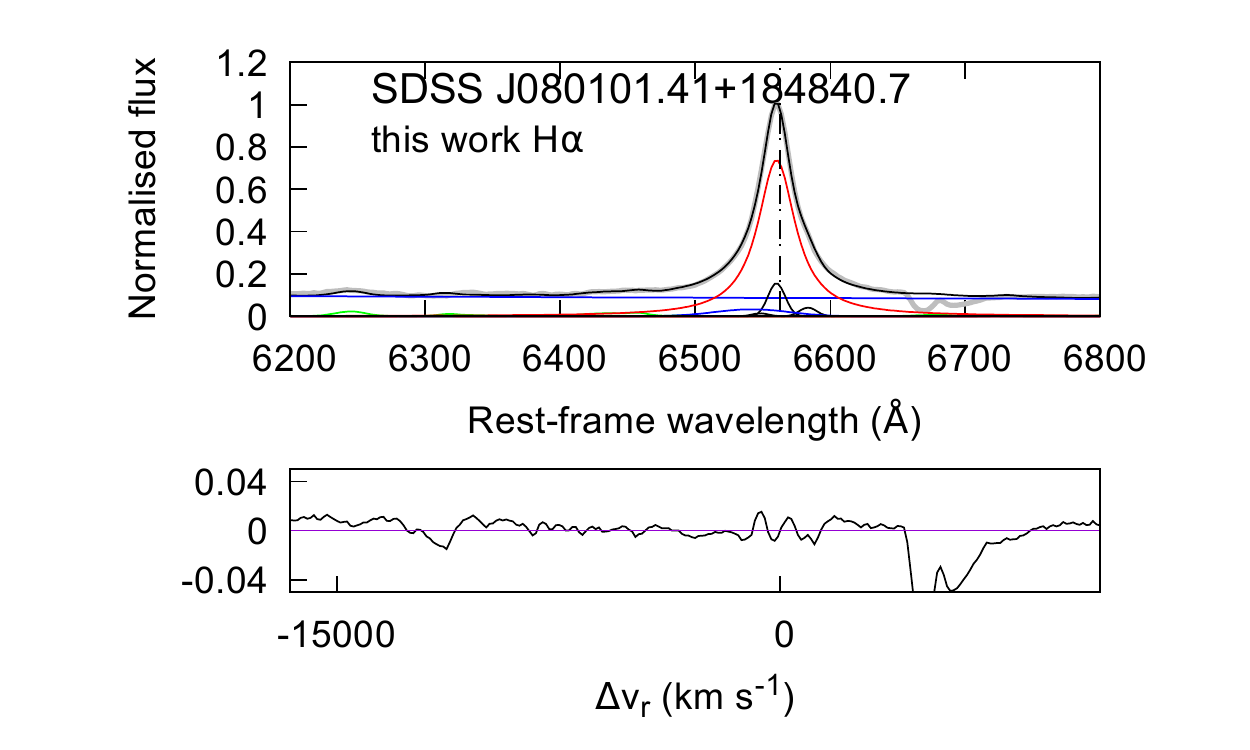}\\
\includegraphics[scale=0.65]{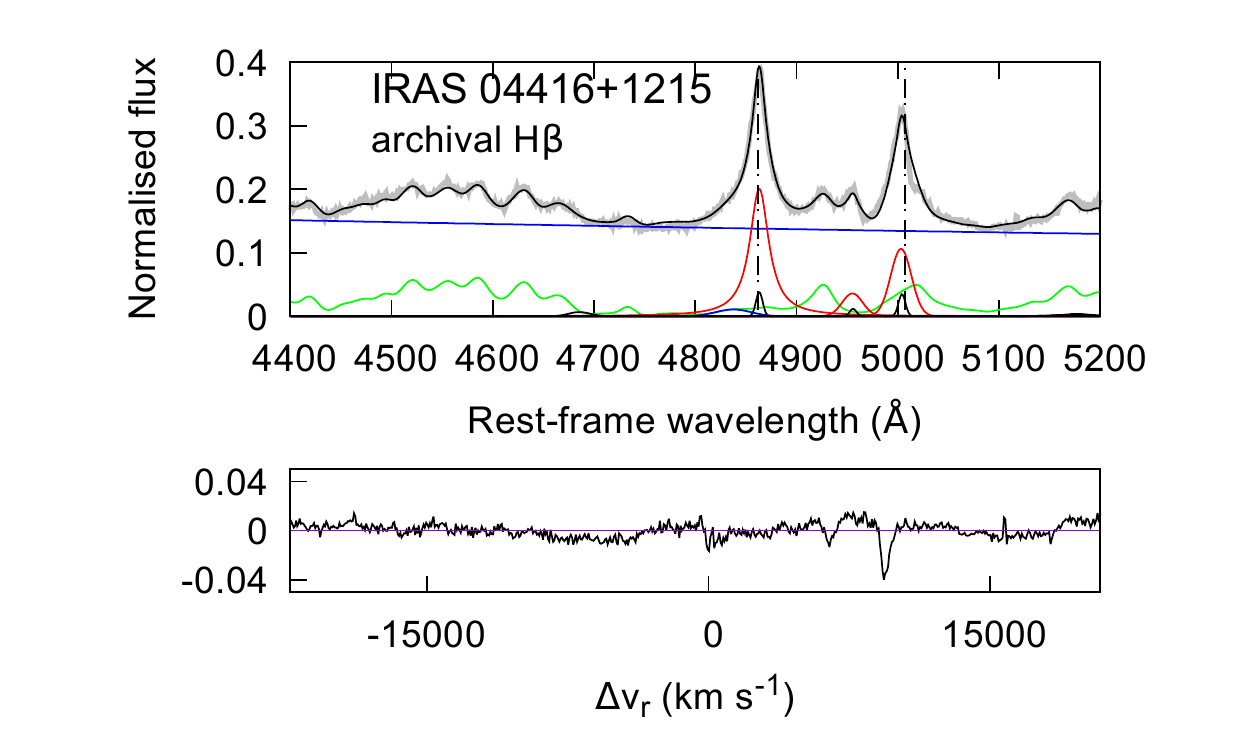}\hspace{-0.1cm}
\includegraphics[scale=0.65]{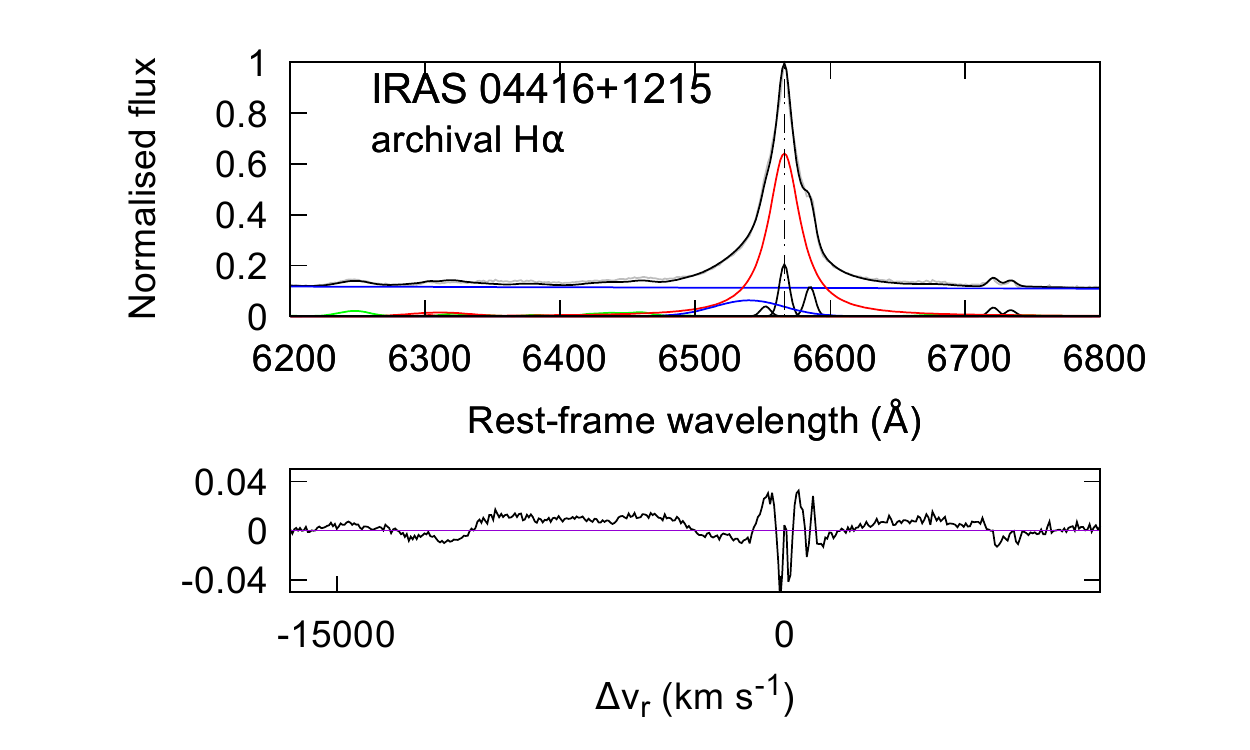}
\includegraphics[scale=0.65]{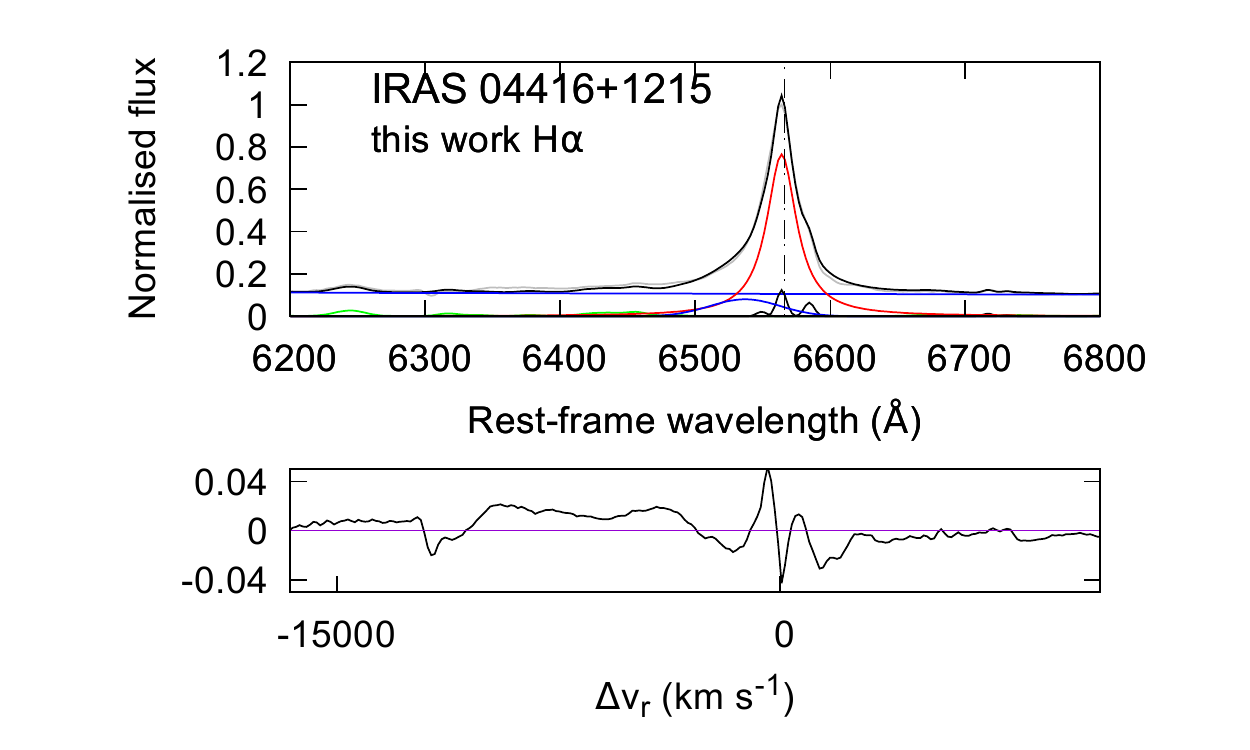}

\caption{Natural light spectra for Mrk 1044 (top row), SDSS J080101.41+184840.7 (middle row), and IRAS 04416+1215 (bottom row).
Top row: Normalized 6dF spectrum of Mrk 1044 from \citep{2009MNRAS.399..683J} with the fitted \hb{} (left) and \ha{} (middle) part. The right plot shows the normalized unpolarized spectrum obtained with FORS2/VLT.
The middle and lower rows show as follows for SDSS J080101.41+184840.7 and IRAS 04416+1215, respectively: Fitted \hb \ and \ha{} part from SDSS normalized archival spectrum and normalized unpolarized spectrum obtained with FORS2/VLT.
Data are marked in gray, and the model is marked in black.  The fit components are given in the following: For \hb{} (left panel), the  \hb{} BC and [O{\sc iii}] semi-broad component doublet are red. The NCs of the \hb{} and [O{\sc iii}] doublet are black. The \hb{} blue component is marked in blue. The power-law component is marked in blue, and the Fe {\sc ii} pseudo continuum is green. For \ha{} (middle and right panel), the \ha{} NC and [N{\sc ii}] and [S{\sc ii}] doublets are marked in black. The \ha{} BC is marked as red. The  \ha{} blue component is marked in blue,  the Fe {\sc ii} pseudo continuum is marked in green, and the power-law component is marked in blue. The lower panels of each plot correspond to the residuals, in radial velocity units km s$^{-1}$. All spectra are normalized to \ha{} intensity. \label{fig:specfit-unpol}}
\end{sidewaysfigure*}

\subsection{Polarization measurements}
\label{sec:polarization-measurements}
We present the spectropolarimetric measurements {for binned data} in Table \ref{tab:polarization-measurements}. For each source, $P_{mean}$ is less than 1\%. We calculated $P_{mean}$ (column 2)
from the mean of the Stokes parameters $Q$ and $U$ of the whole wavelength range of the spectra using the following formula:
\begin{equation}
   {P_{mean}} = \sqrt{{Q_{mean}^2 + U_{mean}^2}}.
\end{equation}
The polarization of the continuum region (column 4) and line region (column 6) is calculated similarly, including different wavelength ranges of the spectra. Additionally, the polarization of the lines includes the "continuum polarization correction," which we explained in Section \ref{sec:spectropol-corrections}. The "continuum polarization correction" does not change the physical interpretation of the results. For the SDSS J080101.41+184840.7, we obtained a $P_{line}$ of {0.20\%} with the correction, and a $P_{line}$ of 0.42\% without this correction. Both values are lower than the $P_{mean}$ and $P_{cont}$ for this source, but by including this correction, the fraction of linearly polarized radiation was reduced.

The polarization position angle (which is the angle of maximum polarization) was calculated using the following formulas \cite[Eq. 8 in][]{2009PASP..121..993B}:

\begin{equation}
\begin{aligned}
& \chi = \frac{1}{2}tan^{-1}\left ( \frac{U}{Q}\right ) \;(\mathrm{if}\;Q>0\ \mathrm{and}\ U \geq 0);\\
& \chi = \frac{1}{2}tan^{-1}\left ( \frac{U}{Q}\right ) + 180^{\circ} \;(\mathrm{if}\;Q>0\ \mathrm{and}\ U < 0);\\
& \chi = \frac{1}{2}tan^{-1}\left ( \frac{U}{Q}\right ) + 90^{\circ} \;(\mathrm{if}\;Q<0)
\end{aligned}
\label{pol-ang}
\end{equation}
We used the mean $Q$ and $U$ from the whole region (column 3), the continuum region (column 5), and the line region (column 7). 
We show the intensity of the \ha{}, the polarization percentage,  the Stokes parameters, and the polarization angle for each object in Figure \ref{fig:polarization-info}. We show the windows for continuum and the line polarization for objects from our sample in red. We used the same windows for each object.
Our results are not sensitive to the exact width of the continuum. We tested a few cases between 100\AA\ and 400\AA\ and did not notice a significant change in the obtained values.

Overall, the level of polarization in our three sources is very low, less than {$1\%$, which tends to be positively biased \citep[see][]{1985A&A...142..100S}}. The measurement of the
change of the polarization angle across the broad emission line is hard to measure for sources with low polarization. Therefore, we concentrated on the estimates of the viewing angle to improve the black hole mass measurement.

In addition, we noticed that the polarized flux profile has a peak-like appearance, with a width not much larger than the one measured in natural light, at least for Mrk 1044 and SJ080101.41+184840.7 (the case of IRAS 04416+1215 is less clear due to its pattern in degree of polarization being dominated by noise). This can be an indication of a relatively high viewing angle (as discussed below in \S\ \ref{sec:viewing-angle}) or as evidence of a polar scatterer (\S \ref{sec:STOKES}).
{The evidence of a polar scatterer is the strongest in the case of Mrk 1044. In the continuum not corrected for polarization, the  degree of polarization of the line (0.33\%) is higher than the one of the adjacent continuum (0.22 \%), an increase by a factor of 1.45. With the correction, the continuum and line polarization degrees are on a similar level.Moreover, $\chi$ changes by approximately $90^{\circ}$, which is a typical signature of polar scattering as well.}

\begin{sidewaysfigure*}[htp!]
  \centering

\includegraphics[width=8cm]{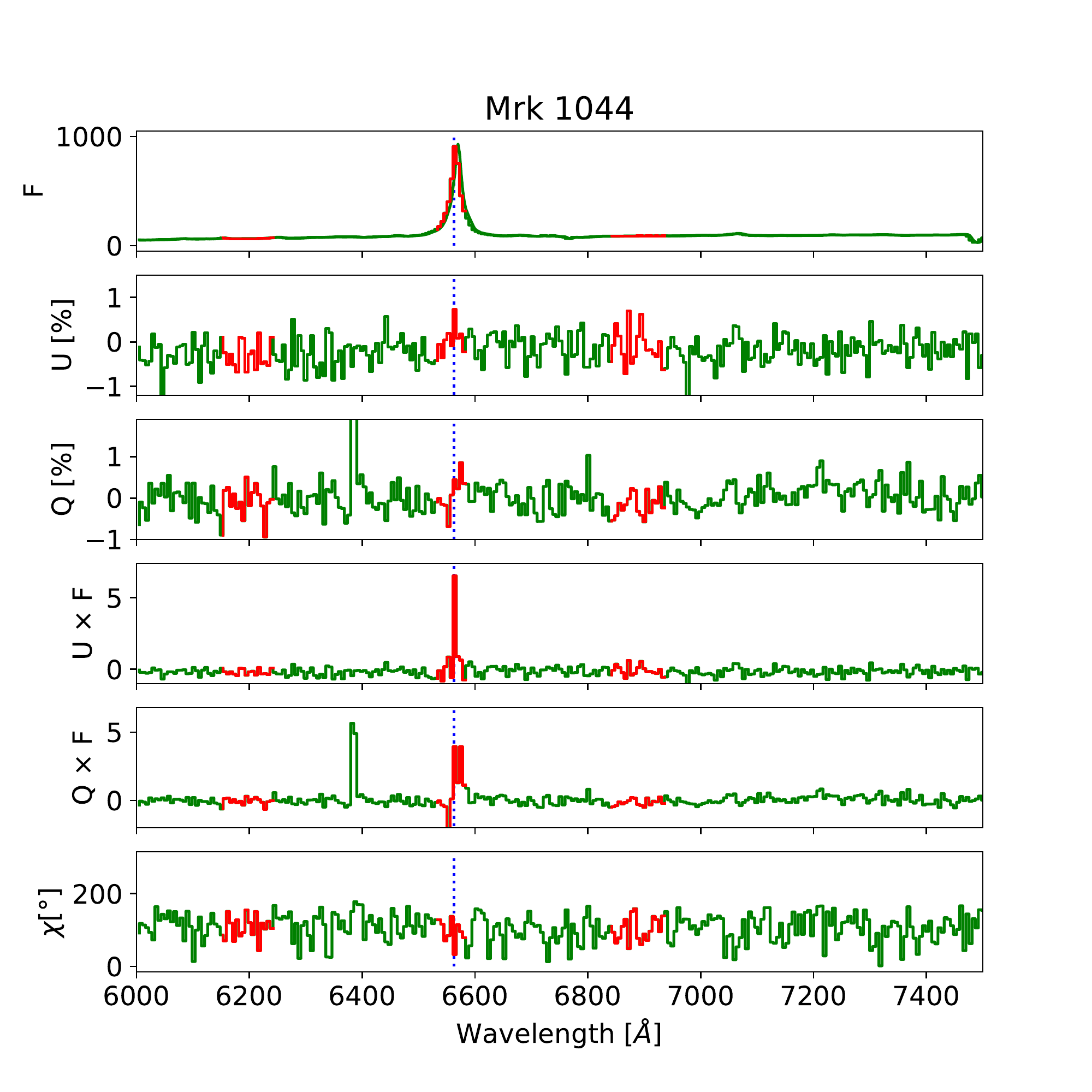}
\includegraphics[width=8cm]{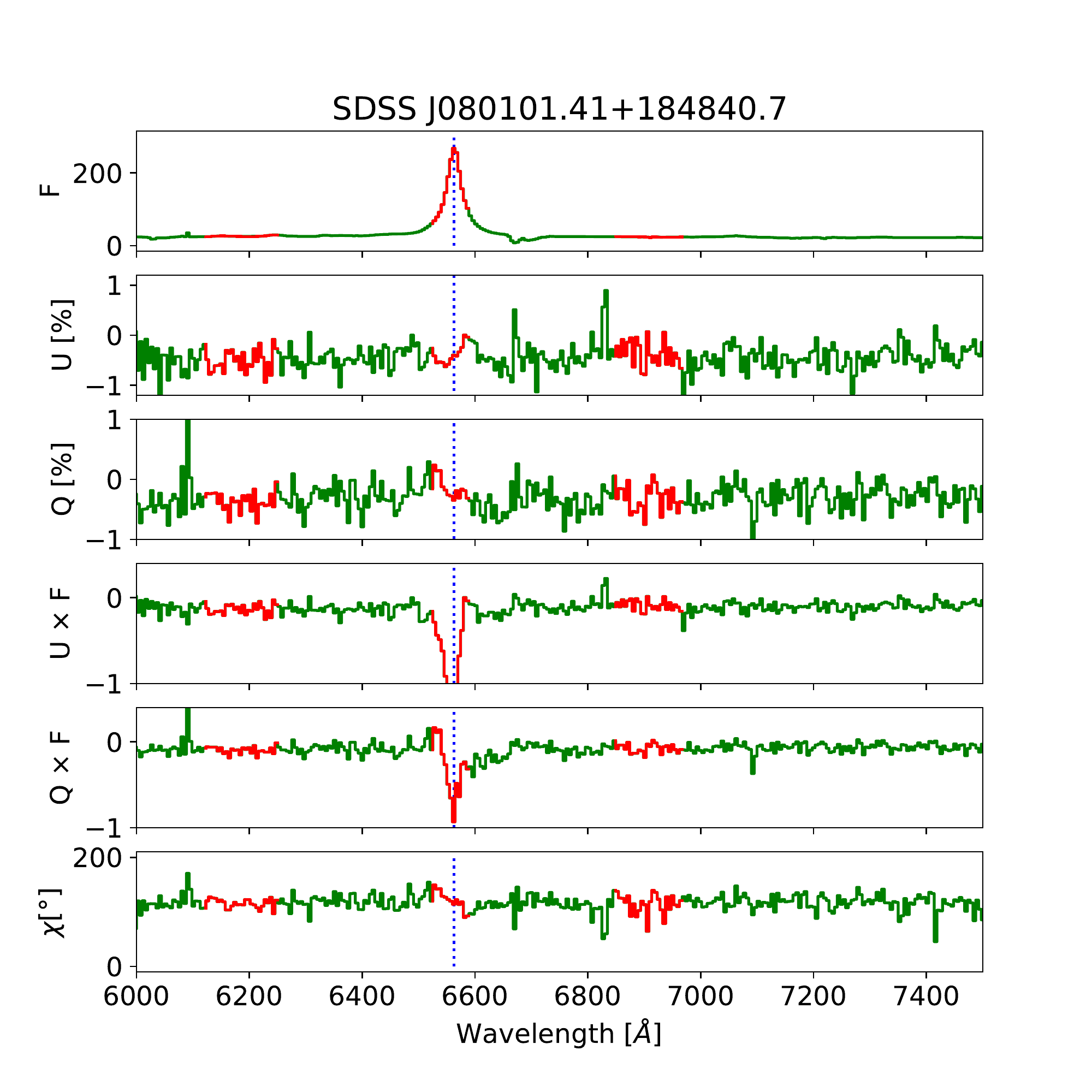}
\includegraphics[width=8cm]{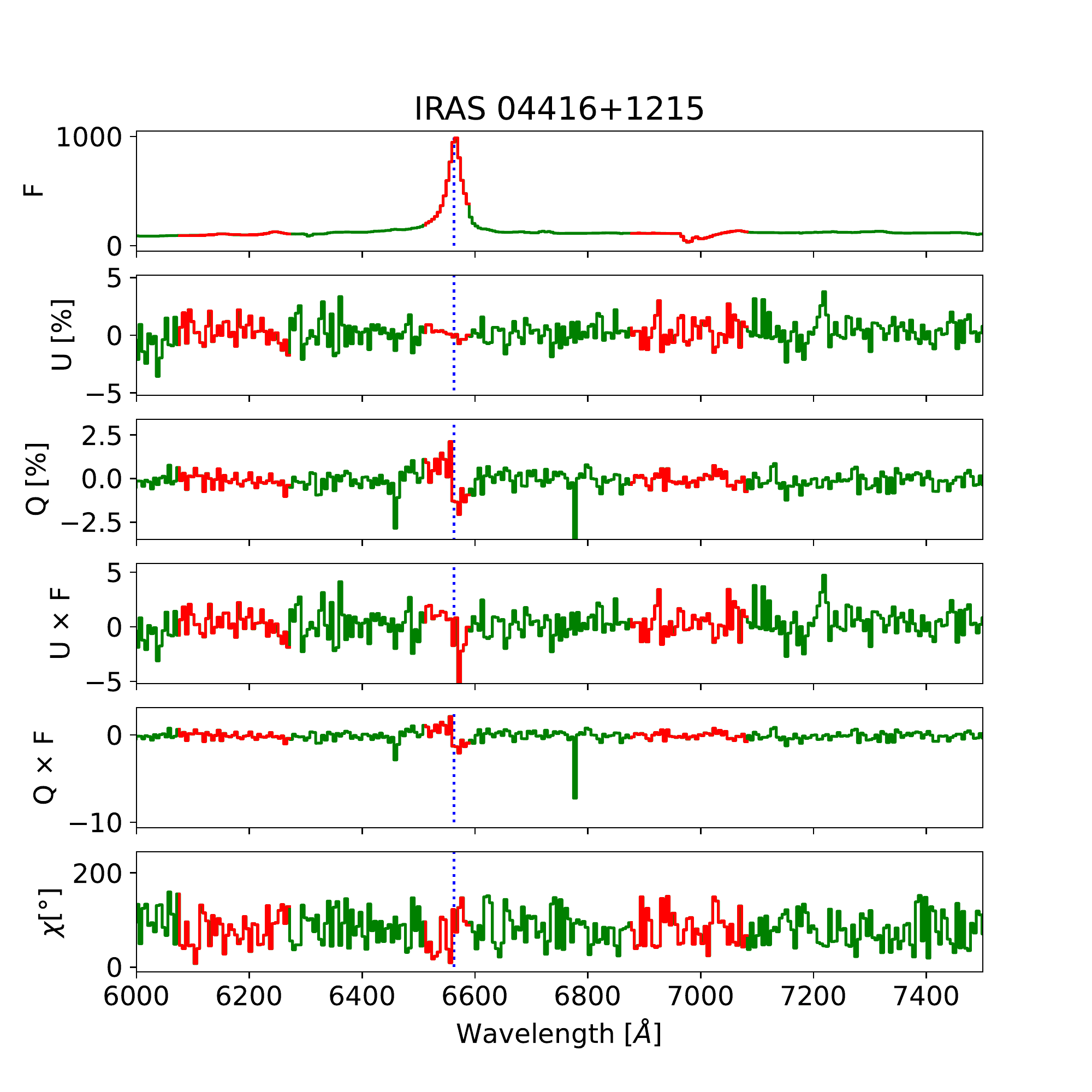}
 \caption{Data of Mrk 1044, SDSS J080101.41+184840.7, and IRAS 04416+1215. Plots, from top to bottom: Total flux, Stokes parameters U and Q, flux of U and Q, and the polarization angle. We mark in red the windows in which we calculated polarization. For continuum polarization, we considered two surrounding \ha{} from its blue and red side, each window has a width of  $\approx$100$\AA$, and {they are at a distance of at least 300$\AA$} from the center of \ha{}. For the polarization of the line, we considered the red window mark on the line. Each central window has width of $\approx$50$\AA$. The vertical blue dotted line marks the central wavelength for \ha{}. The data were rebinned with a 2\AA\ binning.}
\label{fig:polarization-info}
\end{sidewaysfigure*}

\begin{table*}[]
\center
\caption{{Measurements for binned spectra of degree of polarization ($P$) and the polarization angle ($\chi$) for our sample. } }
\label{tab:polarization-measurements}
\begin{tabular}{lllllll}
\hline
\hline
NAME                     & $P_{mean}$ & $\chi_{mean}$ & $P_{cont}$  & $\chi_{cont}$ & $P_{line}$    &  $\chi_{line}$   \\
                  &  {[}\%{]} &[$^{\circ}$] & {[}\%{]} &[$^{\circ}$] &  {[}\%{]}   &  [$^{\circ}$]  \\\hline
Mrk 1044	&	0.15$\pm$0.02	&	142$\pm$1	&	0.23$\pm$0.04	&	120$\pm$8	&	0.17$\pm$0.06	&	12$\pm$10		\\
SDSS J080101.41+184840.7	&	0.52$\pm$0.02	&	117$\pm$1	&	0.48$\pm$0.04	&	116$\pm$4	&	0.20$\pm$0.04	&	173$\pm$4	\\
IRAS 04416+1215         &	0.17$\pm$0.03	&	58$\pm$1	&	0.15$\pm$0.1	&	64$\pm$6	&	0.09$\pm$0.03	&	73$\pm$11	\\\hline  
\end{tabular}
\tablefoot{Columns show {(1)} the name of the source, {(2)} the averaged {$P$} from the whole spectrum, {(3)} the averaged $\chi$ from the whole spectrum ($\chi_{mean}$), the averaged $P$ and $\chi$ from continuum windows {(4 and 5)}, and the averaged $P$ and $\chi$ measured from \ha{} emission line {(6 and 7)}.}

\end{table*}

\subsection{NLSy1 spectra decomposition in polarized light}
In Figure \ref{fig:specfit-pol-three-obj}, we show the spectral decomposition of VLT \ha{} emission line profiles in the polarized light. The expected equatorial placed scatterer was modeled as a single BC broader than the \ha{} in natural light. %We plot the broad component using a red line.
The case in which just a single component is enough to model the line profile is visible in the {bottom} panel of the figure (IRAS 04416+1215). The cases of Mrk 1044 and SDSS J080101.41+184840.7 are different. Modeling the full line profile using just a BC was not possible. Thus, we decided to add the NCs marked in black. The NCs are most probably associated with a non-equatorial scatterer, and we investigated Mrk 1044 and SDSS J080101.41+184840.7 using the STOKES radiative transfer code in this regard (see Sec. \ref{sec:polarization-stokes}).
The residuals for Mrk 1044 and IRAS 04416+1215 are higher than for SDSS J080101.41+184840.7. Due to low S/N in our data, we were not able to obtain better quality fits; however, there is no specific pattern in the residuals, which supports the reliability of our performed fits.

\subsection{Viewing angles}
\label{sec:viewing-angle}
To estimate the viewing angles, we used Equation (8) from \citealt{2006collin}: 
%we assume H/R as 1/3 :
\begin{equation}
    \Delta V_{obs} \approx V_{Kep} [ \,(H/R)^2 + sin^2i  ]^{1/2}.
    \label{eq:viewing-angle}
\end{equation}
We assumed that the FWHM of the line in polarized light measures the true Keplerian velocity at the distance R (i.e.,  $V_{Kep}$), while the observed value of $\Delta V_{obs}$ is measured as the FWHM of the line in non-polarized natural light, and $i$ is the viewing angle of the system toward us. With H/R, we denote the aspect ratio of the disk at any radius. From the observer’s line of sight, the ``full'' Keplerian velocities from the rotating disk are not visible, as the equatorial scattering region and the rotating disk are on the same plane \citep[see Figure 9][]{smith2005}. However, the equatorial scattering region is well positioned for observations of unprojected velocities from the rotating disk.

 The ratio of the thickness of the BLR region to its radius is indicated by H/R, and it represents the random motion of BLR clouds. 
%The assumption of 1/3 means that at each radius the height is 1/3 of the distance to the black hole. 
The physical justification for an H/R higher than 0.1 comes from \citealt{2006collin}. As the authors underlined, the illumination required for the BLR implies a large opening angle of the BLR.  The most typical H/R ratio found for the BLR is 1:3 \citep[e.g.,][]{2012ApJ...748...49S}, and we adopt this value in further considerations. The results are reported in Table~\ref{tab:viewing-angle}.

\begin{figure}[htp!]
\centering
\includegraphics[scale=0.70]{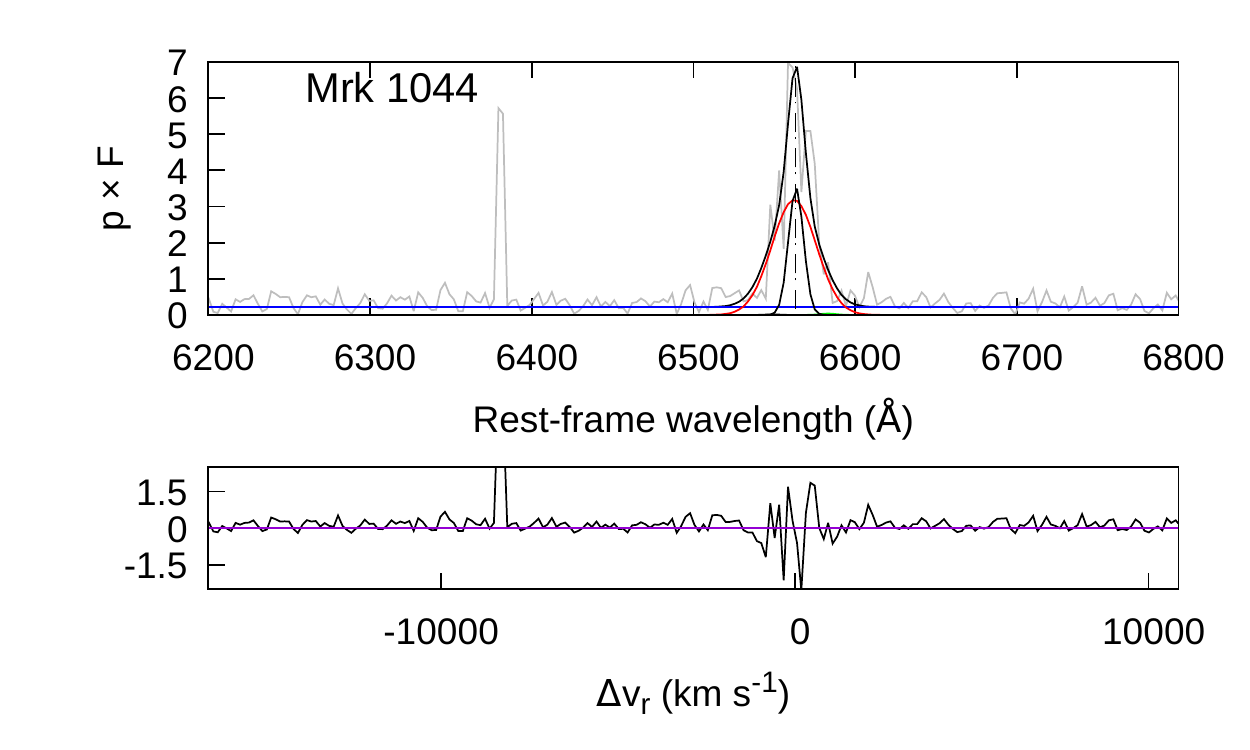}\hspace{-0.2cm}
\includegraphics[scale=0.70]{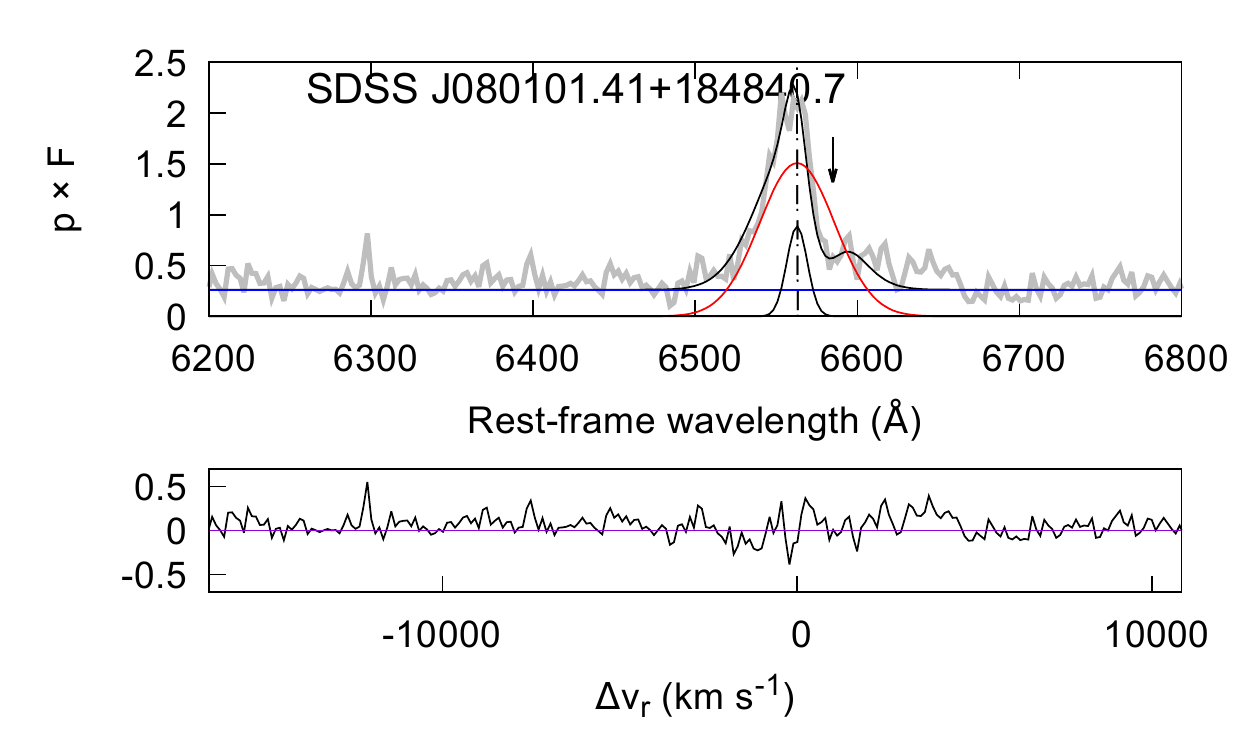}
\includegraphics[scale=0.70]{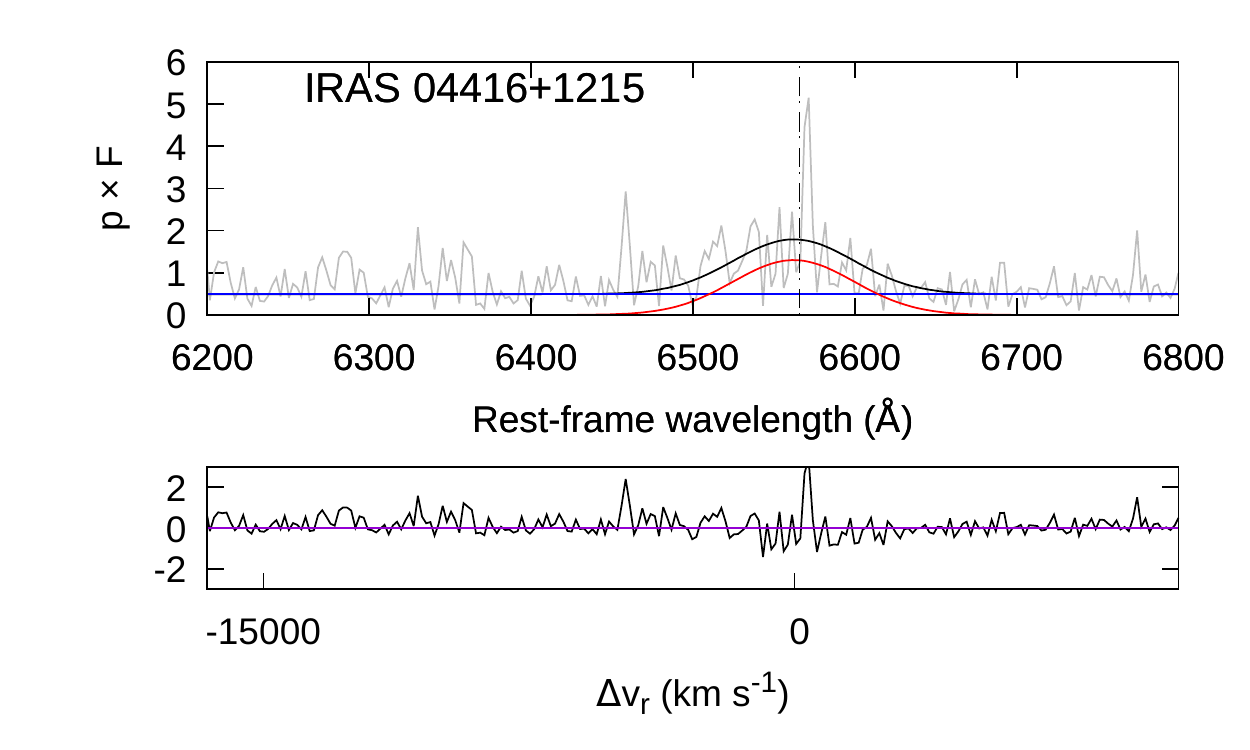}
\caption{Fitted \ha{} part from the polarized spectrum obtained with FORS/VLT. The data are marked in gray, and the model is marked in black. For each fit, we considered two components of \ha: a BC (red) and an NC (black). The blue line corresponds to the power-law continuum. The lower panels of each plot correspond to the residuals, in radial velocity {in units of} km s$^{-1}$. With an arrow, we mark the presence of an "absorption" in the fit of SDSS J080101. The "absorption" is meant to account for the low S/N of the data and to ease the fit of a symmetric Gaussian constrained by the blue side of the polarized flux profile.}

\label{fig:specfit-pol-three-obj}
\end{figure}

\begin{table}[]
\center
\caption{Estimation of viewing angles for our sample using the formula described in Section \ref{sec:viewing-angle}. }
\label{tab:viewing-angle}
\resizebox{\linewidth}{!}{%
\begin{tabular}{lccl}
\hline
\hline
NAME                     & FWHM(\ha{}) BC  &  FWHM(\ha{}) BC    & $i$ \\
                    & non-polarized & polarized    &  [$^{\circ}$]\\\hline
Mrk 1044                 & 1290$\pm$20 & 1480$\pm$20  & 54$\pm$2\\
SDSS J080101 & 1530$\pm$30 & 2500$\pm$60  & 31$\pm$2    \\
IRAS 04416+1215          & 1300$\pm$20 & 3810$\pm$140 & 4$\pm$4    \\\hline
\end{tabular}}
\tablefoot{Columns are: name of the source, estimation of FWHM based on FORS2/PMOS spectra: the FWHM of non-polarized broad component of \ha{}, the FWHM of polarized broad component of \ha{}, and the {estimated} viewing angle of the source.}
\end{table}

\subsection{Estimation of  M$_{BH}$}
The reverberation mapping technique allowed us to determine the black hole mass via a virial relationship:
\begin{equation}
    M_{BH} = f\frac{R_{\rm in} FWHM^2}{G},
    \label{eq:mbh}
\end{equation}
where $G$ is the gravitational constant, and $f$ is the (dimensionless) virial factor that depends on the structure, kinematic, and viewing angle of the BLR. The emissivity-averaged radius of the BLR (also known by the size of the BLR) is denoted as R$_{\rm in}$, and FWHM is the full width at half maximum of the emission line profile
\citep[for details, see][]{2019FrASS...6...75P}. The radius of the BLR is determined from the observed time-lag between the continuum and the broad \hb{} emission line, and the FWHM is determined from the spectral modeling of the \hb{}. The virial factor is more puzzling since it depends on more than one property, and mostly, it is assumed as a fixed factor.
With the spectropolarimetry technique, it is possible to estimate the black hole mass using the FWHM of the polarized line in the formula.

The analytical formula for the virial factor in the case of the disk-like structure of thickness H (assumed $\ll R_{\rm in}$) is given by:
\begin{equation}
    f = [4( \sin(i)^2 + (H/R_{\rm in})^2)]^{-1} 
    \label{eq:factor}
\end{equation}
If we use a polarized line to estimate the black hole mass, we assume that we are observing the line-emitting medium locating the observer on the disk-plane. Thus, %$H/R_{\rm in} \rightarrow 0$} and 
$\sin(i) = 1$ (i.e., $i = 90^{\circ}$).

Including the \textit{f} factor with polarized line assumptions, we used the following equation to compute the black hole mass for our sources,
\begin{equation}
     M_{BH} = \frac{R_{\rm in} FWHM^2}{4G}.
    \label{eq:mbh1}
\end{equation}

Here, FWHM is the full width at half maximum of the BC of the polarized \ha{} line obtained in this work.

In Table \ref{tab:mass-calculation}, we present a comparison between the estimation of the black hole mass for our sample from \citet{dupu2016_alt} using the reverberation mapping technique (column 2) {and using Eq. \ref{eq:mbh} (column 3).} 
We note that in Eq. \ref{eq:mbh1} we still used values obtained with reverberation mapping (i.e., the inner radius of the BLR). {In this work, the inner radius of the BLR is assumed to be equivalent to the emissivity-weighted radius of the BLR, which we estimated using reverberation mapping studies.}

{After using our method, the mass estimations for IRAS 04416+1215 and  SDSS J080101 are both within the (one sigma) error bars in comparison to the reverberation mapping method} \citep{dupu2016_alt}.
 The difference in the mass estimations is only significant for Mrk 1044, as we obtained a lower value from our measurement. However, the Mrk 1044 spectra quality is the best, and we did not have the same problem estimating the viewing angle as we had for IRAS 04416+1215.

\begin{table}[]
\centering
\caption{Comparison of mass estimations.}
\label{tab:mass-calculation}
\begin{tabular}{lll}
\hline
\hline
NAME & log  M$_{BH}$ (M$_{\odot}$) & log  M$_{BH}$  (M$_{\odot}$) \\
                   & RM & this work \\ 
\hline
Mrk 1044                 & 6.45$^{+0.12}_{-0.13}$           & 6.05$\pm$0.13                       \\
SDSS J080101 & 6.78$^{+0.34}_{-0.17}$            & 6.40$\pm$0.19                        \\
IRAS 04416+1215          & 6.78$^{+0.31}_{-0.06}$            & 6.97$\pm$0.42     \\\hline                   
\end{tabular}
\tablefoot{Columns are: {(1) Object's name,}  (2) {the black hole} mass is calculated using reverberation mapping technique using broad \hb{} emission line \citep{dupu2016_alt}, (3) the estimation of black hole mass is done based on this work using broad \ha{} emission line.}
\end{table}

\subsection{Determination of $R_\mathrm{FeII}$}
The strength of the optical Fe {\sc ii} emission, that is, the ratio of the Fe {\sc ii} optical flux measured between 4434–4684\AA\ range to the broad \hb{} flux (also known as $R_\mathrm{FeII}$), seems to be an indirect tracer of the Eddington ratio \citep[e.g.,][]{2014Natur.513..210S,marziani_etal_2018,panda_etal_2018, panda_etal_2019a, panda_etal_2019b,cafe_pca}. Since we had already modeled the optical spectra around the \hb{} region, we decided to also explore our sample in the context of $R_\mathrm{FeII}$. We show equivalent widths of Fe {\sc ii}, \hb, and $R_\mathrm{FeII}$ in Table \ref{tab:rfe}. The source IRAS 04416+1215 has the highest $R_\mathrm{FeII}$ value in our sample and may be called an extreme case, as $R_\mathrm{FeII}$ $>$ 1.3, according to \citet{2018A&A...613A..38S}, and $R_\mathrm{FeII}$ $>$ 1, according to \citet{2015sulenticmarziani}. From the latter criterion, Mrk 1044 is a borderline high-accreting source.  We expected relatively high values of $R_\mathrm{FeII}$ ($ \approx 1$) for all sources in our sample,
 %\textcolor{green}{since high Eddington ratio implies high $R_\mathrm{FeII}$ \citep{2014Natur.513..210S, marziani_etal_2018,panda_etal_2018, panda_etal_2019b},
 since a high Eddington ratio was one of our criteria to create the sample. The two other sources, Mrk 1044 and SDSS J080101 are not extreme Fe II emitters, but they do have significantly above average $R_\mathrm{FeII}$ values of 0.64. The median value is 0.38 in the  \citet{shen2011} catalog if only the measurements with errors below 20\% are considered \citep{2018A&A...613A..38S}. 

\begin{table}[]
\center
\caption{Estimation of $R_\mathrm{FeII}$ for our sample.}
\label{tab:rfe}
\begin{tabular}{llll}
\hline
\hline
\multicolumn{1}{l}{NAME}     & \multicolumn{1}{l}{EW({Fe{\sc ii}})} & \multicolumn{1}{l}{EW(\hb{}) BC} & \multicolumn{1}{l}{$R_\mathrm{FeII}$} \\ 

               & [\AA]                     & [\AA]                          &                     \\ \hline
Mrk 1044                 & 40.17$\pm$0.52                       & 42.79$\pm$1.91                           & 0.94$\pm$0.06                    \\
SDSS J080101 & 60.89$\pm$0.42                        & 74.41$\pm$3.22                            & 0.82$\pm$0.07                    \\
IRAS 04416+1215          & 60.88$\pm$0.24                      & 45.79$\pm$0.92                           & 1.33$\pm$0.08  \\ \hline                 
\end{tabular}
\tablefoot{Columns show {(1)} the name of the source, {(2)} Equivalent widths of {Fe{\sc ii}}  and  {(3)} \hb{}, and {(4)} $R_\mathrm{FeII}$.}
\end{table}

\section{STOKES modeling}
\label{sec:STOKES}

Our analysis of the polarized and unpolarized spectra shows that our sources, selected as highly accretion objects, indeed have intrinsically narrow lines and small black hole masses. Therefore, high accretion rate values suggested by \citet{wang2014} are not an artifact of the special orientation of these sources. Such highly accreting sources are expected to have very intense outflows. For example, Mrk 1044 shows clear signatures of ultra-fast outflows (UFOs; \citealt{krongold2021}). Such strong, high density, highly ionized wind with column density of $\sim 10^{23}$ cm$^{-2}$ and a velocity of 0.15 light speed suggests the presence of a copious amount of material surrounding the black hole, which could lead to efficient polarization while our observations imply very low polarization in these sources. To address the potential discrepancy between strong outflow and low polarization, we performed simulations of the polarization properties of our sources. With those simulations, we intended to independently constrain the amount of scattering material surrounding the AGN.

We used the radiative transfer code {\sc STOKES} \citep{goosmann_gaskell_2007,marin2018,savic_etal_2018} to understand the polarization in the \ha{} emitting BLR of these three objects. The {\sc STOKES} code is a 3D radiative transfer code that incorporates the Monte Carlo approach in order to trace every single photon emitted from a source, of which a considerable number are scattered by the intervening media. The code includes physical processes such as the electron or dust scattering until the photons get absorbed or eventually escape and reach the distant observer. We used the latest version (v1.2) of the code.\footnote{publicly available at \href{http://astro.u-strasbg.fr/~marin/STOKES\_web/index.html}{http://astro.u-strasbg.fr/marin/STOKES\_web/index.html}}

\subsection{Parameters of the model}
In our modeling, we assumed the continuum source to be point-like located at the center, and emitting isotropic unpolarized radiation. The generated flux thus generated takes the form of a power-law spectrum $F_{\rm cont} \propto \nu^{-\alpha}$ with $\alpha$ = 2. This assumption of the value for the spectral index is justified based on prior works \citep{savic_etal_2018,2021bowei} where the focus was to reproduce the polarization properties around a specific emission line (in our case \ha{}) and the continuum around that line. The continuum source is surrounded by the line-emitting BLR of cylindrical geometry that is specified by three spatial parameters, namely (a) the distance between the continuum source to the center of the BLR (R$_{\rm mid}$); (b) the height of extension of the BLR above (or below) the mid-plane joining the source and the BLR ($a$); and (c) the half-width of the BLR ($b$). These three spatial terms can be estimated using the information of the BLR's inner (R$_{\rm in}$) and outer (R$_{\rm out}$) radii, which we extracted from previous studies. Thus, for the inner radius of the BLR, we used the reverberation mapped estimates compiled by \citep{dupu2016_alt} for the three objects considered in this work, while we estimated the outer radius of the BLR using the analytical relation by \citet{netzer_laor_1993} for the dust sublimation radius, which has the following form:
\begin{equation}
R_{\rm out}^{\rm BLR} = 0.2L_{\rm bol, 46}^{0.5}
\end{equation}
where $L_{\rm bol, 46}$ is the bolometric luminosity in units of $10^{46}$ erg s$^{-1}$. The $L_{\rm bol}$ for our objects were estimated by scaling the observed monochromatic luminosity at 5100\AA~ compiled in \citet{dupu2016_alt} by a luminosity-dependent bolometric correction factor \citep{2019netzer}. The assumption of a slightly smaller inner radius than the reverberation mapped estimate (or the emissivity-weighted radius) retrieves nearly identical results for the Stokes parameters, suggesting that, for these sources, assuming the emissivity-weighted radius as a proxy for the inner radius is justified. This is due to the fact that the vertical extension of the BLR scales with the radial location of the onset of the BLR which is estimated based on a fixed half-opening angle of 15$^{\circ}$. We tested two scenarios where the R$_{\rm in}$ was assumed to be smaller by a factor of two and three with respect to the emissivity-weighted radius. The R$_{\rm mid}$ (which is computed as an average of the R$_{\rm in}$ and R$_{\rm out}$) changes from $\sim$26 ld to 22.5 ld and 23.37 ld, for the two alternate cases, respectively. This corresponds to an increase by $\sim$50 \kms{} in the average velocity estimated at R$_{\rm mid}$. The results from the STOKES modeling show very little change in the total and polarized spectra, while we noticed marginal variations in the polarization angle due to the change in the location of the inner radius. The remaining parameter to complete the information for the BLR is the velocity distribution for the \ha{} line. This requires the value of the average velocity ($v_{\rm avg}$) along the x-y direction (or $\phi$ direction; here we assumed the z-axis to be aligned with the BH spin axis). We estimated the average velocity by assuming a Keplerian velocity distribution at a given radius (here, the radius considered is the mean value between the BLR's inner and outer radii) from the source with a predetermined BH mass, that is, $v_{\rm avg} = \sqrt{GM/R}$, where $G$ is the gravitational constant.

For the scattering region that represents the torus, we assumed a geometry of a flared disk characterized by inner and outer radii as well as a half-opening angle. The light becomes scattered predominantly due to free electrons (i.e., Thomson scattering), where the content of this medium is regulated by the number density parameter. In addition to these assumptions, the velocity distribution was again set similarly to the scenario for the BLR but assuming the appropriate inner and outer radii consistent for the torus. The inner radius for the scattering region was set based on the R-L relation from the infrared reverberation mapping \citep{kishimoto_etal_2007,koshida_etal_2014}. We adopted the prescription of \citet{savic_etal_2018} for the outer radius of this region, that is, the location of the outer radius of the torus was set to be at the location where the outer radius of the BLR subtends a half angle of 25$^{\rm o}$. The combined information from the inner and outer radius of the scattering region plus the electron number density allowed us to estimate the optical depth of this region. We obtained a grid of electron number density solutions to arrive at a preferred solution that fits best the spectral information gathered from the observed data. Table \ref{tab:stokes-parameters} lists the various input parameters used in our modeling for the three objects.

\begin{figure}[htp!]
\includegraphics[width=\columnwidth]{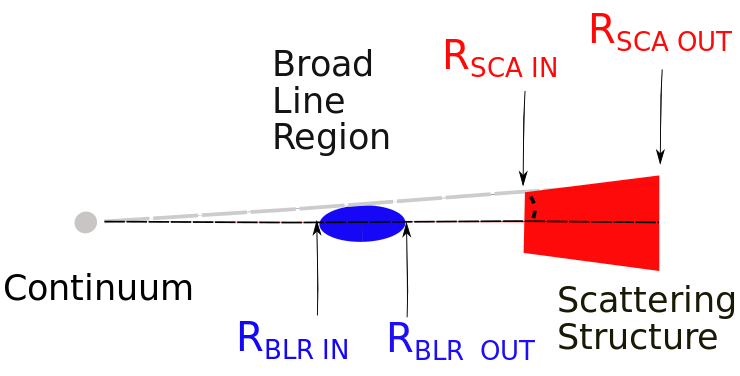}
\caption{Schematic view of the models with the equatorial scatterer. The continuum source is marked in gray, the BLR is marked in blue, and the equatorial scattering region is marked in red. We marked the half-opening angle of the scattering structure with a dashed curve between the {gray} and black dashed lines.}  
\label{fig:cartoon-no-polar}
\end{figure}

In Figure \ref{fig:cartoon-no-polar}, we show the schematic illustrations of the model setup for our equatorial scatterer, which includes the continuum source, the BLR, and the equatorial scattering region. We utilized the equatorial scatterer option, as in \citet{2021bowei}, but we also considered the option of the polar scatterer \citep[see e.g.,][]{smith2005}. To model the polar scatterers, we assumed the scatterers to be located at the same distance from the continuum source and analogous in size (R$_{\rm out}$ - R$_{\rm in}$) to the equatorial scattering region. We also kept the structure of the polar scatterers identical to their equatorial counterparts. We modulated the net density of the polar scatterer until we recovered a good fit for the polarized spectrum and the polarization fraction. This setup is illustrated in Figure  \ref{fig:cartoon-with-polar}.

\begin{figure}[htp!]
\includegraphics[width=\columnwidth]{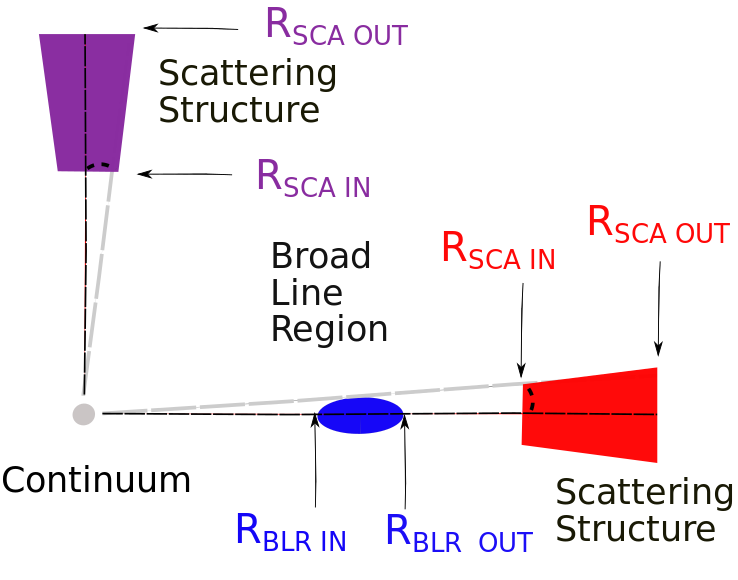}
\caption{Schematic view of the models with two scatterers: polar and equatorial. The various components are colored identically to Figure \ref{fig:cartoon-no-polar}. The polar scattering region is marked in purple.} 
\label{fig:cartoon-with-polar}
\end{figure}

\subsection{Comparison with the observed polarization properties}
\label{sec:polarization-stokes}
To reconstruct the observed spectrum and the corresponding polarized emission of the \ha{} emission line and the continuum around the line, we generated a simulated spectrum using {\sc STOKES} for each of the three objects. Specifically, we considered the spectrum in natural light ($F_{\lambda}$), the polarized spectrum ($p \times F_{\lambda}$), the polarization fraction ($p$) and the polarization angle ($\chi$). We ran a simulation of these four entities using the parameterization as listed in Table \ref{tab:stokes-parameters}, and we compared our results to those from observations.

In addition to the above setup, we also included a polar scattering region that helps recover the peak distribution in the polarized spectra and fractional polarization for Mrk 1044 and SDSS J080101.41+184840.7. These two objects are the ones where we recovered a viewing angle value that is slightly on the higher side (i.e., $\sim$54$^{\rm o}$ for Mrk 1044 and $\sim$31$^{\rm o}$ for the SDSS J080101.41+184840.7). However, for IRAS 04416+1215, which is the source for which we recovered the smallest viewing angle (i.e., $\sim$4$^{\rm o}$), we could reconstruct the polarized emission using the model with only the equatorial scattering region.

The polarization fraction (p\%) for the three sources (Mrk 1044, SDSS J080101.41+184840.7, and IRAS 04416+1215) including the polar scattering region are 0.40, 0.36, and 0.3, respectively. For IRAS 04416+1215, we found the p\% for the case with only the equatorial scatterer to be similar (i.e., 0.28). We note that the parameterization for the polar scattering region is unrestrained, as we lacked observational pieces of evidence to break the degeneracy between the optical depths, the density, the location, and the extension of these media. In addition, the half-opening angle for the polar scatterers was assumed to be identical to their counterparts in the equatorial region. With the assumption that the density of the medium remains rather unaffected there is a coupling between the opening angle and the optical depth of the scattering medium, i.e., an increase/decrease in the opening angle will lead to a decrease/increase in the optical depth along a given line of sight. The parameters used to model this polar scattering region for the two objects are tabulated in Table \ref{tab:stokes-parameters-polar}.

%%%%%%%%%%%%%%%%%%%%%%%%%%%%%% PLOTS

\begin{figure}[htb!]
\includegraphics[width=\columnwidth]{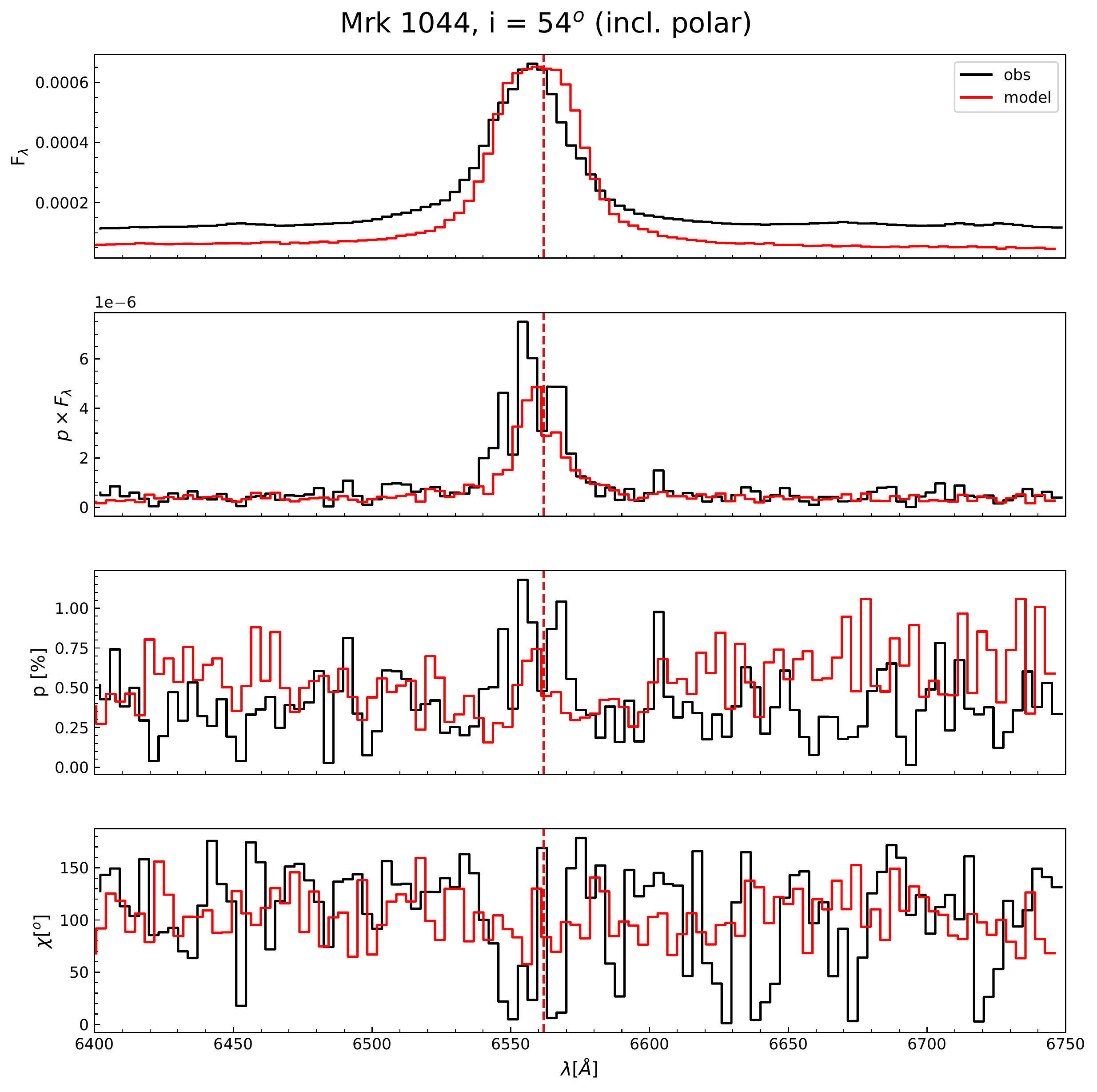}
\caption{STOKES modeling for Mrk 1044 with the equatorial plus polar scattering region and comparison with observational estimates. The case shown has a viewing angle of 54$^{\circ}$. {From top to bottom}: Spectrum in natural light ($F_{\lambda}$), polarized spectrum ($p \times F_{\lambda}$), polarization fraction ($p$), and polarization angle ($\chi$). The vertical red dashed line marks the central wavelength for \ha{}.
\label{fig:stokes-bc-only-mrk}}
\end{figure}

\begin{figure}[htb!]
\includegraphics[width=\columnwidth]{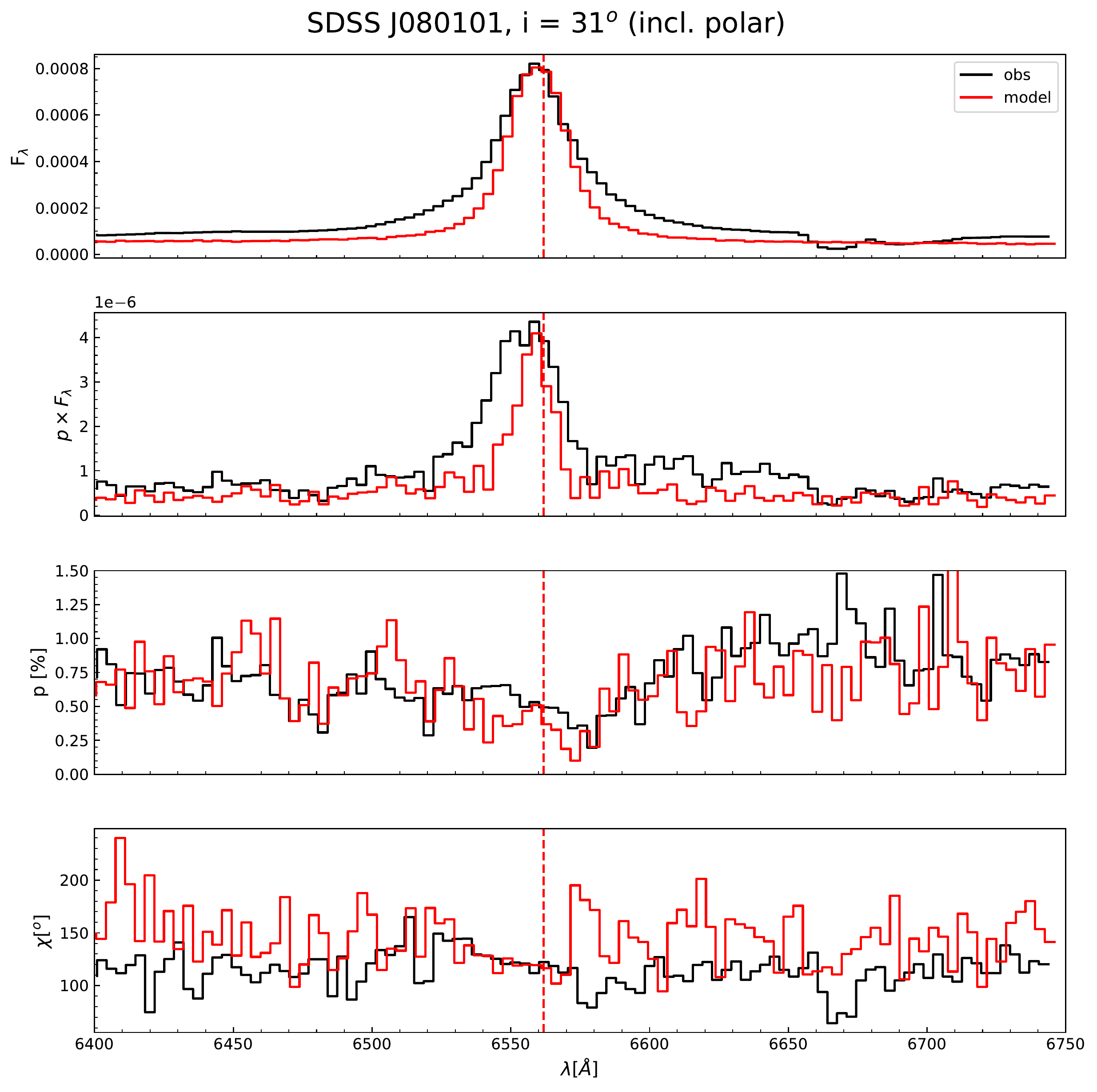}
\caption{STOKES modeling for SDSS J080101.41+184840.7 with the equatorial plus polar scattering region and comparison with observational estimates. The case shown has a viewing angle of 31$^{\circ}$. The panels depict the same parameters as described in Figure \ref{fig:stokes-bc-only-mrk}.  \label{fig:stokes-sdss}}
\end{figure}

\begin{figure}[htb!]
\includegraphics[width=\columnwidth]{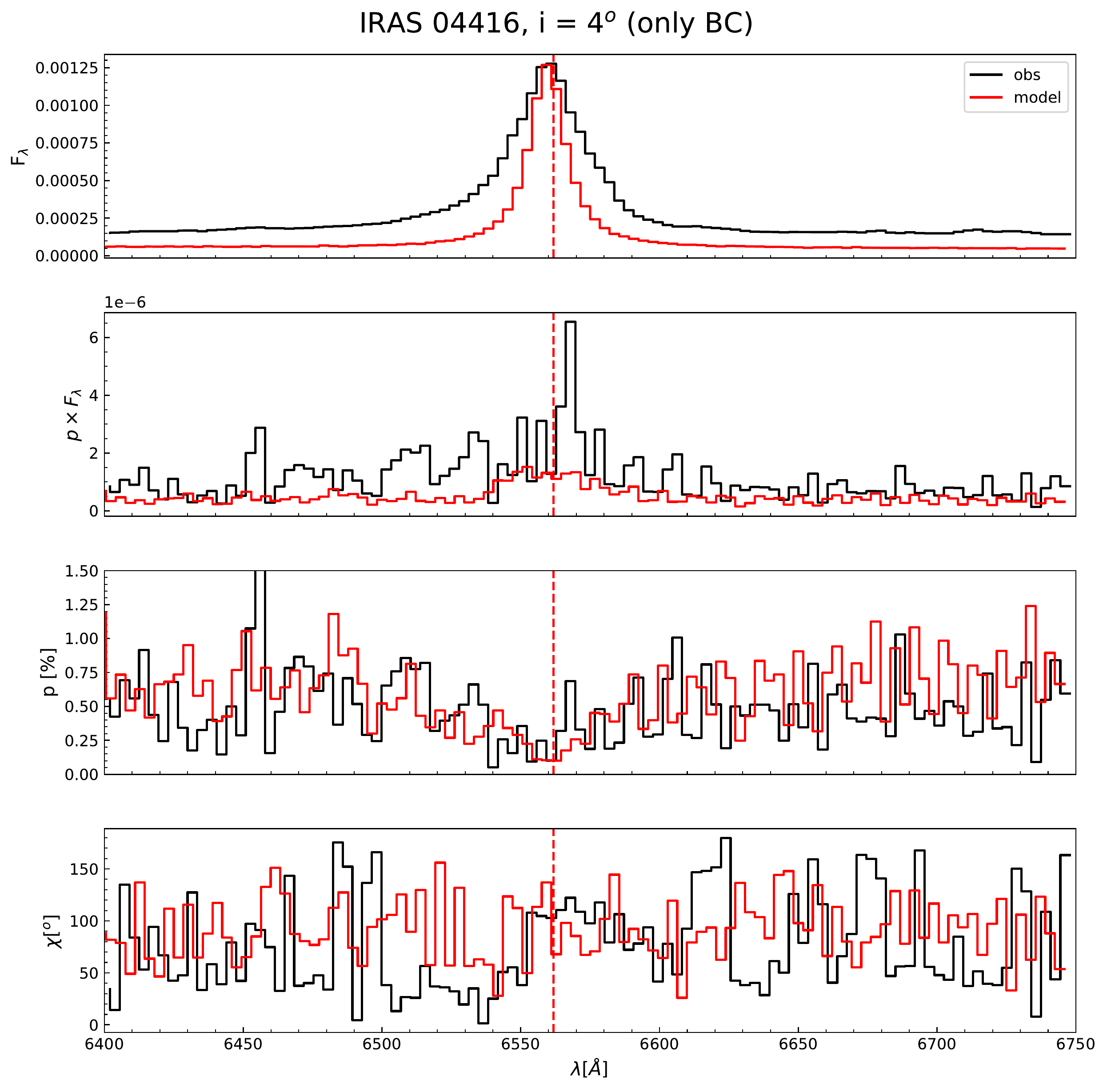}
\caption{STOKES modeling for IRAS 04416+1215 with only the equatorial scattering region and comparison with observational estimates. The case shown has a viewing angle of 4$^{\circ}$. The panels depict the same parameters as described in Figure \ref{fig:stokes-bc-only-mrk}.
\label{fig:stokes-iras}}
\end{figure}

We illustrate the performance of the recovery of the spectrum in natural and polarized light for our three objects, Mrk 1044, SDSS J080101.41+184840.7, and IRAS 04416+1215, in Figures \ref{fig:stokes-bc-only-mrk}, \ref{fig:stokes-sdss} and \ref{fig:stokes-iras}, respectively. 
{The polarization fractions on these plots (third panel from the top in Figures \ref{fig:stokes-bc-only-mrk}, \ref{fig:stokes-sdss} and \ref{fig:stokes-iras}) are notably higher than the mean values reported in Tab. \ref{tab:polarization-measurements}. We note that the mean values of the computed polarization fractions depend on the wavelength range employed and on the order of the executed steps. For example, for Mrk 1044, we obtained 0,67\% when computing the mean value for $P$ using $P$ value for each wavelength point for the full wavelength range (6101-10295\AA). Whereas, when we computed $P$ value using the means of the respective Q and U values (as explained in Sec. \ref{sec:polarization-measurements}), we obtained lower values (as presented in Tab. \ref{tab:polarization-measurements}). {We note that the difference between averaging $P$ and averaging ($Q$,$U$) before computing $P$ is due to the noise that introduces a bias on $P$. In the absence of noise, both the averages would be identical.}}

{The inclusion of the polar scatterers provided an additional flux close to the line center that the equatorial scatterers are not sufficient in providing -- at least in the case of Mrk 1044 and SDSS J080101.41+184840.7. Nevertheless, the equatorial scatterers are necessary to recover wider line profiles.} For comparison purposes, we show the results of our {\sc STOKES} modeling for Mrk 1044 and SDSS J080101.41+184840.7 when excluding a polar scattering region in Figures \ref{fig:stokes-bc-only-mrk-no-polar} and \ref{fig:stokes-bc-only-sdss-no-polar}. Noticeably, the models without the polar scatterers are unable to recover the ``peaky'' profiles obtained in the observed spectrum, both in natural and polarized light. In addition, the polarization fraction is substantially underpredicted in these models, especially at the position marking the line center for \ha{}. We show the case with both the polar and equatorial scattering regions for IRAS 04416+1215 in Figure \ref{fig:stokes-iras-w-polar}. As shown in this figure, the polarization level becomes significantly lower when the polar scatterers are included in the {\sc STOKES} modeling. {We note that the inclination angles used to compare the models to the observations in each of these plots were set as per the estimated value from the observations. For the cases of the SDSS J080101.41+184840.7 and IRAS 04416+1215, the wings in the full profile are much wider than predicted in the models. Larger viewing angles can account for these deficits. We discuss this aspect in the Sec. \ref{sec:pca}.}

%%%%%%%%%%%%%%%%%%%%%%%%%%%%%%% Table
\begin{table*}[]
\centering
\caption{STOKES modeling parameters (without polar scattering region)}
\label{tab:stokes-parameters}
\resizebox{\textwidth}{!}{%
\setlength{\tabcolsep}{0.5em} % for the horizontal padding
\renewcommand{\arraystretch}{1.5}% for the vertical padding
\begin{tabular}{c|c|cccc|cccccccc}
\hline\hline\noalign{\vskip 0.1cm}
\multirow{3}{*}{Object} & \multirow{2}{*}{viewing angle} & \multicolumn{4}{c|}{Broad-line region} & \multicolumn{8}{c}{Scattering region (Equatorial)} \\
 &  & Geometry & R$_{in}^{BLR}$ & R$_{out}^{BLR}$ & $v_{\rm avg}^{BLR}$ & Geometry & R$_{in}^{Sca}$ & R$_{out}^{Sca}$ & half-opening angle & $v_{\rm avg}^{Sca}$ & type & density & optical depth \\
 & [deg.] &  & [light days] & [light days] & [km s$^{-1}$] &  & [light days] & [light days] & [deg.] & [km s$^{-1}$] &  & [cm$^{-3}$] &  \\
 \hline\hline\noalign{\vskip 0.1cm}
Mrk 1044 & 54 & Cylindrical & 10.5 & 41.503 & 745.394 & Flared-disk & 57.018 & 89.0 & 35 & 373.266 & electron & 2.53$\times 10^{5}$ & 0.229 \\
SDSS J080101.41+184840.7 & 31 & Cylindrical & 8.3 & 121.920 & 688.783 & Flared-disk & 219.287 & 549.947 & 35 & 283.395 & electron & 1.06$\times 10^{5}$ & 0.34 \\
IRAS 04416+1215 & 4 & Cylindrical & 13.3 & 146.580 & 621.37 & Flared-disk & 276.068 & 661.182 & 35 & 256.727 & electron & 1.06$\times 10^{5}$ & 0.396 \\
\hline
\end{tabular}%
}
\end{table*}
%%%%%%%%%%%%%%%%%%%%%%%%%%%%%%%

%%%%%%%%%%%%%%%%%%%%%%%%%%%%%%% Table
\begin{table*}[]
\centering
\caption{STOKES modeling parameters for the polar scattering region for Mrk 1044 and SDSS J080101+184840.7}
\label{tab:stokes-parameters-polar}
\resizebox{\textwidth}{!}{%
\setlength{\tabcolsep}{0.5em} % for the horizontal padding
\renewcommand{\arraystretch}{1.5}% for the vertical padding
\begin{tabular}{c|cccccccc}
\hline\hline\noalign{\vskip 0.1cm}
\multirow{3}{*}{Object} & \multicolumn{8}{c}{Scattering region (Polar)} \\
 & Geometry & R$_{in}^{Sca}$ & R$_{out}^{Sca}$ & half-opening angle & $v_{\rm avg}^{Sca}$ & type & density & optical depth \\
 &  & [light days] & [light days] & [deg.] & [km s$^{-1}$] &  & [cm$^{-3}$] &  \\
 \hline\hline\noalign{\vskip 0.1cm}
Mrk 1044 & Double-cone & 57.018 & 89.0 & 35 & 373.266 & electron & 1.11$\times 10^{6}$ & $\lesssim$1.0 \\
SDSS J080101.41+184840.7 & Double-cone & 219.287 & 549.947 & 35 & 283.395 & electron & 6.24$\times 10^{5}$ & $\lesssim$2.0 \\
IRAS 04416+1215 & Double-cone & 276.068 & 661.182 & 35 & 256.727 & electron & 1.06$\times 10^{5}$ & $\lesssim$0.396 \\
\hline
\end{tabular}%
}
\end{table*}

%

%%%%%% STOKES NEW w chi2 analysis

\begin{figure*}
    \centering
    \includegraphics[width=\textwidth]{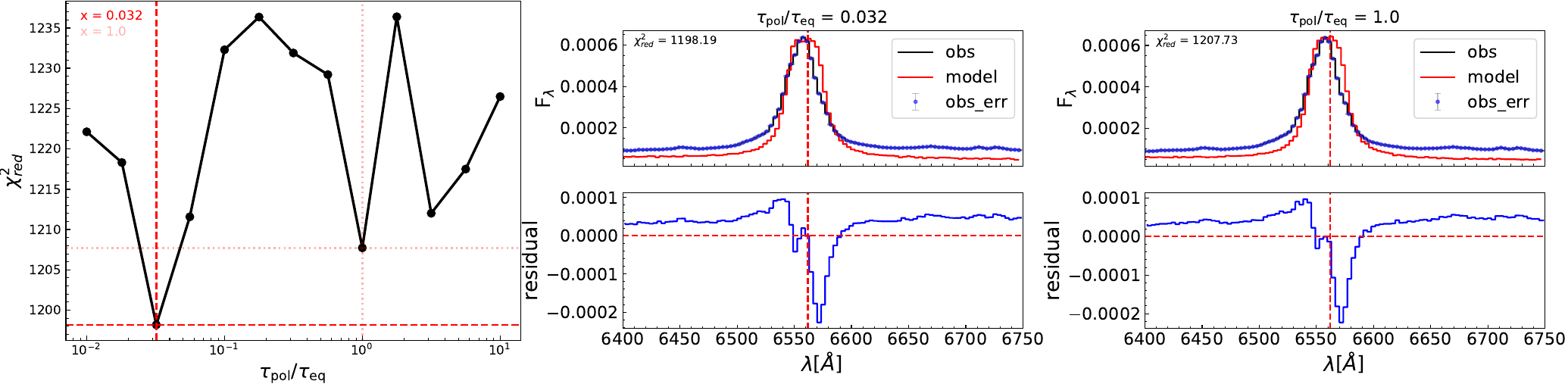}\\
    \includegraphics[width=\textwidth]{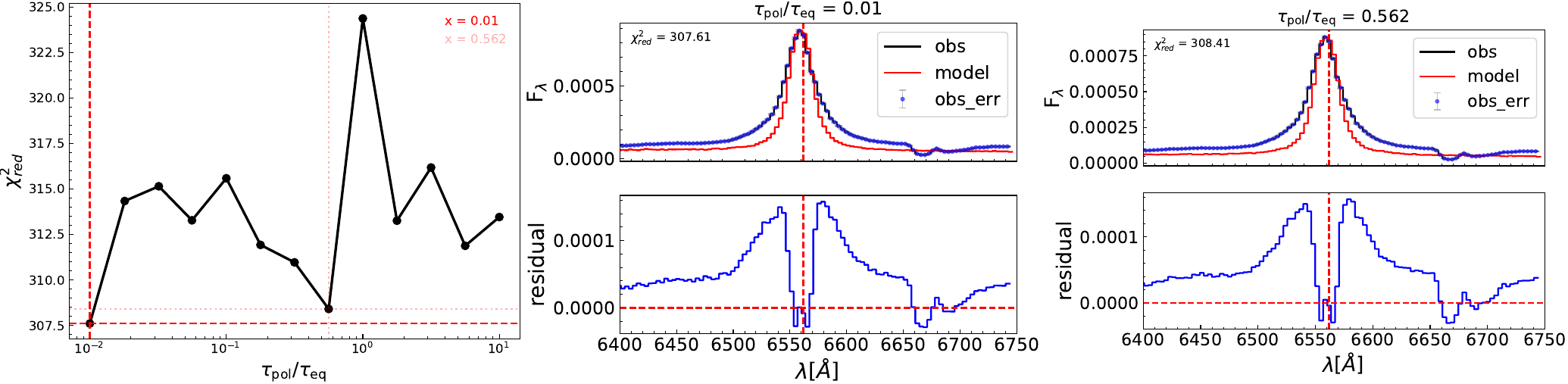}\\
    \includegraphics[width=\textwidth]{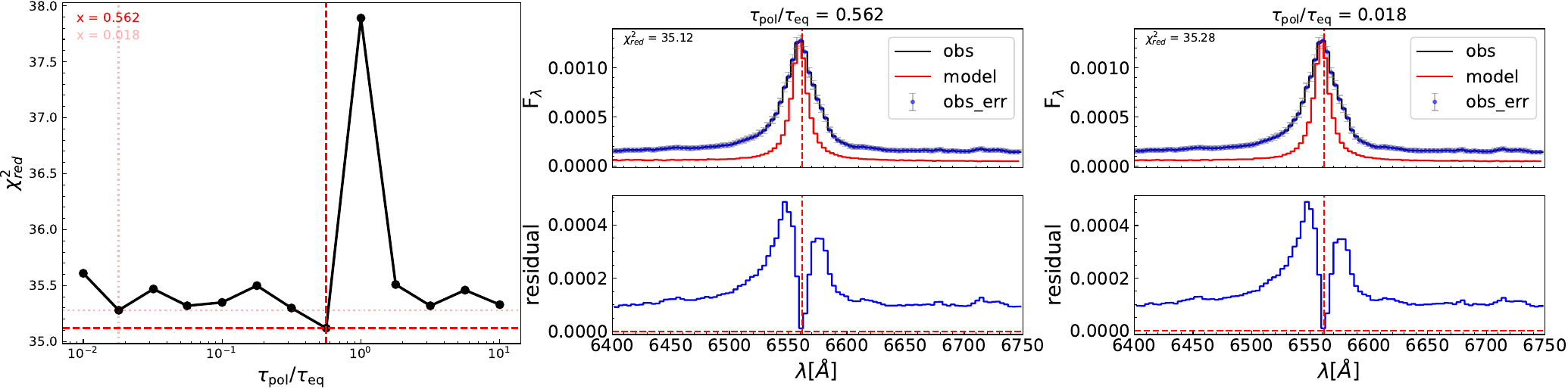}    
    \caption{{Distribution of $\chi^2$ for a range of $\tau_{\rm pol}$/$\tau_{\rm eq}$ for our three sources, Mrk 1044 (top rows), SDSS J080101 (middle rows), and IRAS 04416 (bottom rows). This range is the ratio of the optical depth in the polar region to the equatorial region. The location and the corresponding value of the ratio ($\tau_{\rm pol}$/$\tau_{\rm eq}$) of the minimum and second-to-minimum $\chi^2$ values are marked with dashed and dotted lines, respectively. Middle and right columns: Fit for the total flux for each source corresponding to the minimum $\chi^2$ and the second-to-minimum $\chi^2$.}}
    \label{fig:chi2-analyses}
\end{figure*}

{To disentangle the contribution from the polar scatterers to our modeling, we performed an additional test. Keeping the optical depth for the equatorial scattering region unchanged for our three sources (see values in the last column in Table \ref{tab:stokes-parameters}), we prepared STOKES models for a range of $\tau_{\rm pol}$/$\tau_{\rm eq}$ = [10$^{-2}$, 10] (i.e., the ratio of the optical depth in the polar scattering region to the optical depth in the equatorial scattering region.\footnote{other input parameters for the STOKES modeling are kept identical to our previous setup.} The results are shown in Figure \ref{fig:chi2-analyses}. The left panels in this figure show the reduced-$\chi^2$ distributions.\footnote{here, the total degrees of freedom is 101, i.e., the length of the wavelength interval considered for the \ha{} region (100) plus 1 degree of freedom for the optical depth.} These reduced-$\chi^2$ distributions show multiple minima. We mark the location of the minimum $\chi^2$ and the second-to-minimum $\chi^2$ for each case. The adjacent panels show the fit for the total flux ($F_{\lambda}$) corresponding to the two cases of $\tau_{\rm pol}$/$\tau_{\rm eq}$ corresponding to the two minimum $\chi^2$ results. The remaining parameters ($p \times F_{\lambda}$, $P$ and the $\chi$) show similar behavior. The two minima are comparable, suggesting that the best fit can be obtained with a polar region whose optical depth is either low ($\sim$ 0.01-0.032) or moderate ($\sim$ 0.562-1.0, where 1.0 indicates the same optical depth as that of the equatorial scattering region). Therefore, in summary, the presence of a polar scatterer is quite certain since the representative solutions indicate a non-zero value for the $\tau_{\rm pol}$/$\tau_{\rm eq}$. We however note that obtaining a good fit for all the observables for our sources is not trivial and is affected by the noise in the observed spectra.} 

%%%%%%%%%%%%%%%%%%%%%%%%%%%%%%%

\section{Discussion and conclusions}
\label{sec:discussion}

We performed VLT-FORS2 spectropolarimetric observations of three AGNs that belong to the NLSy1 class and were claimed to accrete at super-Eddington rates. Our goal was to check whether the Eddington rates are not overestimated by the specific orientation of the sources. Polarimetric measurements, allow a unique way to estimate the viewing angle for the observed sources, even for top-view sources, and can reveal the true range of the Keplerian velocities of the BLR. 

The viewing angle of 54$^{\circ}$ for Mkn 1044  measured in this paper when comparing the natural light width and the width in the polarized light is comparable to  $i$ = $47.2^{+1.0}_{-2.5}$ obtained by \citet{mallick2018} from the joint fitting of Swift, XMM-Newton, and NuSTAR X-ray spectra for Mrk 1044.  

The viewing angle of SDSS J080101 using our method is 31$^{\circ}$, and we did not find any independent measurements of the viewing angle for this source in the literature. As for IRAS 04416+1215, our determination of the viewing angle gave a very small value of 4$^{\circ}$. Our formal (systematic) error on this value is small, but what we do not include here is a possible systematic error due to uncertainty in the $H/R$ factor in Equation~\ref{eq:viewing-angle}. On the other hand, \cite{tortosa_etal_2022} analyzed XMM-Newton and NuSTAR observations and were only able to get a highly model-dependent upper limit for the viewing angle for this source in the range of 25 to 45 degree. If the viewing angle is indeed higher than 4$^{\circ}$, it would mean that the random motions in this source are less important (i.e., $H/R$ in Equation~\ref{eq:viewing-angle} is smaller than the one-third value adopted in the current paper). We discuss the issue of the viewing angle of IRAS 04416+1215  more in Sec. \ref{sec:pca}. 

Our results confirmed that the sources are high accretors. The black hole mass values we obtained for each of the three sources  from the polarized spectra are not systematically higher than the masses obtained earlier using reverberation mapping (see Table \ref{tab:mass-calculation}). 
The high accretor character of all three sources is also supported by our  analysis of their X-ray spectra, which revealed features characteristic of high Eddington ratio sources: steep X-ray spectra and large soft X-ray excess \citep{tortosa_etal_2022,mallick2018,Liu_Henzen2021}.    

\subsection{Polarization level}

The overall level of polarization in the three sources is low, 0.5\% or less, typical for Type 1 AGNs \citep[see][]{smith2004, Robinson2011ASPC..449..431R, 2019afanasiev}.

The polarization in Mrk 1044 has been measured previously by \citet{grupe1998}, who gave a polarization continuum of { 0.52\%$+/-$0.05 and a polarization angle of 144$^{\circ}$. }This is roughly consistent with our results for the continuum (although our value is even lower, {0.23\%$+/-$0.04)}. Their observations were performed with a 2.1 m telescope at the McDonald Observatory, which did not allow  for specific study of the wavelength-dependent polarization. 
{In general, a higher value for continuum polarization \citep{Goodrich1989} may be due to a wavelength range that is different from our measurements, as we do not have data for the rising blue side of the spectra.}

\citet{Goodrich1989} reports spectropolarimetric observations of Mrk 1044 and the general effect of degree of polarization rising toward the blue side. We did not notice this effect, but our observations have different wavelength ranges (for the Mrk 1044 observation in \citet{Goodrich1989}, it is 4436-7210\AA, whereas our measurements start from $\sim$ 6000\AA). The mean  degree of polarization reported for Mrk 1044 in our work is 0.15\% $\pm$ 0.01\%, whereas in \citet{Goodrich1989}, the mean degree of polarization is 0.46\% $\pm$ 0.05\%, which confirms the low level of degree of polarization.

{After analyzing measurements for a sample of 16 NLSy1, \citet{Robinson2011ASPC..449..431R}
concluded that the spectropolarimetric properties have a physical rather than a geometrical origin. 
NLSy1 galaxies tend to have a high Eddington ratio and high accretors that show more extreme physical conditions, such as higher metallicities 
\citep[e.g.,][]{shinetal13,sulenticetal14,pandaetal20} 
Moreover, \citet{Robinson2011ASPC..449..431R} reported a peculiar, prominent red wing in  H$\alpha$ in the polarized flux of Mrk 1239, a characteristic of a polar scattering outflow. We did not notice redshifted asymmetries in the polarized H$\alpha$ spectra of our sample. }

{The polarization measurements made by \citet{Lira2021} using ESO/FORS1 observations of NGC 3783, which has a low mass (2.4-3.5$\times 10^7$ M$_{\odot}$ \citet{onkenpeterson2002}), and Mrk 509, which has a relatively high mass (1.4$\times 10^7$ M$_{\odot}$; \citet{peterson2004}), exhibit polarization levels of $\sim$0.7\% and $\sim$0.9, respectively. These are low Eddington ratio sources that show relatively high levels of polarization, unlike the three sources studied in this work, even after correcting for contamination} from the interstellar medium (ISM).

Older papers by \citet{smith2004} and \citet{smith2005} contained many objects with polarization larger than 0.5\%. The recent sample by \citet{2021capetti} shows the mean level of polarization at 0.75\% (the median is at 0.59\%), but their sources do not preferentially belong to the high Eddington ratio class. What seems to be a difference between the sources of previous spectropolarimetry sources studied in previous surveys, such as the ones of \citet{AfaPop2015,2019afanasiev}, is the requirement of an additional polar scatterer. We hypothesize that the occurrence of a polar scatterer could be a genuine difference between sources radiating at high Eddington ratios and sources radiating at lower Eddington ratios (\citealt{sulentic_etal_2000}).

The low polarization in our sample requires the presence of both the equatorial and polar scatterers for the two objects Mrk
1044 and SDSS J080101.41+184840.7. In the third object, the polar scatter is not required, but it is not excluded. 
High Eddington ratio sources in general are expected to have more massive outflows that might lead to higher polarization rather than lower polarization, but this is not what we observed. The upper limits for the optical depth of the polar scatter are quite high, but the geometric location of the scatterer along the symmetry axis apparently does not imprint high polarization.

\subsection{Comparison of sources' properties with the sample from \citealt{2021capetti}}
We decided to compare the spectral properties of our sample to a larger set of spectropolarimetric measurements which contains 25 AGN from the same instrument carried out recently by \citealt{2021capetti} (24 objects) and by \citealt{2021bowei} (1 object). In the paper of
\citealt{2021capetti}, the whole sample is a set of 25 objects. However, the BLR polarization measurements of source J145108 were affected by a cosmic ray, as the authors mentioned, and we did not include this source. For J140700, two observations were performed, and we chose an observation with a higher S/N = 413 {(per resolution element at 6750\AA)}. 

Additionally, the $P_{line}$ from this work is the same quantity as P$_{BLR}$ from \citealt{2021capetti}.
 To illustrate this, we calculated the window from the definition postulated in \citealt{2021capetti} for Mrk 1044. The FWHM(\ha{}) from our spectral fitting is 1290 km/s, which gives 28\AA. After doubling this value, as in \citealt{2021capetti}, we obtained 56\AA, which is comparable with the value of $\approx$50\AA \  in this work. 

 We decided to use the $P_{line}$ term since, as we show for our objects, they may contain a narrow line component in this part of the polarized spectrum. In Figure \ref{fig:pblr-to-ledd}, we show three plots of the $P_{line}/P_{cont}$ - $\lambda_{Edd}$ plane.\footnote{ Corollary plots showing the correlations between the $\lambda_{Edd}$ and the (a) viewing angle, (b) FWHM of the polarized \ha{} emission line, and (c) $R_\mathrm{FeII}$ are shown in Figure \ref{fig:alternate-figures}} Each plot contains a sample from \citealt{2021capetti}, Fairall 9 measurements from \citep{2021bowei}, and measurements for our sources. For Fairall 9 and our sources, we labeled the points. We color coded the $P_{line}/P_{cont}$  Eddington ratio ($\lambda_{Edd}$) to identify three quantities: the viewing angle, FWHM \ha,{}  and  $R_\mathrm{FeII}$.
To calculate $P_{line}/P_{cont}$ from our sample, we used measurements from Table \ref{tab:polarization-measurements}.

In order to calculate the Eddington ratio of the sources from our sample, we used the following equations. For bolometric luminosity, we used the bolometric correction of \citet{2019netzer} (Eq. 3 therein), which in our case (for \l5100{}) is
    \begin{equation}
        k_{BOL}=40 \left( \frac{L_{5100\AA}}{10^{42}\,{\rm erg s^{-1}}}\right)^{-0.2}.
    \end{equation}
    For Eddington luminosity, we used the masses that we calculated and present in Table \ref{tab:mass-calculation}.

For color coding, we used the viewing angle and FWHM(\ha{}) BC from Table \ref{tab:viewing-angle} and the $R_\mathrm{FeII}$ from Table \ref{tab:rfe}.

We plot in Fig. \ref{fig:pblr-to-ledd} the sample from \citealt{2021capetti} using the  $P_{line}/P_{cont}$ values from Table 3 in their paper (ratio of column 5 and column 2). For our sample, we used values from Table \ref{tab:polarization-measurements} (for Mrk 1044, $P_{line}/P_{cont}$ = 1.59; for SDSS, J080101.41+184840.7 $P_{line}/P_{cont}$ = 0.55; and for IRAS 04416+1215, $P_{line}/P_{cont}$ = 0.19).

For Fairall 9, we used values from the analysis of \citep{2021bowei}: $P_{line}/P_{cont}$ = 0.89 ($P_{line}$ = 0.94, $P_{cont}$ = 1.06).
To calculate the Eddington ratio of the sources using the same method as our sample, we used \mbh{} and \l5100{}\ from Table 1 from \citealt{2021capetti}. For the color coding, we used the viewing angle we calculated from Eq. \ref{eq:viewing-angle}, wherein for the $V_{Kep}$ we input values of inter-quartile width (between 25\% and 75\%) of the BLR in total intensity from \citealt{2021capetti} (Table 3, column 8 in their paper), and for $\Delta V_{obs}$, we input the FWHM of non-polarized \ha{} BC (\citealt{2021capetti}; Table 1, column 7 in their paper). We also used the FWHM of polarized \ha{} BC (Table 3, column 9 from their paper). Finally, we calculated $R_\mathrm{FeII}$ from EW(Fe$_{II}$) and EW(\hb{}) BC from the catalog of \citep{s11} for each source from the sample of \citealt{2021capetti}.

For Fairall 9, \citet{2021bowei} determined the viewing angle of the source as 50$^{\circ}$ based on the polarization level and polarization angle fit to the data. For our sources, we determined this viewing angle from a simple comparison of the FWHMs in natural and polarized light, following Equation~\ref{eq:viewing-angle}. To achieve consistency with our sample and to gain insight into the accuracy of the viewing angle determination, we used the data from \citep{2021bowei} and adopted our method, based on Equation \ref{eq:viewing-angle}. In this case, we obtained the value $\sim$27$^{\circ}$ for the viewing angle, which is smaller compared to the original determination by \citet{2021bowei}. We used both values in the plot, marking them as Fairall 9, our value, and Fairall (II), the original value.  However, as \cite{2021bowei} have commented, if the object has a more complex structure (i.e., a warped structure), a single value of the viewing angle may only be illusory.
%here i finished
We computed the Eddington ratio with the method used for the rest of the sample.
We used values for \l5100{} and \mbh{} from \citealt{dupu2015}. This value of the Eddington ratio is plotted in the middle and right panels of Figure \ref{fig:pblr-to-ledd}.

The outlying source on the $P_{line}/P_{cont}$ - $\lambda_{Edd}$ diagram with the highest  $P_{line}/P_{cont}$ ratio is J110538. This high ratio is caused by the extremely low continuum polarization in comparison to the rest of the \citep{2021capetti} sample.
In the left panel, color coded with the viewing angle, we do not have the viewing angle calculation for J140336 since the FWHM of non-polarized \ha{} is broader than the polarized FWHM, thus making Eq. \ref{eq:viewing-angle} unusable for this case. We mark this source as a black dot in this plot.
For the right panel, we color-coded with $R_\mathrm{FeII}$, and after cross match g with the catalog of \citep{s11}, we noticed no EW(Fe {\sc ii}) measurement for J154743 and no EW(\hb{}) BC  measurement for J084600.
After visual inspection of those spectra, we saw no strong Fe {\sc ii} emission for J154743 nor for J084600, and we decided not to fit the Fe {\sc ii} template in order to avoid biasing the sample.

A quick look at those three panels does not reveal any obvious trend for the color-coded values. We suppose there is a weak trend with decreasing non-polarized FWHM of \ha~ along the x-axis (i.e., Eddington ratio). However, to confirm this trend (and others), we performed a principal component analysis for the full sample, which is  shown in the next section. 
The increasing $R_\mathrm{FeII}$ along the Eddington ratio is a confirmation of the expected behavior from many previous works \citep{2015sulenticmarziani, panda_etal_2019b}.

\begin{figure*}[htp!]
\includegraphics[scale=0.3]{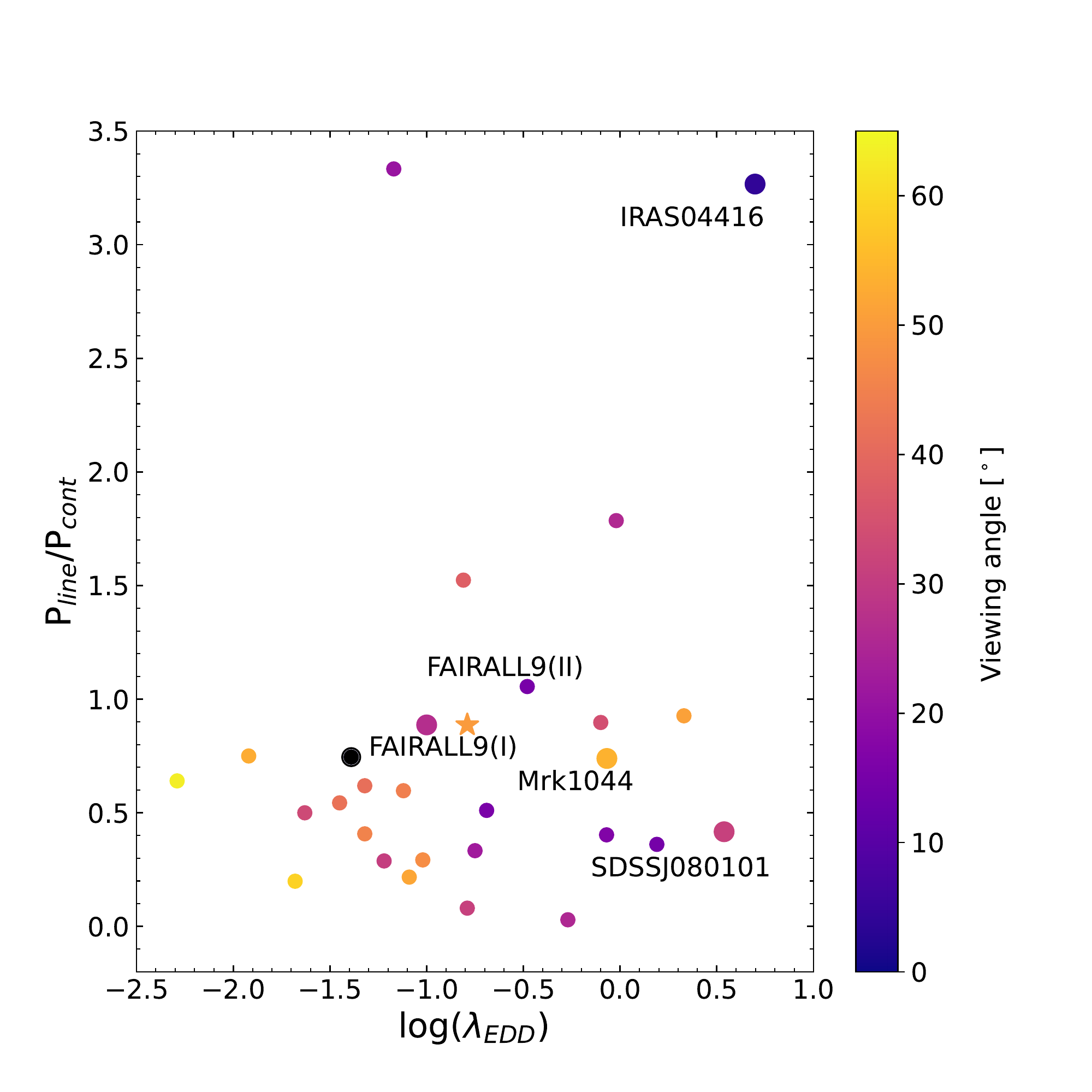}
\includegraphics[scale=0.3]{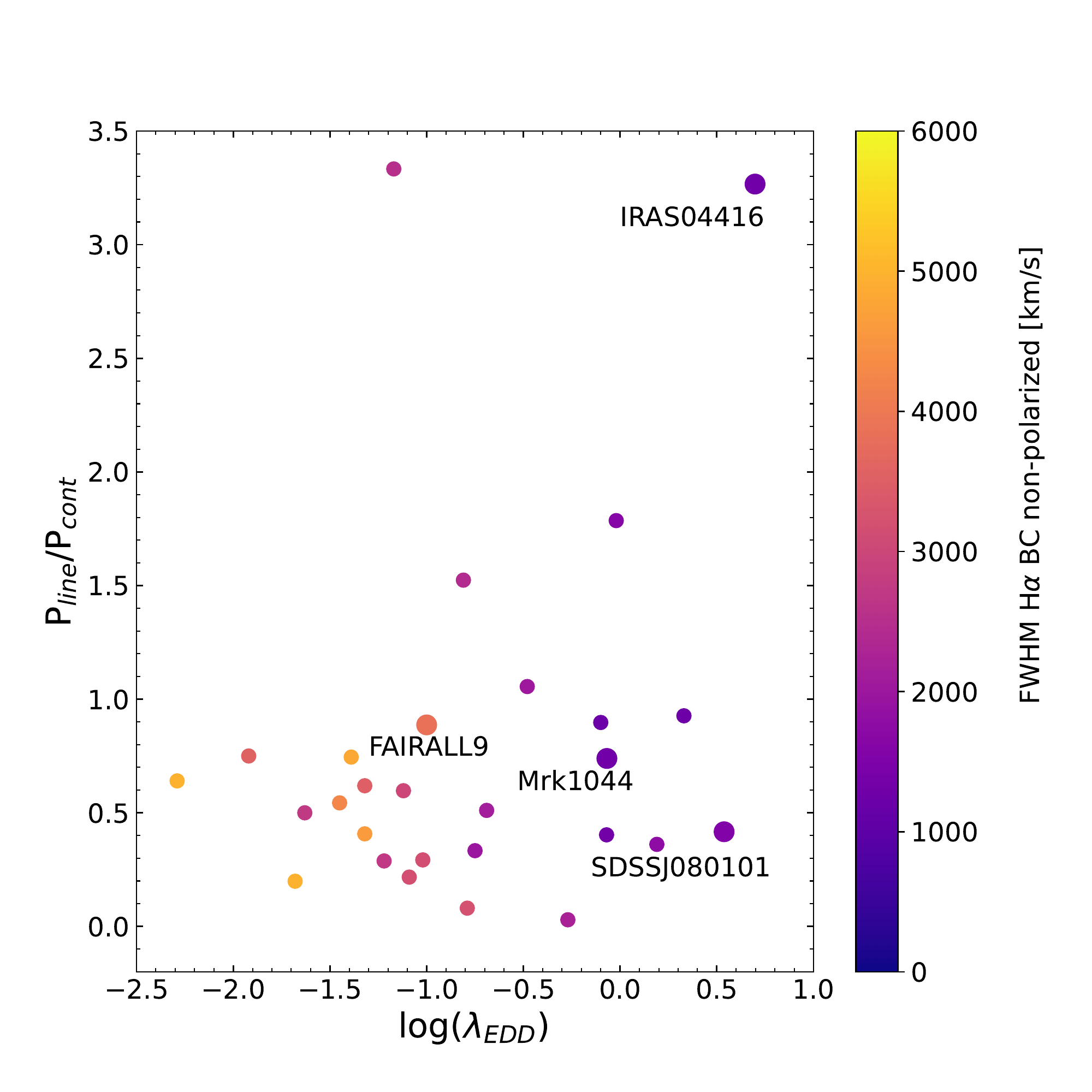}
\includegraphics[scale=0.3]{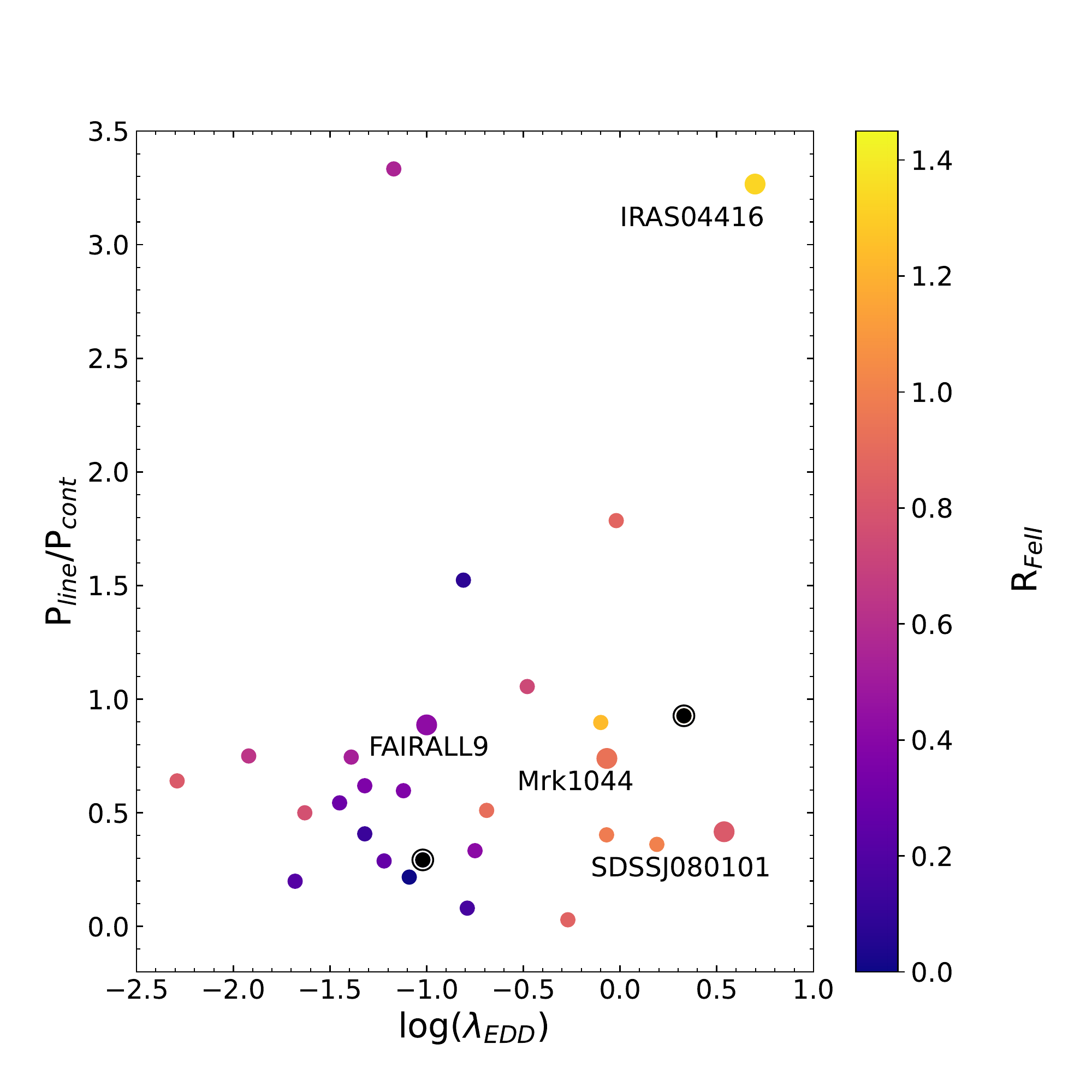}

\caption{Plane of $P_{line}/P_{cont}$ - $\lambda_{Edd}$ for sources from  \cite{2021capetti} (marked as dots), Fairall 9 (labeled and marked as a star), and source from this work (with labels). %As gray dashed, 'valley', we marked plane less preferable for this sample. 
Left panel: The color code is based on the viewing angle. We indicate Fairall 9 with the viewing angle measurement from \cite{2021bowei} with a star and label it as Fairall (II). We indicate Fairall 9 with the viewing angle measurement based on the method from this work as a dot and label it as Fairall (I).
The black point is {the source} J140336 from the sample of \cite{2021capetti}. We do not provide the viewing angle {for this source since the FWHM$_{\rm unpolarized}$ $>$ FWHM$_{\rm polarized}$ (see Equation \ref{eq:viewing-angle}}). Middle: The same sample color coded with the FWHM of a broad \ha{}. The sources from \cite{2021capetti} are plotted with a broad \ha{} measurement from \cite{2021capetti}, and Fairall 9 is plotted with a broad \ha{} measurement from \cite{2021bowei}. In the case of our sources (the three sources observed with VLT), we used measurements from Table \ref{tab:viewing-angle}.
Right panel: The same sample color coded with $R_\mathrm{FeII}$. The sources from \cite{2021capetti} with $R_\mathrm{FeII}$ were calculated from \cite{s11} values. 
The black points are sources: J154743 (no EW(Fe {\sc ii}) measurement in \citealt{s11}) and J084600 (no EW(\hb) BC and no EW(Fe {\sc ii}) measurement in \citealt{s11}).
\label{fig:pblr-to-ledd}}
\end{figure*}

\subsection{Searching for the driver with principal component analysis}
\label{sec:pca}
Principal component analysis (hereafter, PCA) is a "linear" dimensionality reduction technique where the dataset is projected onto a set of orthogonal axes that explain the maximum amount of variability in the dataset. The PCA technique works by initially finding the principal axis along which the variance in the multidimensional space (corresponding to all recorded properties) is maximized. This axis is known as "eigenvector 1." Subsequent orthogonal eigenvectors, in order of decreasing variance along their respective directions, are found, until the entire parameter space is spanned \citep[see, for example,][]{borosongreen1992,kur09,wildy2019,cafe_pca}. The PCA method is particularly useful when the variables within a dataset are highly correlated. Correlation indicates that there is redundancy in the data. Due to this redundancy, PCA can be used to reduce the original variables into a smaller number of new variables (principal components, or PCs), explaining most of the variance in the original variables. This approach allows us to determine correlated parameters, and in the context of our work, we utilized this technique to determine the physical parameter(s) that lead to the correlations shown in Figure \ref{fig:pblr-to-ledd}.

\begin{table*}[]
\caption{Values used for the PCA. }
\label{tab:pca_table}
\begin{tabular}{llccccccccc}
\hline
\hline
\# & NAME                    & FWHM(\ha{}) & FWHM(\ha{}) &  EW(\hb{})& EW(Fe{\sc ii})&\l5100{}  &$i$  &    M$_{\rm BH}$& $R_\mathrm{FeII}$               & $\lambda_{\rm Edd}$         \\
 &                   & unpolarized & polarized & &  &   &   &   &                 &            \\
  &                   & [km s$^{-1}$] & [km s$^{-1}$] & [$\AA$] &[$\AA$] &  [erg s$^{-1}$]&   [deg.]  &  [M$_{\odot}$] &                 &            \\
\hline
1  & SDSSJ031027.82-004950.7  & 1974 & 3890 & 42.0  & 17.6  & 44.1  & 22.50 & 7.93 & 0.42 & -0.75 \\
2  & SDSSJ074352.02+271239.5  & 2148 & 5020 & 110.2 & 100.7 & 45    & 15.56 & 8.59 & 0.91 & -0.69 \\
3  & SDSSJ083535.80+245940.1  & 2239 & 4120 & 100.6 & 87.1  & 45.3  & 25.42 & 8.41 & 0.87 & -0.27 \\
   & SDSSJ084600.42+130812.0* & 3186 & 3940 & -     & -     & 45    & 47.45 & 8.92 & -    & -1.02 \\
4  & SDSSJ100402.61+285535.3  & 1773 & 4220 & 51.3  & 51.6  & 45.4  & 14.82 & 8.03 & 1.01 & 0.19  \\
5  & SDSSJ100447.60+144645.5  & 3551 & 4110 & 42.8  & 27.2  & 43.9  & 52.85 & 8.94 & 0.64 & -1.92 \\
6  & SDSSJ100726.10+124856.2  & 4603 & 5860 & 38.1  & 4.6   & 45.4  & 45.34 & 9.54 & 0.12 & -1.32 \\
7  & SDSSJ105151.44-005117.6  & 3222 & 5230 & 78.6  & 12.9  & 45.6  & 31.20 & 9.17 & 0.16 & -0.79 \\
8  & SDSSJ110205.92+084435.7  & 1618 & 2960 & 56.9  & 49.9  & 45.1  & 25.67 & 8.0  & 0.88 & -0.02 \\
9  & SDSSJ110538.99+020257.3  & 2523 & 5170 & 85.9  & 47.4  & 44    & 20.88 & 8.27 & 0.55 & -1.17 \\
10 & SDSSJ113422.47+041127.7  & 2710 & 4460 & 76.5  & 20.7  & 44.1  & 30.53 & 8.4  & 0.27 & -1.22 \\
11 & SDSSJ114306.02+184342.9  & 2422 & 3460 & 78.6  & 6.0   & 45.1  & 37.99 & 8.79 & 0.08 & -0.81 \\
   & SDSSJ140336.43+174136.1* & 4808 & 4380 & 40.5  & 21.5  & 44.9  & -     & 9.21 & 0.53 & -1.39 \\
12 & SDSSJ140621.89+222346.5  & 1225 & 1870 & 51.0  & 63.1  & 44.2  & 34.33 & 7.36 & 1.24 & -0.1  \\
13 & SDSSJ140700.40+282714.6  & 4255 & 5700 & 66.6  & 19.8  & 44.3  & 41.90 & 8.79 & 0.30 & -1.45 \\
14 & SDSSJ142613.31+195524.6  & 2020 & 4740 & 93.7  & 68.7  & 44.6  & 15.40 & 8.06 & 0.73 & -0.48 \\
15 & SDSSJ142725.04+194952.2  & 3515 & 4750 & 70.3  & 25.6  & 44.7  & 41.35 & 8.98 & 0.36 & -1.32 \\
16 & SDSSJ142735.60+263214.5  & 4982 & 5410 & 123.7 & 26.4  & 45.1  & 59.14 & 9.66 & 0.21 & -1.68 \\
17 & SDSSJ154007.84+141137.0  & 3003 & 3870 & 74.9  & 27.7  & 44.3  & 44.49 & 8.46 & 0.37 & -1.12 \\
18 & SDSSJ154019.56-020505.4  & 4978 & 5220 & 79.2  & 64.9  & 44.1  & 63.31 & 9.47 & 0.82 & -2.29 \\
   & SDSSJ154743.53+205216.6* & 3186 & 3730 & 91.4  & -     & 45.1  & 51.85 & 9.07 & -    & -1.09 \\
19 & SDSSJ155444.57+082221.4  & 1252 & 1480 & 63.7  & 79.0  & 45.2  & 51.03 & 7.73 & 1.24 & 0.33  \\
20 & SDSSJ214054.55+002538.1  & 1325 & 2990 & 40.9  & 40.2  & 44.4  & 16.98 & 7.49 & 0.98 & -0.07 \\
21 & SDSSJ222024.58+010931.2  & 2719 & 4250 & 45.9  & 35.4  & 44    & 33.10 & 8.73 & 0.77 & -1.63 \\
22 & Fairall 9                & 3848 & 6857 & 47.6  & 110.8 & 43.98 & 26.84 & 8.09 & 0.43 & -1.0  \\
23 & Mrk 1044                 & 1290 & 1480 & 42.79 & 40.17 & 43.1  & 53.90 & 6.05 & 0.94 & -0.07 \\
24 & SDSSJ080101.41+184840.7  & 1530 & 2500 & 74.41 & 60.89 & 44.27 & 31.10 & 6.40 & 0.82 & 0.54  \\
25 & IRAS 04416+1215          & 1300 & 3810 & 45.79 & 60.88 & 44.47 & 4.20  & 6.97 & 1.33 & 0.70 
\\  
\hline 
\end{tabular}
\tablefoot{With an asterisk, we marked sources excluded from PCA, since not all values are available. The AGN luminosity at 5100\AA\ (\l5100{}), the BH mass (\mbh{}), and the Eddington ratio ($\lambda_{\rm Edd}$) are reported in log-scale.}
\end{table*}

\begin{figure*}[htb!]
    \centering
    \includegraphics[width=2\columnwidth]{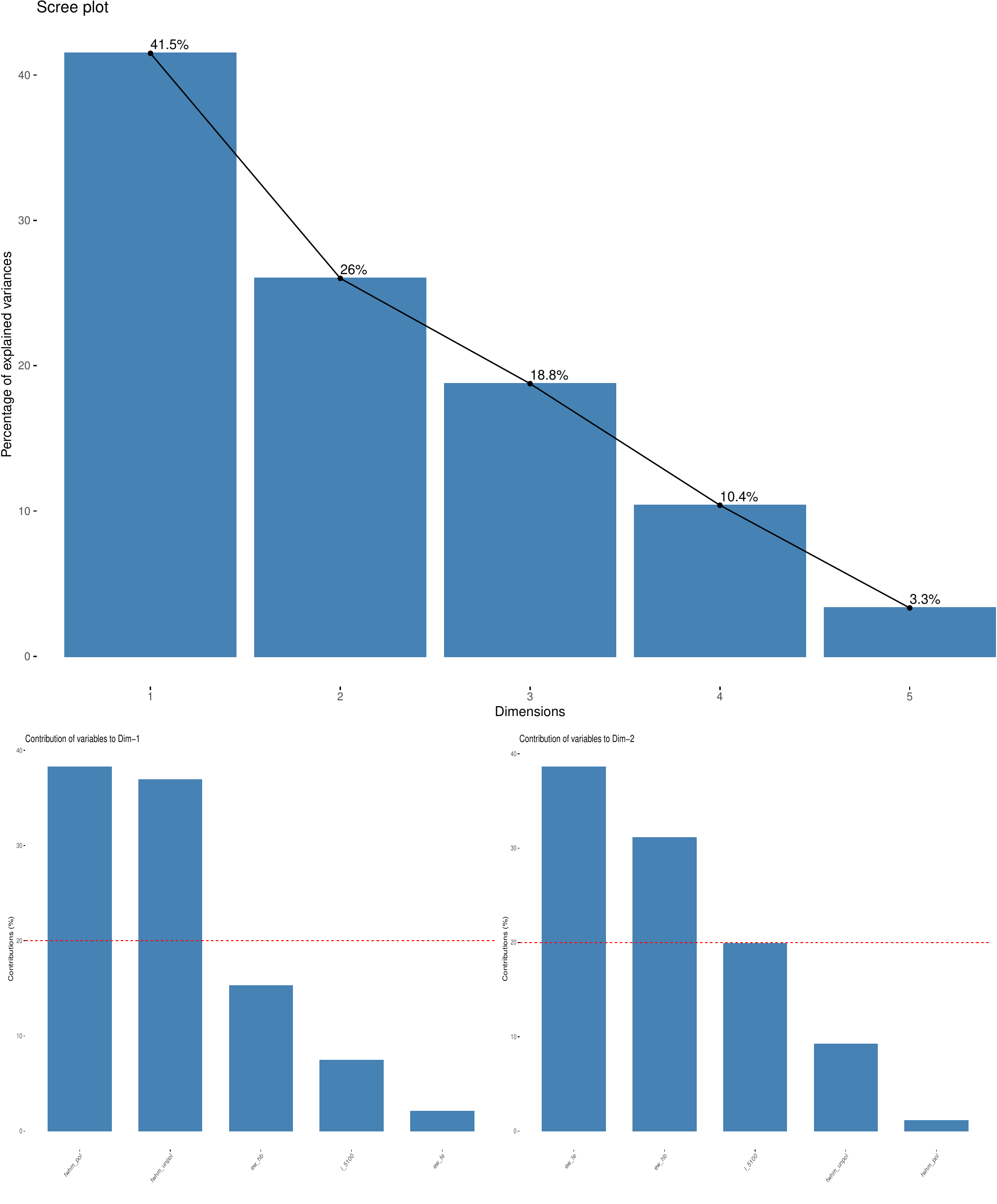}
    \caption{Contribution of principal components. Top panel: Scree plot using the PCA and showing the contribution (as a percentage) of the five principal components (PCs) to the overall variance in the dataset. Bottom panels: Contributions of the original variables: the AGN luminosity at 5100\AA~ (\l5100{}, in units of erg s$^{-1}$); the FWHM of the polarized and unpolarized \ha{} emission line (in units of km s$^{-1}$); and the EW for the corresponding \hb{} and optical Fe {\sc ii} emission within 4434-4684\AA~(in units of \AA) to the first two PCs. The red dashed line indicates the expected average contribution.}
    \label{fig:scree-plot_pca}
\end{figure*}

\begin{figure*}[htb!]
    \centering
    \includegraphics[width=\textwidth]{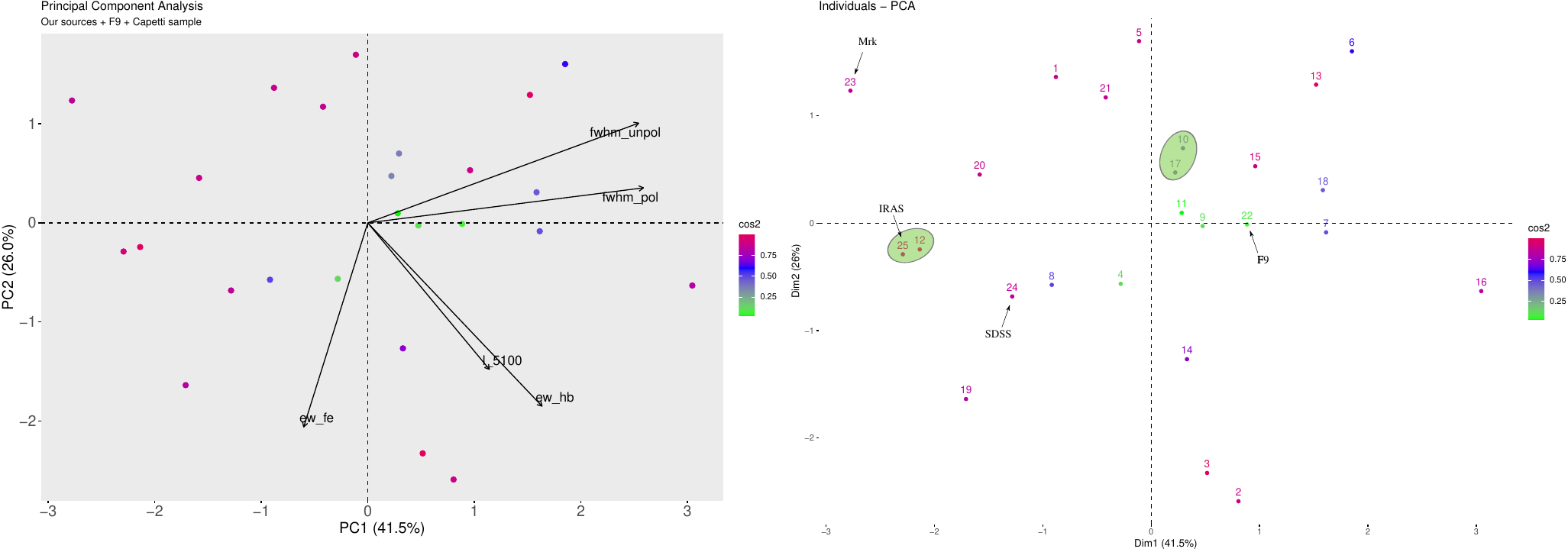}
    \caption{Two-dimensional projections of the first two principal components (PCs) using the PCA for the sources shown in Table \ref{tab:pca_table} after filtering out the sources marked with an asterisk. Only direct observables from the spectra were used in the PCA, namely, the AGN luminosity at 5100\AA~ (\l5100{}, in units of erg s$^{-1}$), the FWHM of the polarized and unpolarized \ha{} emission line (in units of km s$^{-1}$), and the EW for the corresponding \hb{} and optical Fe {\sc ii} emission within 4434-4684\AA~(in units of \AA). The percentages reported in parenthesis mark the contribution of the respective PC to the overall variance in the data. The importance of the direct observables to each PC is shown using vectors. The sources are color coded according to their "squared cosine" values and are a function of the squared distance of the observation to the origin. The sources are numbered as in Table \ref{tab:pca_table}. The location of our three sources (Mrk 1044, SDSS J080101.41+184840.7, and, IRAS 04416+1215) and Fairall 9 are indicated. The two mini-clusters containing pairs of sources that were revealed from our clustering selection are shown with green shaded ovals. The clustering selection was based on the geometrical distance between two data points, $\delta$r $\lesssim$ 0.57, across all the projections.}
    \label{fig:pca_projections}
\end{figure*}

\begin{figure*}[htb!]
    \centering
    \includegraphics[width=\textwidth]{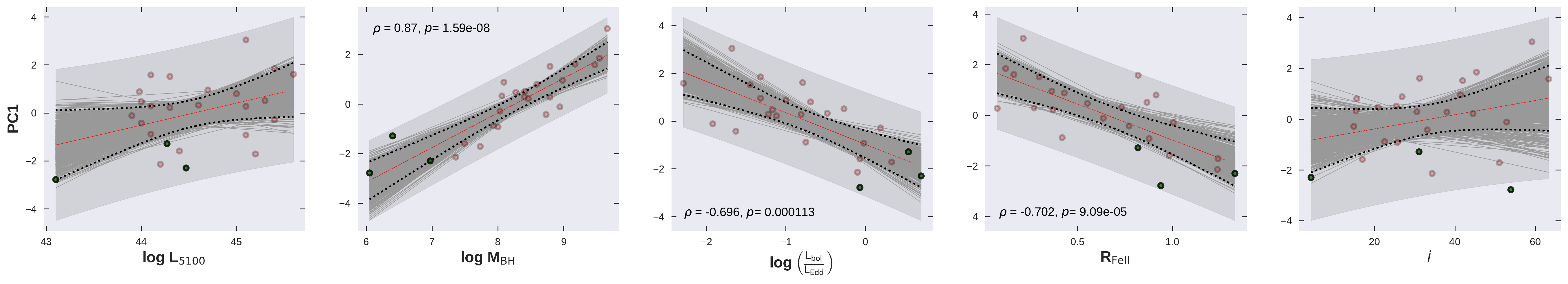}
    \caption{Correlations between the first principal component (PC1) and the physical properties of the sample. From left to right: Correlation with the AGN luminosity at 5100\AA~ (\l5100{}, in units of erg s$^{-1}$), the black hole mass (M$_{\rm BH}$, in units of M$_{\odot}$), the Eddington ratio, the ratio $R_\mathrm{FeII}$, and the viewing angle ($i$). Our three sources (Mrk 1044, SDSS J080101.41+184840.7, and, IRAS 04416+1215) are marked in green, while the other sources are marked in red. The Spearman's rank correlation coefficients ($\rho$) and the $p-$values are reported for the correlations whenever the $p-$value $<$ 0.001. The ordinary least-square fit for each panel is shown using a red solid line. The black dotted lines mark the confidence intervals at 95\% for the 1000 realizations (dark {gray} lines) of the bootstrap analysis. The corresponding prediction intervals are shown in the background using a light {gray} color.}
    \label{fig:pca-correlations}
\end{figure*}

We used the sources shown in Table \ref{tab:pca_table} for our PCA. We administered the PCA on the sample using only the direct observables, that is, the FWHM of the \ha{} emission line for both the polarized and the unpolarized profiles, the AGN monochromatic luminosity at 5100\AA~ (L$_{5100\AA}$), and the equivalent widths (EW) of the \hb{} and the optical Fe{\sc ii} blend within the 4434-4684\AA~ that is key to the optical plane of the quasar main sequence \citep{borosongreen1992, sulentic_etal_2000, 2014Natur.513..210S, marziani_etal_2018, panda_etal_2018, panda_etal_2019a, panda_etal_2019b}. This choice of using only the direct observables allowed us to remove redundancies in the PCA that enter when the parameters that are derived (e.g., viewing angle, black hole mass, and Eddington ratio) using the direct observables are incorporated within the analysis \citep[see][for a detailed study on this issue]{cafe_pca}. Thus, in our PCA, we used 25 sources with five properties.

Similar to \citet{wildy2019, cafe_pca}, we used the \textmyfont{prcomp} module in the \textit{R} statistical programming software. In addition to \textmyfont{prcomp}, we used the \textmyfont{factoextra\footnote{\href{https://cloud.r-project.org/web/packages/factoextra/index.html}{https://cloud.r-project.org/web/packages/factoextra/index.html}}} package to visualize the multivariate data at hand and especially to extract and visualize the eigenvalues and variances of the dimensions.

There is no well-accepted way to decide how many principal components are enough, we used the "scree plot" method to evaluate the number of principal components that best describe the variance in our dataset. A scree plot shows the variances (in percentages) against the {numbered} principal component and allows the number of significant principal components in the data to be visualized. The number of components is determined at the point beyond which the remaining eigenvalues are all relatively small and of comparable size \citep{PERESNETO2005974,Jolliffe2011}.

The top panel in Figure \ref{fig:scree-plot_pca} shows the scree plot obtained from our PCA. It can be seen that the first four principal components explain 96.7\% of the total variance in our dataset, with the first principal component (PC1) contributing 41.5\% to this. The subsequent principal components (PC2, PC3, and PC4) account for 26\%, 18.8\%, and 10.4\%, respectively. The first two principal components are sufficient to explain a major fraction of the variance (together they amount to 67.5\%), yet we considered the first four principal components to obtain a more consistent picture for the sample using the PCA. The lower panels of Figure \ref{fig:scree-plot_pca} show the contributions of the five input properties to the first two principal components. The red dashed line on each of these sub-panels indicates the threshold (i.e., the expected average contribution). If the contribution of the variables were uniform, the expected value would be 1/length(variables) = 1/5 (i.e., 20\%). For a given component, a variable with a contribution larger than this cutoff could be considered important and as contributing to the component. We thus note that the two FWHMs (polarized and then unpolarized) are the two dominant parameters contributing $\sim$75\% to the PC1. The remaining parameters are below the threshold. Similarly, for the PC2, the EW(Fe{\sc ii}) dominates, followed by EW(\hb{}). 

We show the 2D projection maps from our PCA in Figure \ref{fig:pca_projections}. We only show the projection map from the first two PCs (PC1 and PC2). The left panels in this figure show the projection maps with the corresponding PCs where their contribution to the total variance is also reported within parentheses. The 25 sources are shown along with the five input properties that are represented in the form of vectors. The magnitude and direction of these vectors show their importance to the respective PCs. The sources are colored as a function of the "squared cosine" (\textmyfont{cos2}) that shows the importance of a principal component for a given observation. It indicates the contribution of a component to the squared distance of the observation to the origin and corresponds to the square of the cosine of the angle from the right triangle made with the origin, the observation, and its projection on the component \citep[see][for an in-depth review on PCA]{abdi_williams_pca_2010}. The panels on the right show the same projections but mark the source number and locate our three sources (Mrk 1044, SDSS J080101.41+184840.7, and, IRAS 04416+1215) and Fairall 9. The PCA also serves as a clustering technique that can allow for the identification of groups of sources that have similar physical properties in multidimensional space. We derived the geometrical distance between every pair of observations on each projection map (PC1 to PC5) and retrieved two unambiguous mini-clusters, namely, (a) \#10 (SDSS J113422.47+041127.7) and \#17 (SDSS J154007.84+141137.0), and (b) \#25 (IRAS 04416+1215) and (blue) \#12 (SDSS J140621.89+222346.5). We set an upper limit for this distance ($\delta$r) at 0.57 to identify the mini-clusters and discard the imposters (see Fig. \ref{fig:pca_projections}). These projection maps allowed us to scrutinize the real mini-clusters of sources from the imposters. What we mean by this is that if one focuses only on the dominant projection map, that is the PC1-PC2 plane, one may end up marking more mini-clusters. But the majority of these mini-clusters were found to be imposters -- their relative positions vary significantly when traced on the remaining projections. The two pairs of sources have many identical/similar properties. For example, the first pair (\#10 and \#17) have identical luminosities, log \l5100{} = 42.9 (in erg s$^{-1}$) and almost identical black hole masses, log M$_{\rm BH}$ = 8.4 and 8.46 (in M$_{\odot}$), in addition to similar EW(\hb{}) and EW(Fe{\sc ii}). For the second pair (\#12 and \#25) which includes one of our sources (i.e., IRAS 04416+1215), we found similar luminosities, log \l5100{} = 44.2 and 44.47; black holes masses close to each other, log M$_{\rm BH}$ = 7.36 and 6.97; and similar EW(\hb{}) and EW(Fe{\sc ii}). The sources in each pair also have similar Eddington ratios as well as similar $R_\mathrm{FeII}$ values (see Table \ref{tab:pca_table}).

We show the spectra of both pairs of objects in Fig. \ref{fig:pca-similar}. They indeed show similar spectral properties, such as the shapes of the continuum, Balmer lines (\ha{}, \hb{}, H$\gamma$), [OIII] doublet, and Fe {\sc ii} pseudo-continuum emission. Also, they have a similar redshift, namely,  z = 0.119 for J154007 and z = 0.108 for J113422. For the second pair, z = 0.097 for J140621 and z = 0.089 for IRAS 04416. The overall similarity of the spectra and the derived parameters thus suggests that an upper limit for the viewing angle for IRAS 04416+1215 can be $\sim$34$^{\circ}$. {We note that assuming this value of the inclination angle gives us a BH mass, log \mbh{} $\sim$ 6.397, which is smaller by a factor of three relative to the BH mass estimated assuming inclination angle of 4$^{\circ}$ (see Table \ref{tab:mass-calculation}).} Such an analysis can be useful for larger samples to identify sources with similar spectral and physical properties. In Figure \ref{fig:stokes-iras-35deg}, we show the {\sc STOKES} modeling and a comparison with the observed spectra for IRAS 04416+1215 assuming an inclination angle of 35$^{\circ}$ {and including the polar scattering region. We note that the model with only the equatorial scattering region gives qualitatively similar results to Figure \ref{fig:stokes-iras-35deg}, although the polarized flux has a significantly lower flux level at the line center (highlighted with the red dashed line).} The modeled spectrum in the natural light has a better agreement relative to the 4$^{\circ}$ case (see Figure \ref{fig:stokes-iras}), although we need to assume an inclination closer to 54$^{\circ}$ to have the wings modeled with the lowest residuals. In our {\sc STOKES} modeling, since we assume the FWHM that is estimated from the spectral fitting of the observed spectrum in the polarized light, which is quite noisy, the FWHM can be larger than the assumed value. {Thus, we have a degeneracy between the estimation of the inclination angle and the FWHM of the polarized \ha{} spectrum. We need better-quality data to break this degeneracy.}}

\begin{figure*}[htb!]
\center
\includegraphics[width=\columnwidth]{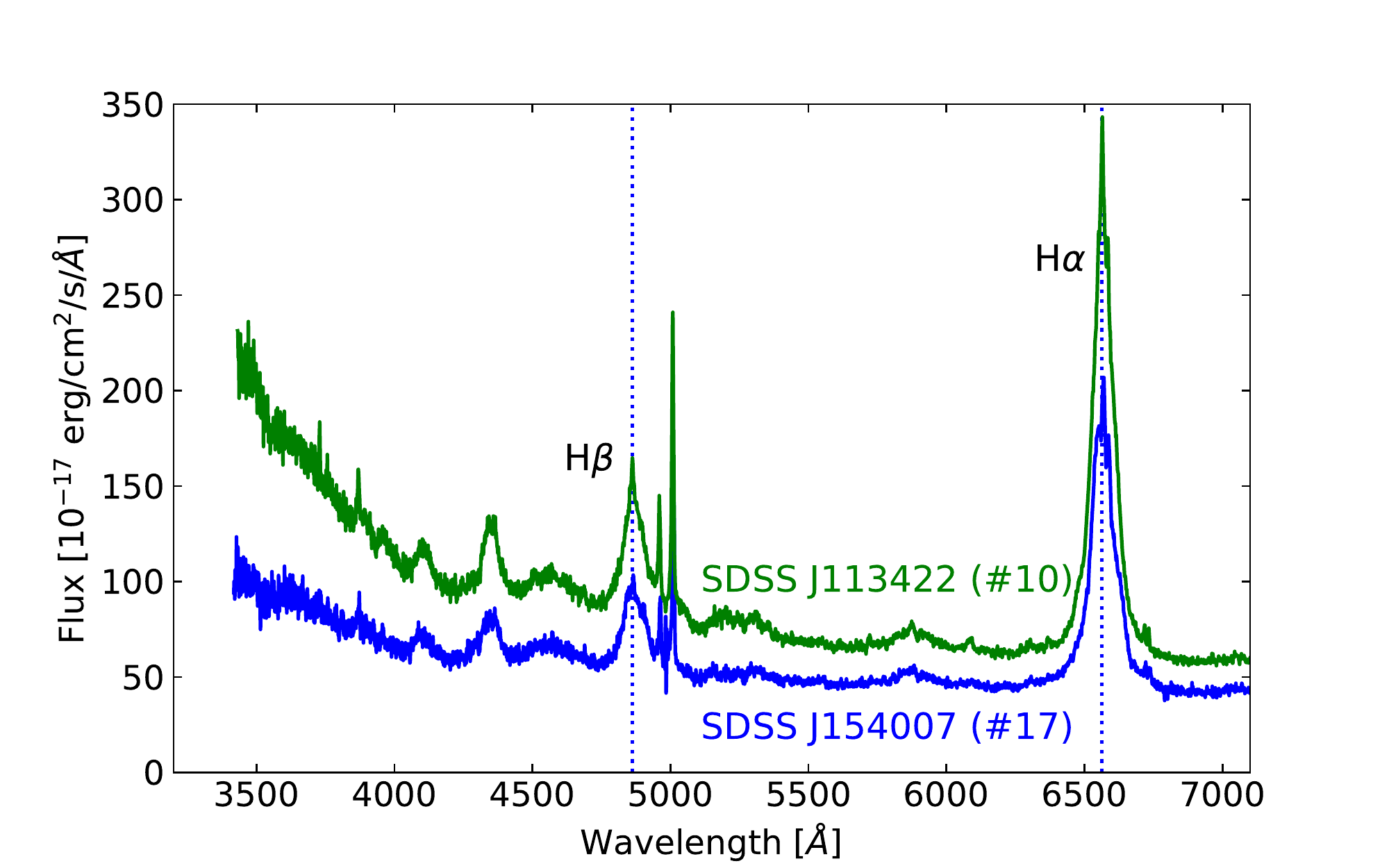}
\includegraphics[width=\columnwidth]{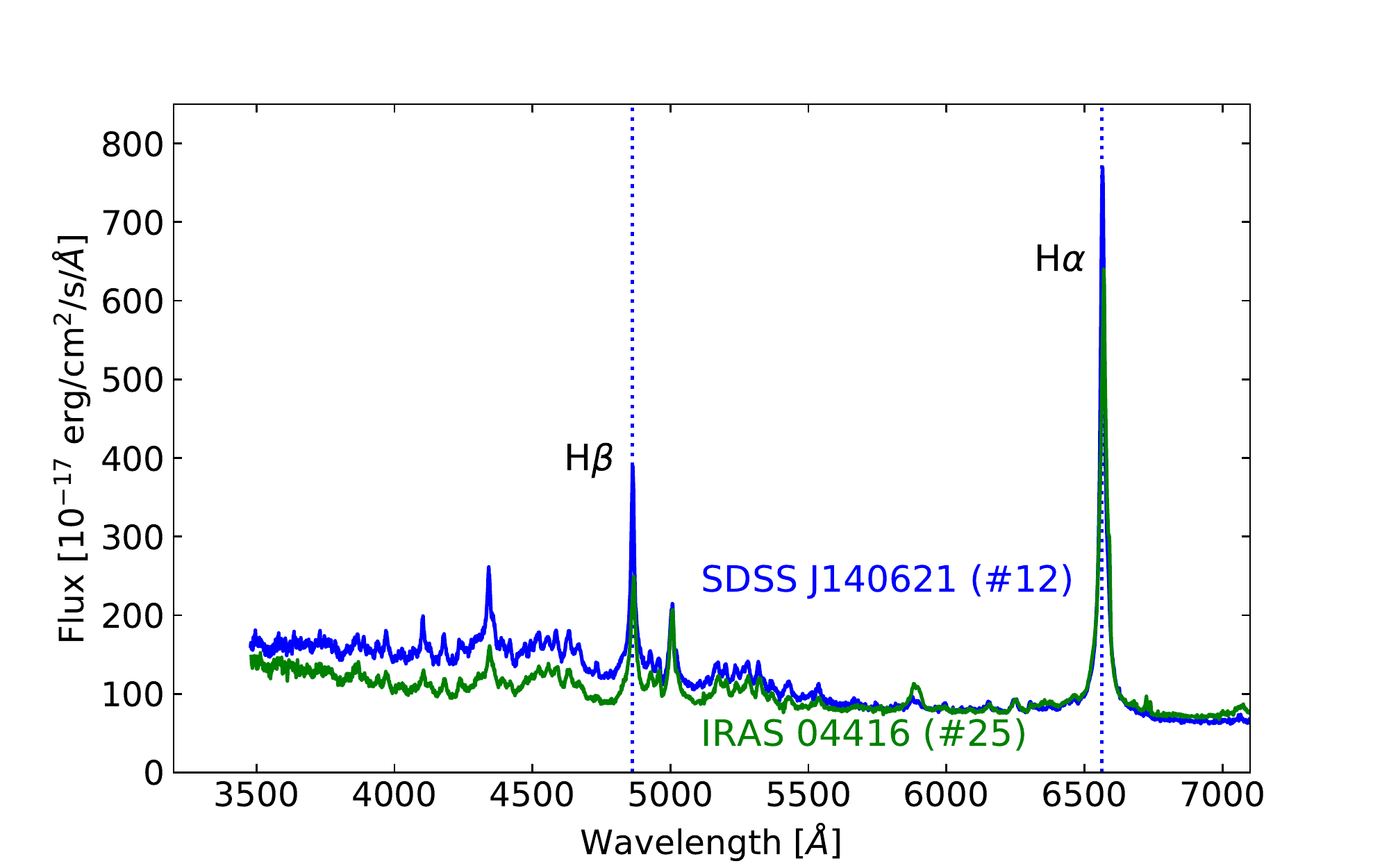}
\caption{Spectra of the grouped objects due to our PCA analysis. In the left panel, we show (green) \#10 (SDSS J113422.47+041127.7) and (blue) \#17 (SDSS J154007.84+141137.0). In the right panel we show (green) \#25 (IRAS 04416+1215) and (blue) \#12 (SDSS J140621.89+222346.5). The vertical dotted line represents the \hb{} and \ha{} rest-framed position. \label{fig:pca-similar}}
\end{figure*}

Finally, we show the correlations of the important derived parameters for the first principal component to check the primary driver(s) of this sample. In Figure \ref{fig:pca-correlations}, we show AGN luminosity at 5100\AA~ in addition to the derived parameters (i.e., black hole mass, Eddington ratio, $R_\mathrm{FeII}$, and the viewing angle). We performed an ordinary least-square fit for each panel and report the Spearman's rank correlation coefficient and the corresponding p-value for the panels only when the p-value was less than 0.001. This criterion was taken from our previous study \citep{cafe_pca}, and it  allowed us to identify true and robust correlations. We noticed that the first principal component (PC1) is primarily driven by a combination of black hole mass, Eddington ratio, and $R_\mathrm{FeII}$. The remaining PCs do not show any significant correlations and hence are not shown. As noticed in earlier works, the parameter $R_\mathrm{FeII}$ is a primary driver of the eigenvector 1 in the optical plane of the quasar main sequence \citep{borosongreen1992, sulentic_etal_2000, 2014Natur.513..210S, marziani_etal_2018, panda_etal_2018, panda_etal_2019a, panda_etal_2019b} and is a direct tracer of the Eddington ratio. Smaller black hole masses and larger luminosities can push the Eddington ratios to higher values, and we observed this effect in our correlations. Also, such sources show stronger Fe II emissions. The correlation between the PC1 and black hole mass, and the anti-correlation between the PC1 and $R_\mathrm{FeII}$ highlight this trend appropriately. Our three sources are highlighted in green and occupy the region with the lowest black hole masses and highest Eddington ratios. Their \l5100{} and $R_\mathrm{FeII}$ are comparable to some of the sources from \citet{2021capetti}, mostly those occupying the more luminous regions and with higher $R_\mathrm{FeII}$ values. In no situation did we observe the viewing angle (panels in the last column) to be significant in the PCA for this sample. A reason for this result could be the overall similarity of the two FWHMs (unpolarized and polarized) vectors as seen from the projection maps (see Figure \ref{fig:pca_projections}), and since the viewing angle was estimated using the ratio of these two FWHMs, the overall correlation is affected. The FWHMs individually have strong correlations with the PCs, especially with the PC1 (see Figure \ref{fig:scree-plot_pca}), which dampens the correlation with the viewing angle.

\subsection{Summary}
In this work, we present new VLT-FORS2 spectropolarimetric measurements for three objects selected as high Eddington ratio candidates. Our findings from this work are as follows:
\begin{itemize}
     
    \item The viewing angles recovered for the three sources indicate that they cover a large range in the viewing angle values, from an almost face-on orientation (IRAS 04416+1215) through an intermediate case (SDSS J080101.41+184840.7) to a highly inclined  orientation (Mrk 1044). We find it important to highlight that the viewing angle estimation for IRAS 04416+1215 should be taken with caution.
    %The variety of viewing angles thus result in a significant change in the black hole masses relative to those reported in the literature.
    
    \item Despite the large differences in the viewing angles, we confirm the small values of the black hole mass in these sources and their high Eddington nature.
    
    \item We were successful in recovering the observed \ha{} line profile both in the natural and polarized light using the {\sc STOKES} modeling. We recovered the polarization fractions of the order of 0.2-0.5\% for the three sources, although the recovery of the phase angle is sub-optimal, mainly due to the noise in the observed data.
   
   \item Our principal component analysis shows that the sample of the 25 sources, including our sources, is mainly driven by the black hole mass and Eddington ratio. We reaffirm the connection of the strength of the optical Fe {\sc ii} emission with the Eddington ratio, but the dependence on the viewing angle is ordinary and resembles more of a secondary effect.
   
\end{itemize}

\section*{Acknowledgements}
We are very grateful to the Referee for very helpful and constructive remarks which help to improve considerably the content of the manuscript.
MS acknowledges support from Polish Funding Agency National Science Centre, project 2021/41/N/ST9/02280 (PRELUDIUM 20), the European Research Council (ERC) under the European Union’s Horizon 2020 research and innovation program (grant agreement 950533) and from the
Israel Science Foundation, (grant 1849/19). This project has received funding from the European Research Council (ERC) under the European Union’s Horizon 2020 research and innovation program (grant agreement No. [951549]). MS is grateful to S. Bagnulo for the discussion about data reduction and all tutors and organizers of \textit{Torun Summer School 2019: Polarimetry as a diagnostic tool in astronomy} for providing a review of methods and techniques developed for the polarimetry data. We also thank organizers and participants of the conference \textit{AGN and Polarimetry conference: Towards a panchromatic understanding of the polarization of Active Galactic Nuclei} for discussion which helps to improve this work. The project was partially supported by the Polish Funding Agency National Science Centre, project 2017/26/\-A/ST9/\-00756 (MAESTRO  9), and MNiSW grant DIR/WK/2018/12. SP acknowledges the financial support of the Conselho Nacional de Desenvolvimento Científico e Tecnológico (CNPq) Fellowships 164753/2020-6 and 300936/2023-0. \DJ S acknowledges support by the Ministry of Education, Science and Technological Development of the Republic of Serbia through contract no. 451-03-9/2021-14/200002 and by the Science Fund of the Republic of Serbia, PROMIS 6060916, BOWIE.
MLMA acknowledges financial support from Millenium Nucleus NCN19-058 (TITANs). L. \v C. P. is supported by the Ministry of Education, Science and Technological Development of Serbia, the number of the contract is 451-03-68/2020-14/200002 and  L. \v C. P. acknowledges the support by the Chinese Academy of Sciences President’s International Fellowship Initiative (PIFI) for visiting scientist. 

Funding for the Sloan Digital Sky Survey (SDSS) has been provided by the Alfred P. Sloan Foundation, the Participating Institutions, the National Aeronautics and Space Administration, the National Science Foundation, the U.S. Department of Energy, the Japanese Monbukagakusho, and the Max Planck Society. The SDSS Web site is \href{http://www.sdss.org/}{http://www.sdss.org/}.

The SDSS is managed by the Astrophysical Research Consortium (ARC) for the Participating Institutions. The Participating Institutions are The University of Chicago, Fermilab, the Institute for Advanced Study, the Japan Participation Group, The Johns Hopkins University, Los Alamos National Laboratory, the Max-Planck-Institute for Astronomy (MPIA), the Max-Planck-Institute for Astrophysics (MPA), New Mexico State University, University of Pittsburgh, Princeton University, the United States Naval Observatory, and the University of Washington.

\section*{Authors' Contribution}
MS performed the data reduction and spectral fitting of the observational data. SP performed the {\sc STOKES} modeling and the PCA analysis. MS, SP, and BC primarily wrote the manuscript. \DJ S helped with the {\sc STOKES} modeling. MLMA helped with the spectral fitting. PM helped with text writing and computations. JMW, PD, and LCP helped with the discussions. CSS helped with the data analysis of \textit{Planck}-derived products.

\section*{Software}
\textmyfont{MATPLOTLIB}  (\citealt{hunter07}); \textmyfont{NUMPY} (\citealt{numpy}); \textmyfont{SCIPY} (\citealt{scipy}); \textmyfont{SKLEARN} (\citealt{scikit-learn}); \textmyfont{STATSMODELS} (\citealt{seabold2010statsmodels}); \textmyfont{STOKES} (\citealt{marin2018}); \textmyfont{HEALPIX} (\citealt{Gorski2005}); \textmyfont{ASTROPY} (\citealt{astropy})

\bibliographystyle{aa}
\bibliography{sample63}

\appendix

\section{Contamination of the polarized emission of studied sources {with respect to} the interstellar extinction and host galaxy}
\label{sec:contamination}

As clearly discussed and modeled by \citet{marin2018}, in the case of observations of Seyfert 1 galaxies the contamination due to the ISM and dilution of the observed flux by the host can considerably affect the polarization measurements. Dust scattering in our Galaxy along the line of sight from the source imprints its own polarization signature that is predominantly set by the total dust column. The maximum of the effect is seen at $\sim 5450$ \AA~\citep{serkowski1975}. The {ISM polarization in our Galaxy is primarily due to dust dichroic extinction, that is, differential linear extinction for the two waves polarized along and perpendicular to the direction of alignment of the light propagation vector before reaching the ISM}.

The best way to address the role of the {ISM} is to measure the observed polarization for several unpolarized stars located close to the selected object. Such an approach was adopted by \citet{2021bowei} for Fairall 9, which was also observed with VLT/FORS2, and it showed that for this source the effect is unimportant. Unfortunately, for the sources discussed in the present paper, such observations were not performed, and the two comparison stars (polarized and unpolarized) were located far from any of the objects. The other way to estimate the possible role of interstellar polarization is through the measurement of the Galactic extinction in the direction of the source. We give these values,  parameterized by $A_V$ (taken from NED Database) in Table~\ref{tab:my-table-alt}. 

We observed that for two of the sources, Mrk 1044 and SDSS J080101.41+184840.7, the extinction is not considerably higher than for Fairall 9, and the effect of the ISM can be neglected. However, this is not the case for IRAS 04416+1215, {as the Stokes parameter Q changes sign and affects the measurements the most.}

\begin{figure}[!htb]
\centering
\includegraphics[width=0.9\columnwidth]{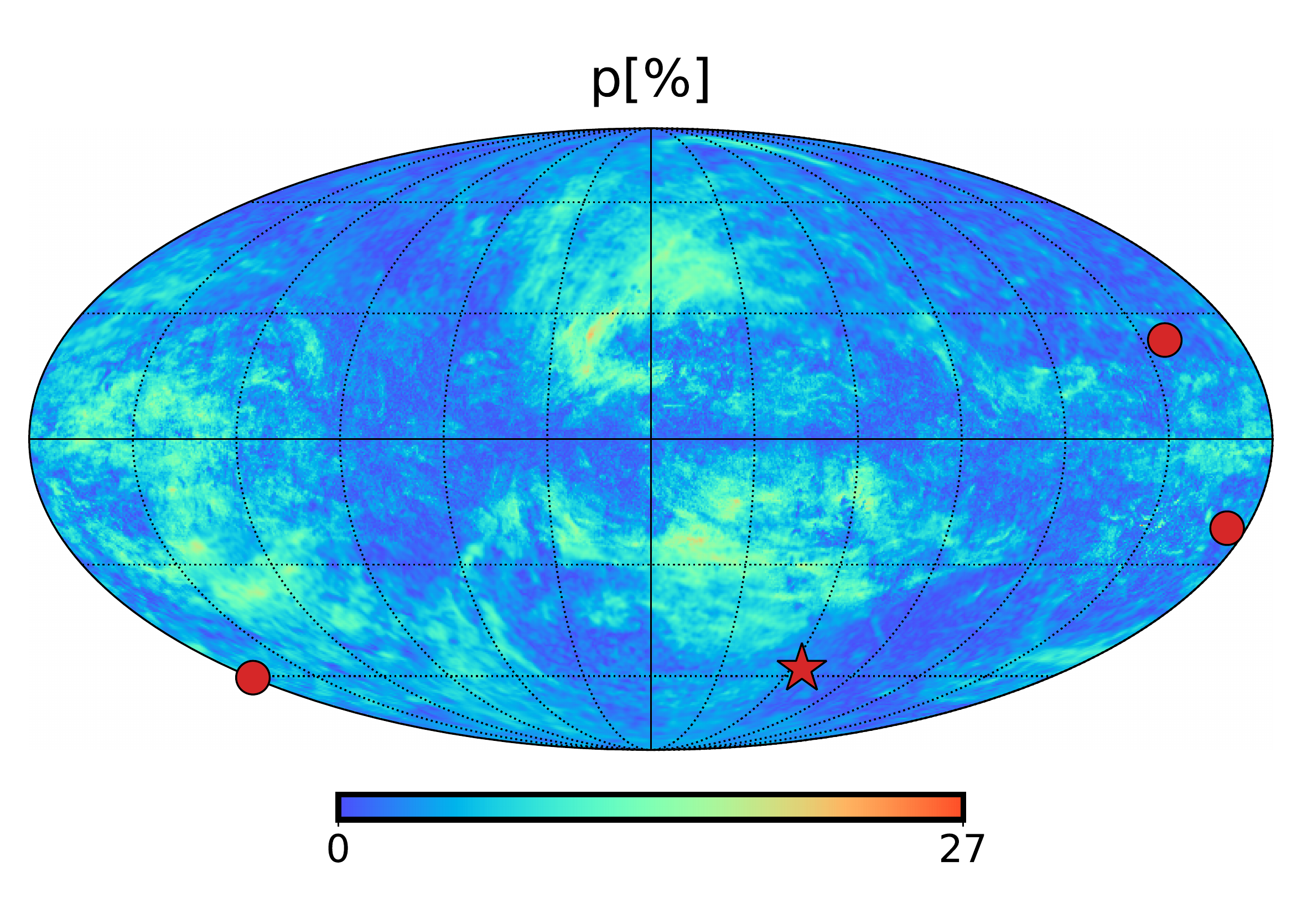}
\caption{Degree of polarization (\textit{P} in \%) computed using the GNILC 353GHz thermal dust map from \citet{Planck_2020A&A...641A...4P} at {\sc healpix} resolution, N$_{\rm side}$ = 2048. The location of our three sources (black dots) and Fairall 9 (black star) are indicated based on their galactic coordinates reported in Table \ref{tab:my-table-alt}. The values for \textit{P} for these sources are reported in Table \ref{tab:my-table-alt-pcorr-chi-corr}}.  
\label{fig:planck-pol-frac}
\end{figure}

\begin{figure}[!htb]
\centering
\includegraphics[width=0.9\columnwidth]{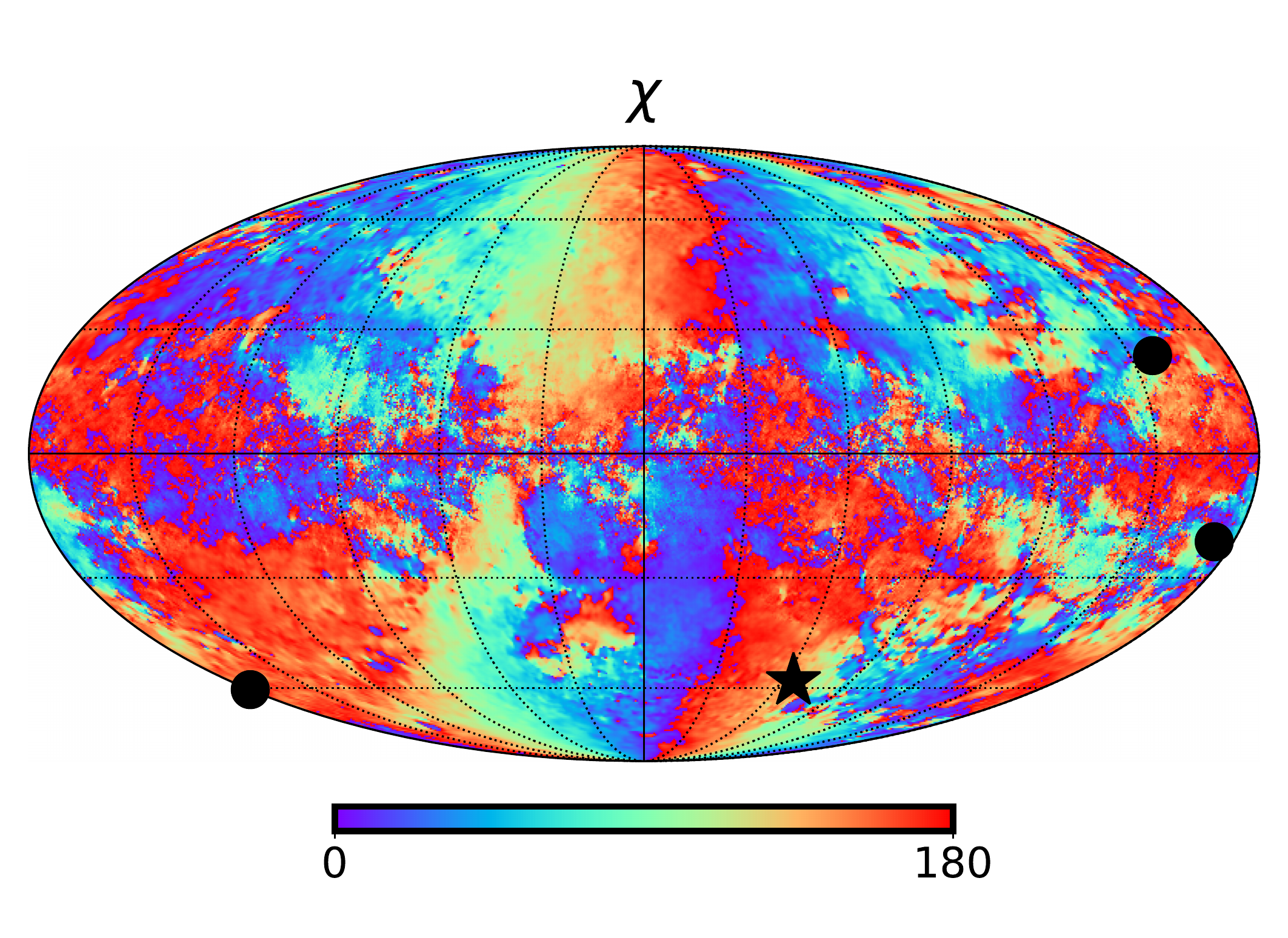}
\caption{Same as Figure \ref{fig:planck-pol-frac} but for the polarization angle ($\chi$) computed using Equation \ref{pol-ang}. The values for $\chi$ for these sources are reported in Table \ref{tab:my-table-alt-pcorr-chi-corr}}.  
\label{fig:planck-pol-ang}
\end{figure}

Therefore, we consulted the \textit{Planck} polarization maps, as recommended by \citet{2021capetti} and \citet{pelgrims2019}.  We used the intensity ($I$) and polarization ($Q$ and $U$) thermal dust maps at the 353 GHz frequency channel from the 2018 \textit{Planck} data release\footnote{\url{https://pla.esac.esa.int/}} described in \cite{Planck_2020A&A...641A...4P}. {These dust maps were produced using the Generalized Needlet Internal Linear Combination (GNILC) component separation method, which uses spectral information as well as the angular power spectrum to disentangle specific diffuse foreground components. The Stokes parameters $I, Q,\text{ and }U$ from the \textit{Planck} datasets are in units of the thermodynamic temperature $K_{\text{CMB}}$\footnote{The $K_{\text{CMB}}$ depends on the bandpass frequency (353 GHz in our case) and the temperature of the CMB radiation at redshift $z=0$.}, in milli-Kelvins. We converted these Stokes parameters to MJy sr$^{-1}$, assuming the CMB temperature at $z=0$ to be 2.73 K.} The values of the polarization level and the polarization angle of the foreground thermal dust contribution at the position of the discussed sources are given in Table~\ref{tab:my-table-alt-pcorr-chi-corr}, and their positions are indicated in Figures~\ref{fig:planck-pol-frac} and \ref{fig:planck-pol-ang}. {We estimated the polarization contribution from the ISM for sources from our sample and for Fairall 9.}
{To correct the Stokes parameters computed from optical observation, we first estimated the contribution of the ISM from submillimeters to V-band using a prescription from \citet{pelgrims2019, 2021capetti} and values from Table \ref{tab:my-table-alt} using the following equation:
\begin{equation}
     Q_{CORR} = -0.238 * A_V * Q_{GNILC}
\end{equation}
which has a similar expression for $U_{CORR}$. Next, we corrected our observed values of the Stokes parameters and computed \textit{P$_{\rm corr}$} and $\chi_{\rm corr}$, which we report in Table \ref{tab:my-table-alt-pcorr-chi-corr}.} For our three sources, Mrk 1044 and SDSS J080101.41+184840.7, and Fairall 9, there is no significant change in the $\chi_{\rm corr}$ and \textit{P$_{\rm corr}$}. {For Mrk 1044, the degree of polarization before correction is 0.15 $\pm$ 0.02\%, and after correction it is 0.19\%. For SDSS J080101.41+184840.7, the degree of polarization before correction is 0.52\%, and after correction it is 0.49\%.} Hence, the ISM correction does not change our original inferences. For IRAS 04416+1215, however, we observed a noticeable change after the correction: 0.17\% before the correction and 0.64\% after the correction. Compounded with other degeneracies, such as problems with the estimation of the viewing angle (see Section \ref{sec:pca} for details), this result makes it difficult to draw any clear conclusion for IRAS 04416+1215 with the available data.

\citet{Robinson2011ASPC..449..431R} argues that variations of  $P$ and $\chi$ across the Balmer lines indicate that sources are also intrinsically polarized. We did not see such signatures in our sources, but it may be due to short exposures.

In the case of the host galaxy's contamination of our results, we compared the shape of the degree of polarization along the wavelength from our measurements to models from \citet{marin2018} (Figure 3 therein). Models from \citet{marin2018} suggest a small decrease in the degree of polarization along the wavelength in the case of contamination from the ISM or host galaxy. For our sources, however, we noticed a rather small increase in the degree of polarization along the wavelength. This small contribution to the measured flux comes from the fact that in our VLT/FORS2 observations, we used a very narrow slit of 0.7'', so the spectrum contains much less starlight than the typical measurements available in the NED Database\footnote{\href{http://ned.ipac.caltech.edu/}{http://ned.ipac.caltech.edu/}} that reflect a red spectrum contaminated by the host.

We performed an additional test using the available acquisition image for Mrk 1044. The exposure time was only $\sim$1s  (in OG590 filter) but the quality of the image was good. We analyzed the image using the method described in \citet{bentz2013} for the starlight subtraction using Hubble Space Telescope images.  We fitted the background and the active galaxy profile assuming a Gaussian profile for the point spread function and the Sersic profile for the host, with a fixed value of the profile index n. After fitting the parameters we integrated the profiles for the long slit used in the data acquisition (0.7 arcsec).  We assumed Sersic profiles for two cases of {Sersic index} n: n = 2 and n = 4. The contamination weakly depended on the value of n and was about 0.1852 in the case of n=2, and 0.1858 in the case of n=4. {The numbers represent the fraction of the host galaxy flux to the AGN flux (which is dimensionless). Hence, the contribution from the host galaxy is not dominant.}

\begin{table*}[!htb]
\centering
\caption{{Source properties and ISM properties (from \textit{Planck}'s GNILC 353GHz thermal dust map) with corrected Stokes parameters.}}
\label{tab:my-table-alt}
\begin{tabular}{lccccccccccc}
\hline
Source                  & Gal. long. & Gal. lat. & \textit{Q$_{obs}$}    & \textit{U$_{obs}$} & \textit{Q$_{\rm GNILC}$}    & \textit{U$_{\rm GNILC}$}     &A$_{\rm V}$   & \textit{Q$_{\rm conv}$}&\textit{U$_{\rm conv}$} &\textit{Q$_{\rm corr}$} &\textit{U$_{\rm corr}$}  \\
\multicolumn{1}{c}{}    & [deg.]       & [deg.] & [\%]   & [\%]     & [\%]   & [\%] &[mag]   & [\%]  &  [\%] & [\%]&  [\%] \\
\multicolumn{1}{c}{}   & (2)       & (3) & (4)   & (5)     & (6)   & (7) & (8) & (9) & (10) & (11) & (12) \\ \hline

Mrk 1044  	&	179.694	&	-60.477	&	0.039	&	-0.148	&	3.307	&	-0.252	&	0.095	&	-0.075	&	0.006	&	0.114	&	-0.153	\\
SDSSJ080101	&	203.015	&	+23.510	&	-0.305	&	-0.422	&	2.761	&	-0.142	&	0.086	&	-0.057	&	0.003	&	-0.249	&	-0.423	\\
IRAS04416&	185.854	&	-21.088	&	-0.125	&	0.149	&	2.623	&	0.098	&	1.187	&	-0.741	&	-0.028	&	0.616	&	0.176	\\
Fairall 9	&	295.073	&	-57.827	&	-1	&	-2.5	&	-0.697	&	-3.412	&	0.071	&	0.012	&	0.058	&	-1.012	&	-2.558	\\ \hline
\end{tabular}%
%}
\tablefoot{The columns show the (1) name of the source, (2-3) sky coordinates in a galactic coordinate system, (4-5) observed values of $Q$ ($Q_{obs}$) and $U$ ($U_{obs}$) from our measurements, (6-7) $Q_{GNILC}$ and $U_{GNILC}$ obtained from \textit{Planck}'s GNILC dust maps, (8) optical extinction coefficient A$_{V}$, (9-10) converted $Q$ ($Q_{conv}$) and $U$ ($U_{conv}$) from sub-mm to optical V-band, and (11-12) corrected values of $Q_{corr}$ and $U_{corr}$.}
\end{table*}

\begin{table}[]
\centering
\caption{{Corrected sources' properties}}
\label{tab:my-table-alt-pcorr-chi-corr}
\begin{tabular}{lcc}
\hline
Source       &\textit{P$_{\rm corr}$}    & $\chi_{\rm corr}$ \\ 
\multicolumn{1}{c}{}    & [\%]       & [deg.] \\ \hline
Mrk 1044  & 0.191              & 153            \\
SDSSJ080101    & 0.492               & 120            \\
IRAS04416     & 0.640               & 8              \\
Fairall 9 & 2.751               & 124    \\ \hline       
\end{tabular}
\end{table}

\section{Supplementary plots} \label{sec:analysis}
%%% OLD MODELS

\begin{figure}[!htb]
\includegraphics[width=\columnwidth]{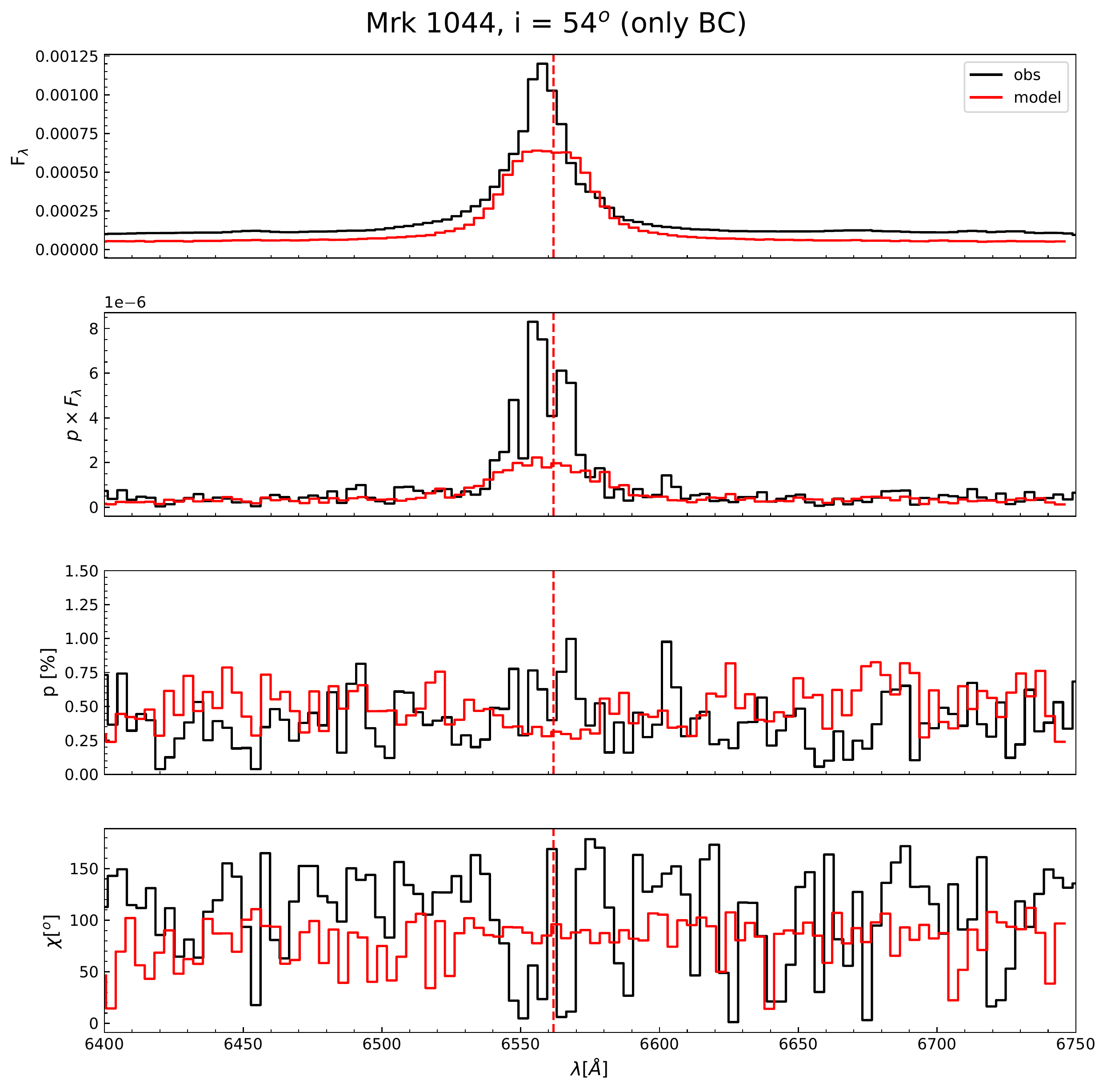}
\caption{STOKES modeling and comparison with observational estimates for Mrk 1044 with only the equatorial scattering region included. The case shown has a viewing angle of 54$^{\circ}$. From top to bottom, the panels show the spectrum in natural light ($F_{\lambda}$), the polarized spectrum ($p \times F_{\lambda}$), the polarization fraction ($p$) and the polarization angle ($\chi$). The vertical red dashed line marks the central wavelength for \ha{}. \label{fig:stokes-bc-only-mrk-no-polar}}
\end{figure}

\begin{figure}[!htb]
\includegraphics[width=\columnwidth]{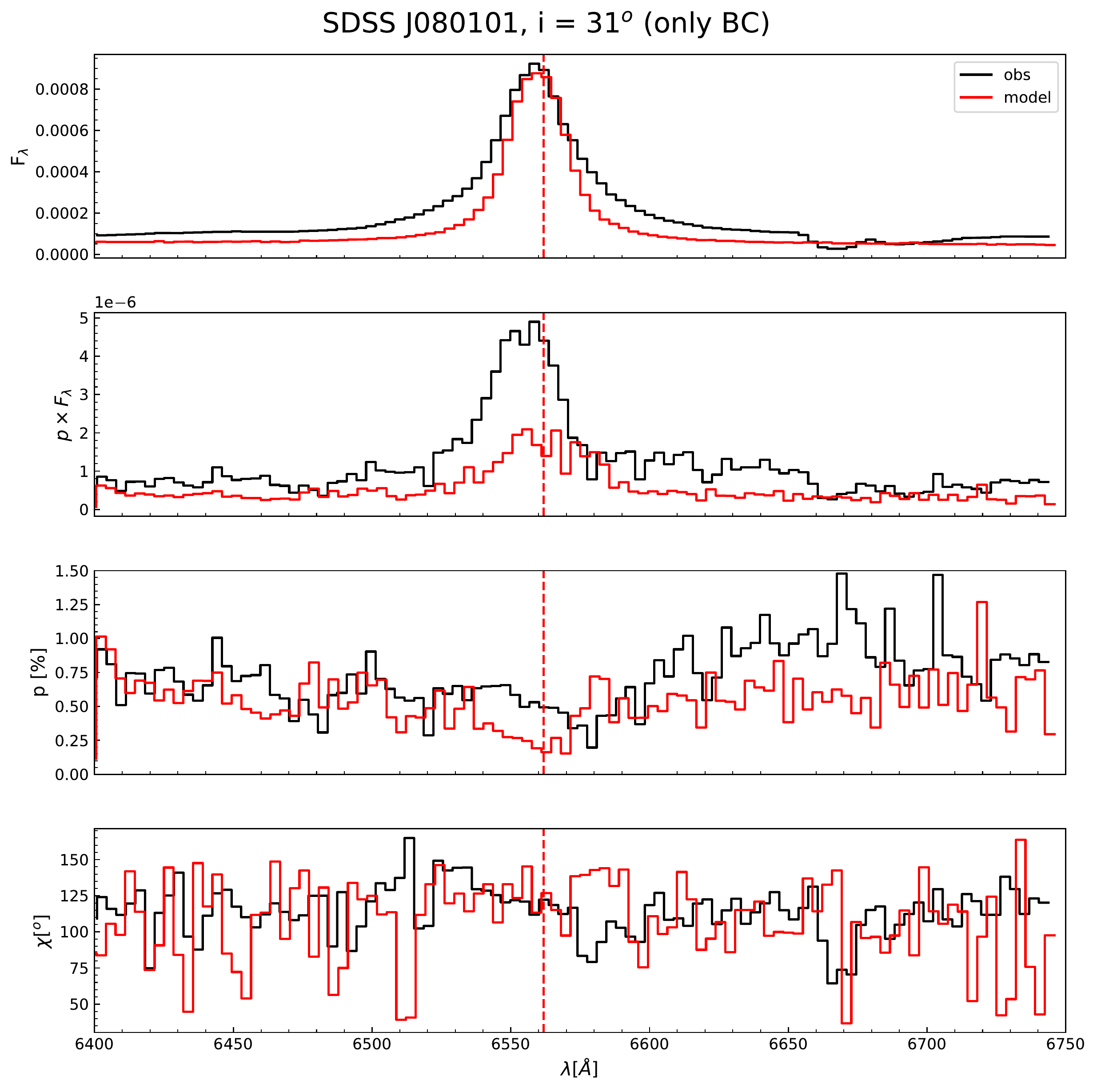}\hspace{-0.2cm}
\caption{STOKES modeling and comparison with observational estimates for SDSS J080101.41+184840.7 with only the equatorial scattering region included. The case shown has a viewing angle of 31$^{\circ}$. The panels are similar to Figure \ref{fig:stokes-bc-only-mrk-no-polar}.  \label{fig:stokes-bc-only-sdss-no-polar}}
\end{figure}

\begin{figure}[!htb]
\includegraphics[width=\columnwidth]{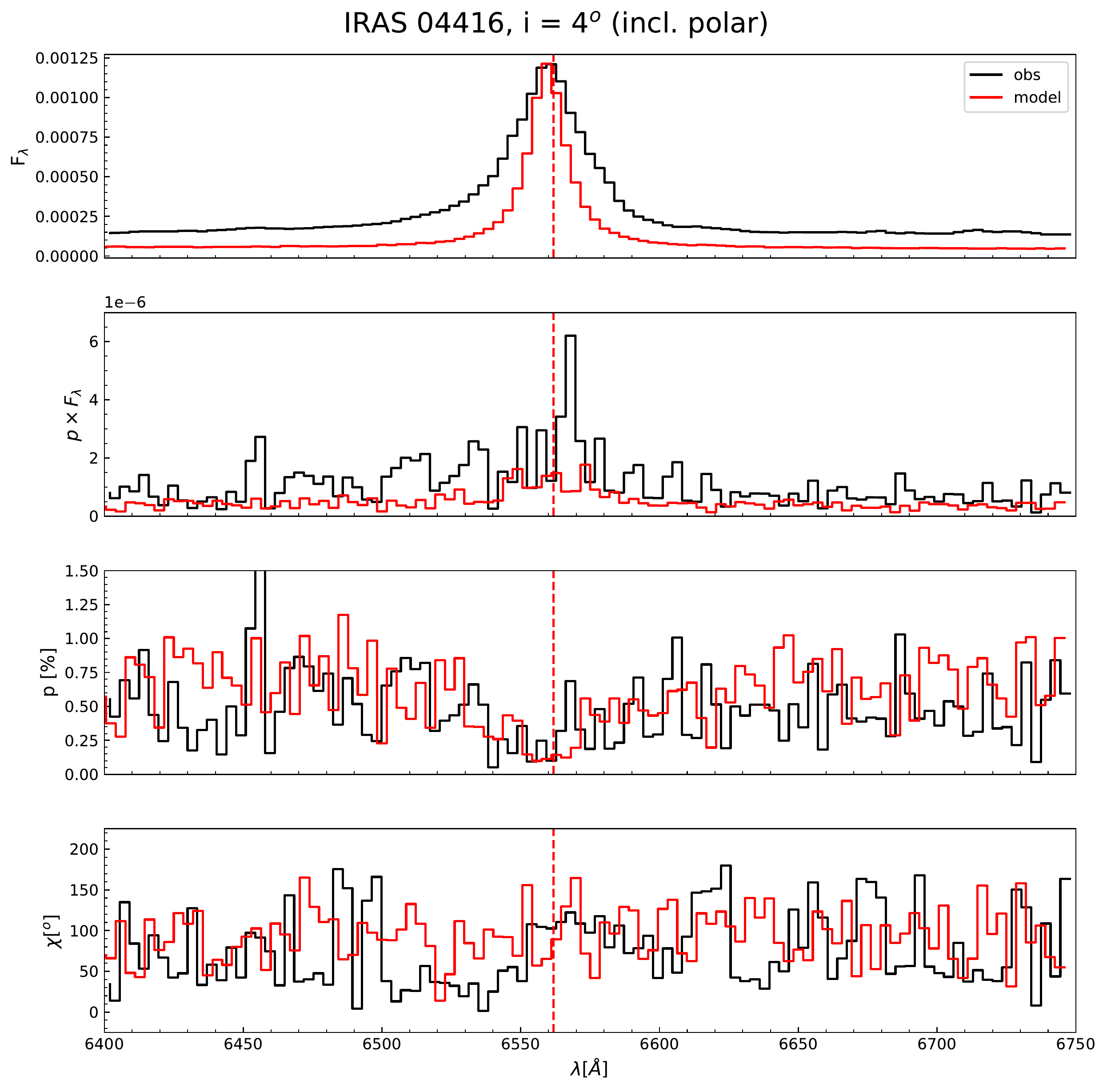}\hspace{-0.2cm}
\caption{STOKES modeling and comparison with observational estimates for IRAS 04416+1215 with the equatorial plus polar scattering region included. The case shown has a viewing angle of 4$^{\circ}$. The panels are similar to Figure \ref{fig:stokes-bc-only-mrk-no-polar}.  \label{fig:stokes-iras-w-polar}}
\end{figure}

\begin{figure}[!htb]
\includegraphics[width=\columnwidth]{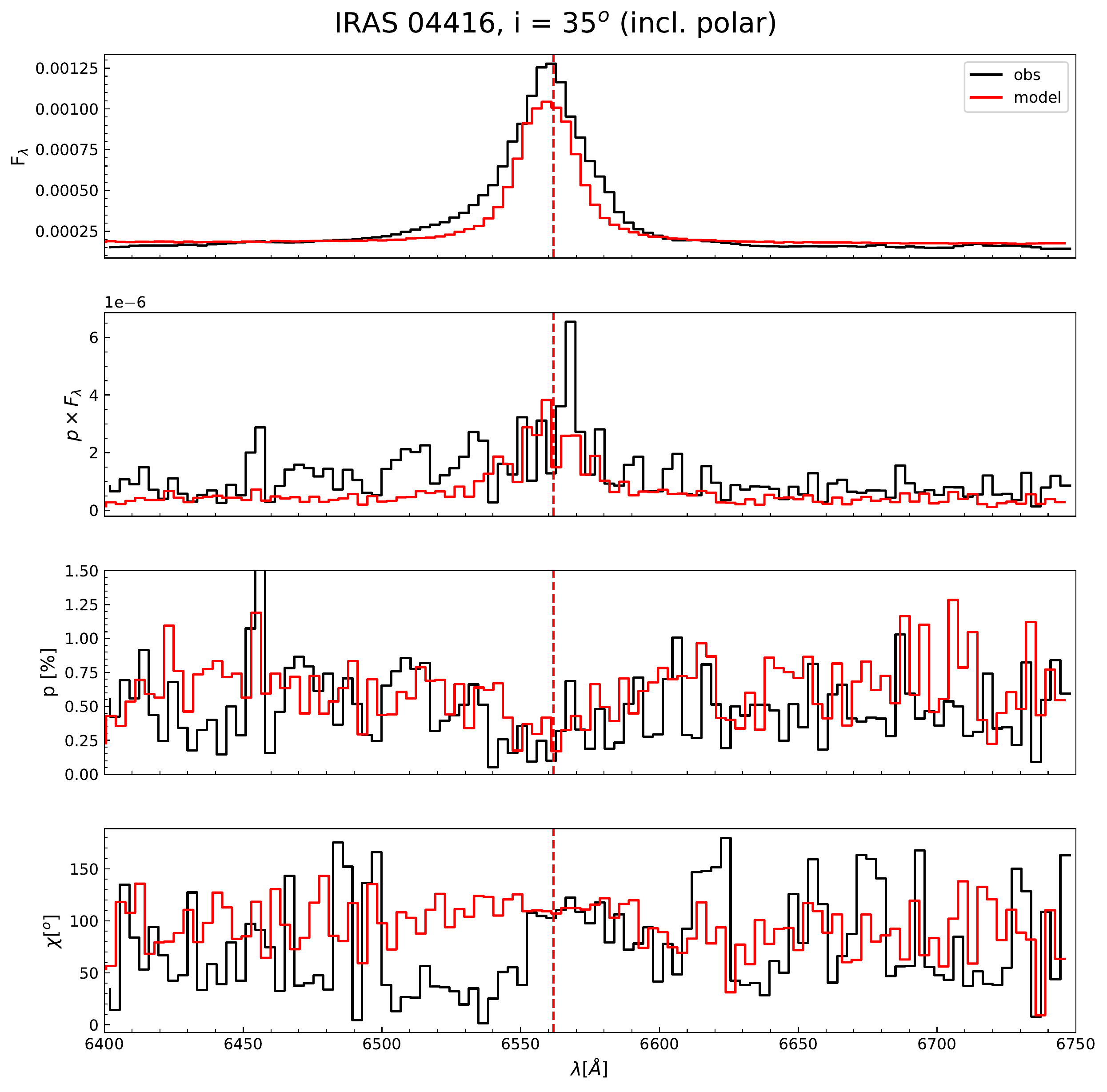}\hspace{-0.2cm}
\caption{{STOKES modeling and comparison with observational estimates for IRAS 04416+1215 {with the equatorial plus polar scattering regions} included. The case shown has a viewing angle of 35$^{\circ}$. The panels are similar to Figure \ref{fig:stokes-bc-only-mrk-no-polar}.  \label{fig:stokes-iras-35deg}}}
\end{figure}

\begin{figure}
    \centering
    \includegraphics[width=0.75\columnwidth]{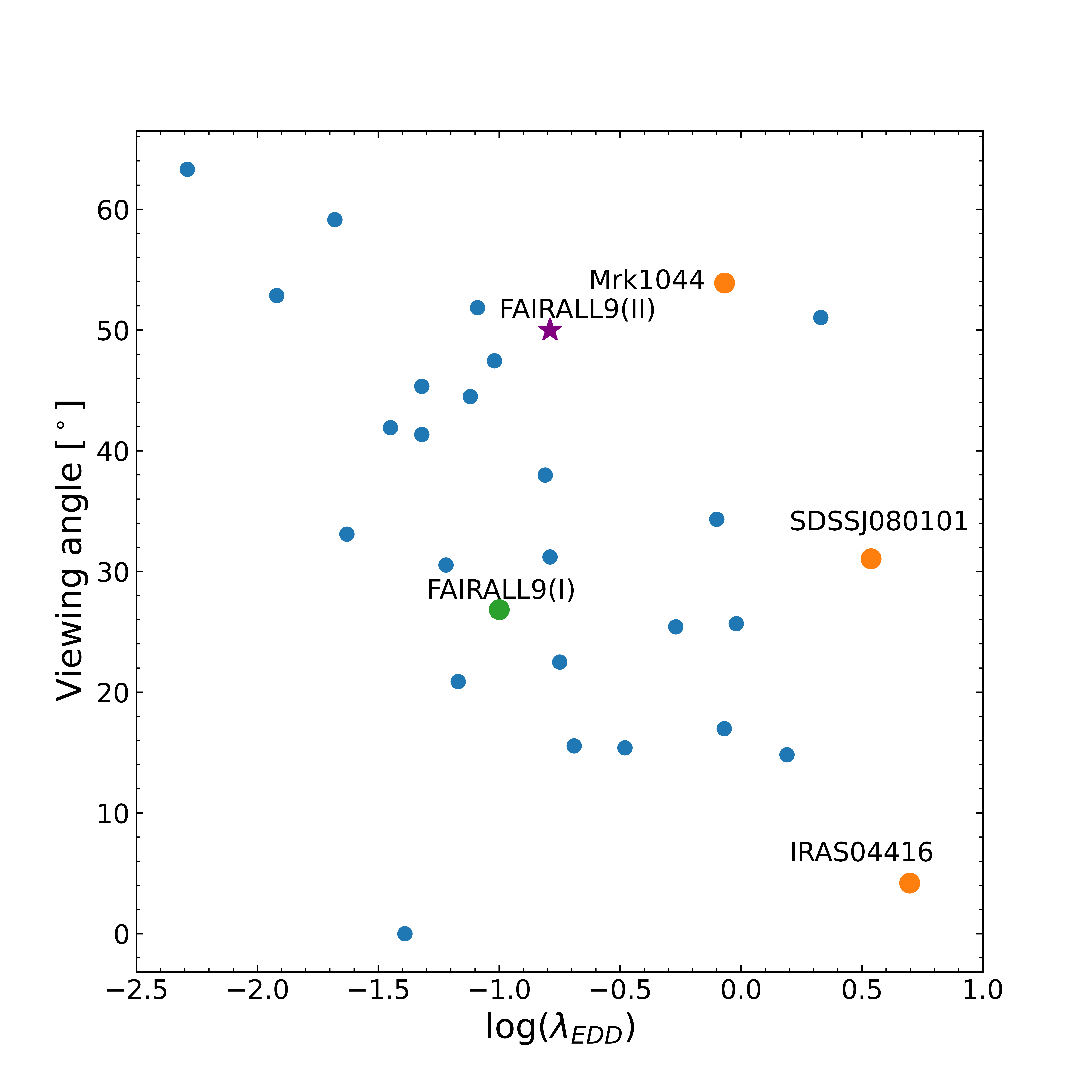}
    \includegraphics[width=0.75\columnwidth]{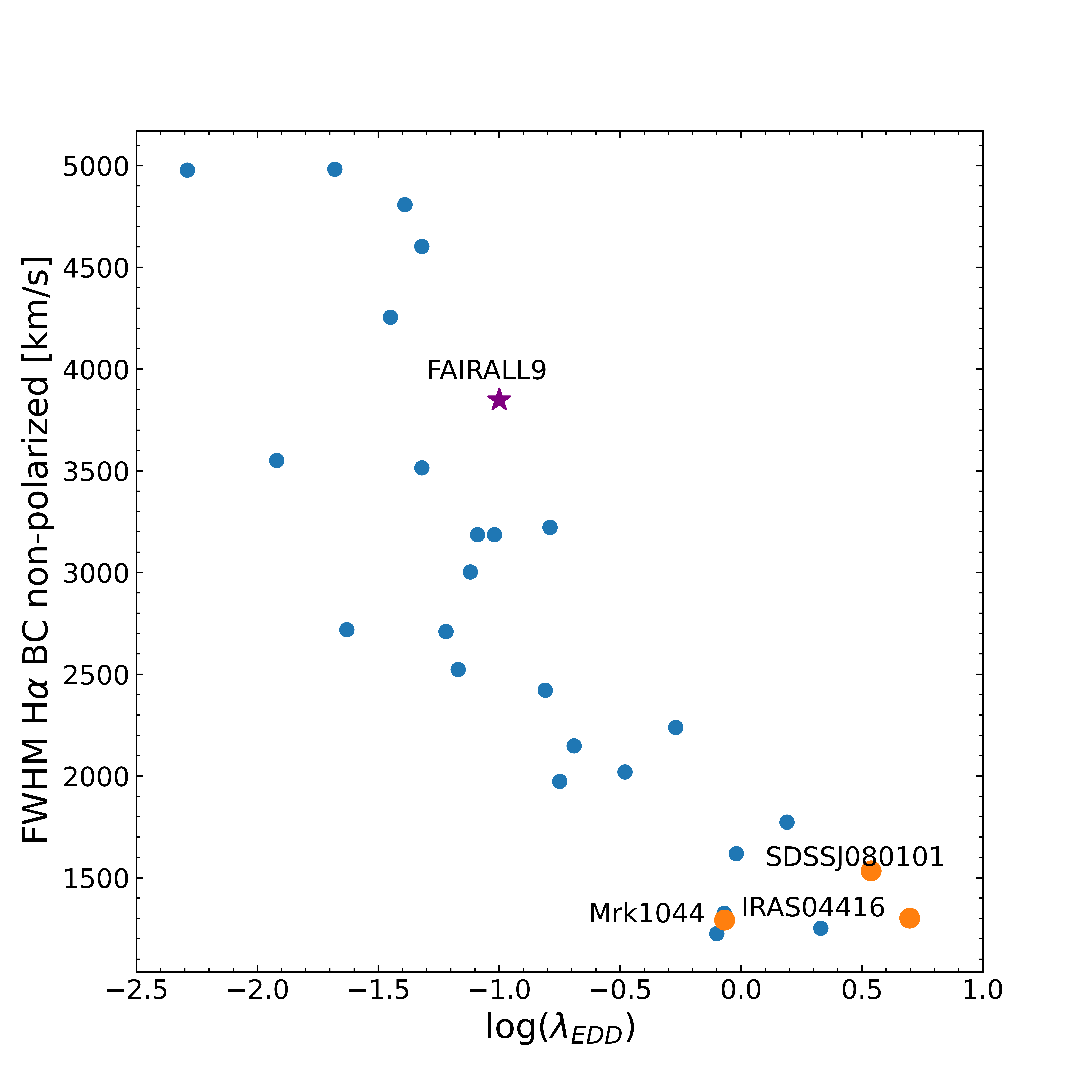}
    \includegraphics[width=0.75\columnwidth]{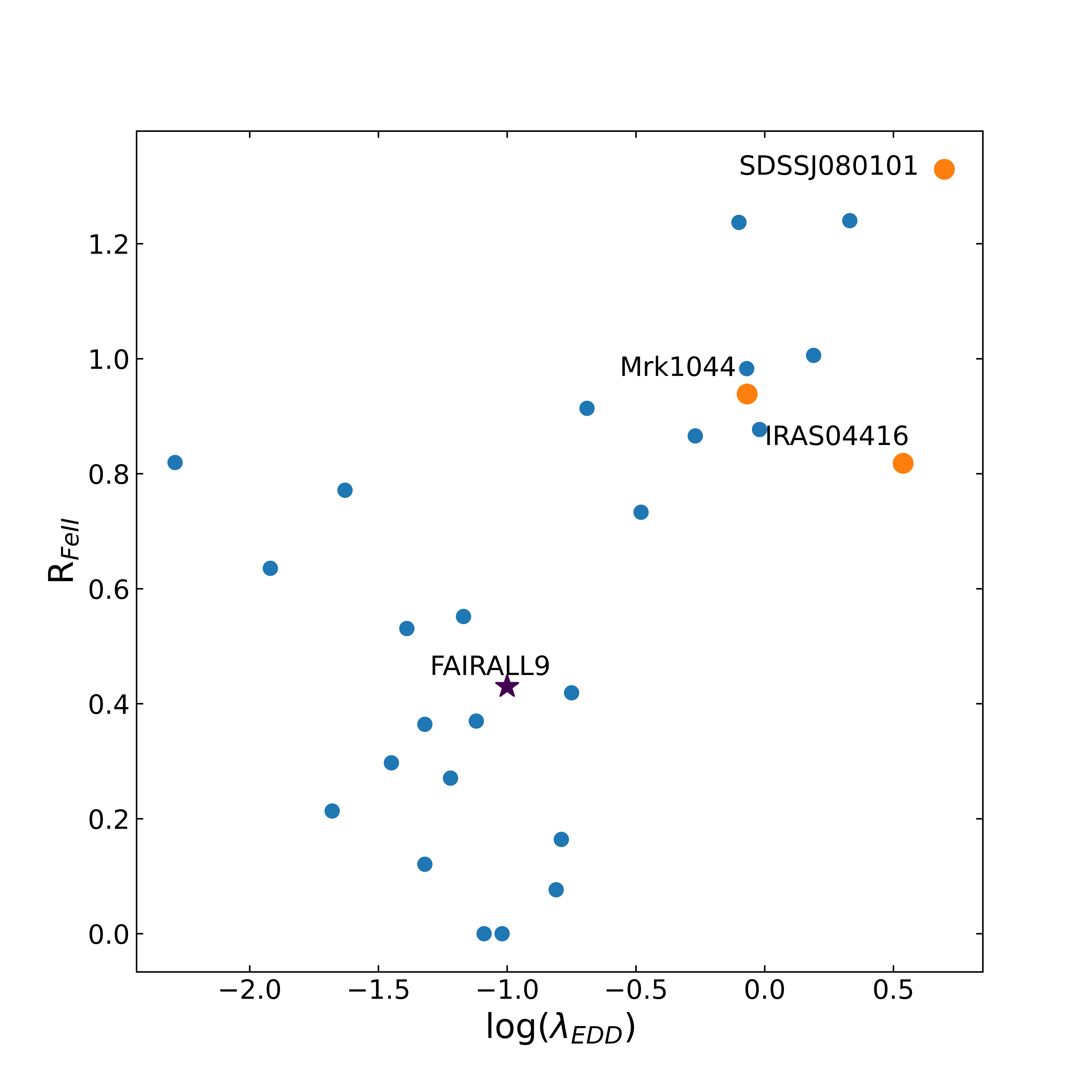}
    \caption{Corollary plots for Figure \ref{fig:pblr-to-ledd}.}
    \label{fig:alternate-figures}
\end{figure}

%%%%%%%%%%%%%%%%%%%%%%%%%%%%%%%

\end{document}